\documentclass[aps,showpacs,superscriptadress,preprint,showkeys]{revtex4}
\usepackage{graphicx}
\usepackage{amsmath}
\usepackage{amssymb}
\usepackage{mathrsfs}
\usepackage{stmaryrd}
\usepackage{amsthm}
\usepackage{bbm}
\usepackage{hyperref}
\usepackage{multirow}
\hypersetup{hypertex=true,
	colorlinks=true,
	linkcolor=red,
	anchorcolor=green,
	citecolor=blue}
\usepackage{caption}
\usepackage{tikz,xcolor,hyperref}

\definecolor{lime}{HTML}{A6CE39}
\DeclareRobustCommand{\orcidicon}{
\begin{tikzpicture}
\draw[lime, fill=lime] (0,0)
circle[radius=0.18]
node[white]{{\fontfamily{qag}\selectfont \tiny \.{I}D}}; 
\end{tikzpicture}
\hspace{-3mm}
}
\foreach \x in {A, ..., Z}{%
\expandafter\xdef\csname orcid\x\endcsname{\noexpand\href{https://orcid.org/\csname orcidauthor\x\endcsname}{\noexpand\orcidicon}}
}

\begin{document}
\title{Constraints on Shadow Radius and Instabilities Arising from Various Perturbations in Spherically Symmetric Black Holes in Einstein-power-Yang-Mills-Gauss-Bonnet Gravity}

\author{
Zening Yan$^{1,2}$\footnote{Corresponding author: z.n.yan.bhtpr@gmail.com or znyan21@m.fudan.edu.cn}\hspace{-2mm}\orcidA{}
}

\affiliation{
\small 1. Center for Field Theory and Particle Physics \& Department of Physics, Fudan University, No.2005 Songhu Road, Yangpu District, Shanghai 200438, China\\
\small 2. Center for Astronomy and Astrophysics, Fudan University, No.2005 Songhu Road, Yangpu District, Shanghai 200438, China\\
}

\date{October 13, 2024}

\begin{abstract}
The space-time geometry under investigation is chosen to be a high-dimensional, static, spherically symmetric solution in an asymptotically flat background within the Einstein-power-Yang-Mills-Gauss-Bonnet (EPYMGB) gravity.
To address the limitations of previous shadow constraints, we construct a standardized framework based on the Schwarzschild-Tangherlini metric to constrain the characteristic parameters.
Additionally, we provide a rigorous derivation of the shadow radius formula for a general high-dimensional spherically symmetric black hole.
Subsequently, we systematically and comprehensively present the effective potential equations governing spin-0, spin-1, $p$-form, and spin-2 perturbations in high-dimensional static spherically symmetric flat space-time.
Our analysis reveals that the Yang-Mills magnetic charge $\mathcal{Q}$ and the power $q$ have a negligible impact on both the shadow radius and perturbations of the black hole when compared to the Gauss-Bonnet coupling constant ${\alpha}_2$ in various dimensions.
Hence, the physical signatures of the parameters $\mathcal{Q}$ and $q$ in the black hole environment remain undetectable through either perturbation analysis or shadow observations.
Cross-validation of the allowable range of ${\alpha}_2$ derived from the high-dimensional constraint on shadow radius and the dynamical stability analysis of gravitational perturbations demonstrates excellent agreement between these independent approaches.
The conclusions derived from the cross-analysis further substantiate the validity of the high-dimensional shadow constraint formula proposed in this work and indicate that the duality between black hole shadows and perturbations may persist in high-dimensional space-times.
\end{abstract}



\maketitle

\tableofcontents

\section{Introduction}
The theoretical foundations for detecting black hole shadow imaging were established in 2000 through the pioneering work of Falcke et al. \cite{Falcke:1999pj}.
These advancements culminated in 2017 when the Event Horizon Telescope (EHT) Collaboration institution achieved a historic milestone by capturing the first resolved images of the supermassive black hole $\mathrm{M87}^{\star}$ \cite{EventHorizonTelescope:2019dse}.
In 2020, the EHT Collaboration showed that the measured shadow size provides a powerful test of Einstein's theory in the strong-field regime and enables quantitative constraints on alternative gravitational theories \cite{EventHorizonTelescope:2020qrl}.
Based on the successful observation of $\mathrm{M87}^{\star}$, the EHT further revealed in May 2022 high-resolution visuals of the supermassive black hole Sagittarius $\mathrm{A}^{\star}$ ($\mathrm{Sgr~A}^{\star}$), located at the center of the Milky Way \cite{EventHorizonTelescope:2022wkp}.
And they further confirmed that the external space-times of all black holes are described by the Kerr metric, which depends on the mass and spin of the black hole \cite{EventHorizonTelescope:2022xqj}.
Afterwards, the EHT successively unveiled polarized views of $\mathrm{M87}^{\star}$ \cite{EventHorizonTelescope:2021bee} and $\mathrm{Sgr~A}^{\star}$ \cite{EventHorizonTelescope:2024rju} pertaining to the magnetic field surrounding them.
In May 2022, Jordy Davelaar and Zolt\'an Haiman explored shadow imaging under the influence of self-lensing formed by binary black hole systems \cite{Davelaar:2021eoi, Davelaar:2021gxx}.
In April 2023, an international team of scientists led by Ru-Sen Lu released a panoramic image capturing both the black hole shadow and its jet in the Messier 87 (M87) galaxy.
This was achieved by combining observations from the Global Millimeter VLBI Array (GMVA), the Atacama Large Millimeter/submillimeter Array (ALMA) and the Greenland Telescope (GLT) \cite{Lu:2023bbn}.
Concurrently, Lia Medeiros and collaborators introduced a new technique called Principal-component Interferometric Modeling (PRIMO), which enhances EHT image data to reconstruct a sharper shadow image of the black hole \cite{Medeiros:2023pns}.
In August 2024, the EHT achieved the highest diffraction-limited angular resolution ever attained on Earth by operating at a frequency of 345 $\text{GHz}$.
This advancement enabled the capture of significantly clearer shadow images of black holes.
It is anticipated that the new results will enhance the clarity of the image by $50\%$.
In September 2025, the new image from the EHT Collaboration revealed a dynamic environment with changing polarization patterns near the $\mathrm{M87}^{\star}$ black hole's magnetic field from 2017 to 2021, providing new insight into how matter and energy behave in the extreme environments near black holes \cite{EventHorizonTelescope:2025vum}.
Furthermore, the EHT Collaboration has established observationally constrained confidence intervals for the shadow radii of black holes $\mathrm{M87}^{\star}$ and $\mathrm{Sgr~A}^{\star}$.
These measurements provide critical empirical benchmarks for testing modified gravity theories, enabling quantitative constraints on space-time metric parameters across multiple theoretical frameworks---including the Reissner-Nordstr\"{o}m (RN), Kerr, Bardeen, Hayward, Frolov, Janis-Newman-Winicour (JNW), Kazakov-Solodhukin (KS), and Einstein-Maxwell-dilaton (EMd-1) \cite{EventHorizonTelescope:2021dqv, EventHorizonTelescope:2020qrl, EventHorizonTelescope:2022xqj}.
Subsequently, the characteristic parameters of many other black holes have been further constrained \cite{Khodadi:2022pqh, Vagnozzi:2022moj, Yan:2023pxj, Yan:2025lzc, Uniyal:2022vdu, Pantig:2022qak, Yan:2024rsx, Xavier:2023exm}.

One of the most important characteristics of compact stars is their quasinormal modes (QNMs), which govern the evolution of perturbations surrounding these objects.
This characteristic oscillation does not originate from the matter inside the black hole but instead primarily depends on the space-time geometry outside the event horizon.
This indicates that space-time itself actively participates in the oscillation process.
This perturbation has now been experimentally confirmed by the Laser Interferometer Gravitational-Wave Observatory (LIGO) Scientific Collaboration and Virgo Collaboration in their observations of gravitational waves (dubbed GW150914) from the merger of two black holes \cite{LIGOScientific:2016aoc}.
Beyond gravitational perturbations, our analysis extends to encompass scalar, electromagnetic and $p$-form perturbations.
These characteristic oscillations are defined through boundary conditions requiring complex frequencies to simultaneously satisfy purely outgoing waves at infinity and purely ingoing waves at the event horizon.
The characteristic oscillation frequencies of black holes typically cannot be derived using analytical methods; it requires numerical or semi-analytical approaches for their determination.
Numerous numerical techniques are currently employed for computing QNMs, including the Wentzel-Kramers-Brillouin (WKB) approximation, the asymptotic iteration method (AIM), the time-domain integration technique and the matrix method, among others \cite{Berti:2009kk, Konoplya:2011qq}.
So far, numerous studies have investigated black hole perturbations, with particular attention given to those in high-dimensional space-times, which have been examined systematically and comprehensively \cite{Konoplya:2007jv, Kodama:2003jz, Ishibashi:2003ap, Kodama:2003kk, Konoplya:2007jv, Konoplya:2008au, Konoplya:2008rq, Kodama:2009rq, Chabab:2016cem, Chakraborty:2017qve, Witek:2010xi, Konoplya:2008ix, Shibata:2010wz}.

Throughout this study, all discussions are based on the exact solution of a high-dimensional, static, spherically symmetric black hole in an asymptotically flat space-time within the framework of Einstein-Gauss-Bonnet gravity theory coupled with power-Yang-Mills fields.
This exact solution, referred to as the Einstein-power-Yang-Mills-Gauss-Bonnet (EPYMGB) black hole, was originally discovered by Mazharimousavi and Halilsoy \cite{Mazharimousavi:2009mb}. 
Given that Einstein's general relativity fails to constitute a complete gravitational description, particularly evident in phenomena observed in extreme gravitational regimes surrounding compact celestial objects, theoretical extensions beyond the standard formulation become imperative.
Gravitational theories incorporating higher-order curvature terms, which emerge from the low-energy limit of string theories, present particularly promising candidates for frameworks beyond Einstein's theory.
A concrete realization is found in the quadratic curvature correction augmenting the Einstein-Hilbert action, leading to a specific case of Lovelock theory.
This modified theory is applicable to higher-dimensional space-times.
This study investigates the nonlinear power extension of Maxwell's theory, incorporating a power Yang-Mills source of the form $\left(F_{\nu \lambda}{}{}^{(a)} F^{\mu \lambda (a)}\right)^q$.
Here, $F_{\nu \lambda}{}{}^{(a)}$ denotes the field strength of a non-Abelian Yang-Mills field and $q$ is an arbitrary positive real number that serves as the scaling exponent.
This system constitutes a highly nonlinear theory: Einstein's inherently nonlinear theory of gravity is coupled with the power-Yang-Mills theory, which itself exhibits nonlinear characteristics, and further combined with nonlinear Gauss-Bonnet terms (i.e., second-order Lovelock).
This sophisticated nonlinear system nevertheless preserves the fundamental second-order structure of the field equations while admitting exact EPYMGB black hole solutions within its geometric framework.
When $q=1$, the solution reduces to the Einstein-Yang-Mills (EYM) black hole solution with the $SO(n-1)$ gauge symmetry, whereas in the absence of charge, it reduces to the Einstein-Gauss-Bonnet (EGB) black hole solution.
Since then, the solutions for four-dimensional EPYMGB black holes in asymptotically flat space-time have been obtained, prompting a series of subsequent investigations \cite{Zubair:2023cep}.
Some papers have systematically investigated the solutions of EYM black holes \cite{Li:2021qim, Rincon:2023hvd, Soroushfar:2025yhr}.
Moreover, a substantial body of research currently concentrates on the solutions of EGB black holes and their derivative solutions, encompassing areas such as thermodynamics, perturbations, geodesics, and related fields \cite{Blazquez-Salcedo:2016enn, Pani:2009wy, Blazquez-Salcedo:2017txk, Cuyubamba:2016cug, Cuyubamba:2020moe, Chowdhury:2022zqg, Konoplya:2020bxa, Witek:2018dmd, Konoplya:2017zwo, Chen:2015fuf, Gonzalez:2017gwa, Deppe:2016dcr, Deppe:2014oua}.

The primary motivations driving this study are outlined below:
\begin{enumerate}
\item[$\bullet$]
We have observed that some studies incorrectly apply the constraint formula for the four-dimensional black hole shadow radius derived from Schwarzschild modeling to certain $n$-dimensional black hole solutions, which is obviously inappropriate \cite{Nozari:2024jiz, Nozari:2023flq}.
Consequently, a principal objective of this study is to derive a shadow radius constraint formula for high-dimensional black holes within the framework of Schwarzschild-Tangherlini metric extensions, thereby establishing a standardized framework for characteristic parameter constraints in high-dimensional black hole solutions.
\item[$\bullet$]
Previous studies on high-dimensional black hole solutions in EGB gravity have indicated that large values of the Gauss-Bonnet coupling constant ${\alpha}_2$ may lead to dynamical instabilities in perturbations \cite{Kanti:2005xa, Beroiz:2007gp, Gleiser:2005ra, Dotti:2005sq}.
This work aims to further investigate this issue by determining explicit and admissible numerical ranges for the parameter ${\alpha}_2$ in each space-time dimension.
Moreover, we will systematically examine the stability of various perturbations while analyzing how the parameter $p$ affects the system's behavior under $p$-form perturbations.
\item[$\bullet$]
This investigation aims to establish rigorous constraints for the Gauss-Bonnet coupling constant ${\alpha}_2$ and the Yang-Mills magnetic charge $\mathcal{Q}$ through the synergistic application of shadow radius analysis and dynamical stability evaluation techniques.
\item[$\bullet$]
This study seeks to systematically evaluate the accuracy and applicability of various methods for computing QNMs, with a particular emphasis on quantitatively analyzing how the Gauss-Bonnet coupling constant ${\alpha}_2$, the Yang-Mills magnetic charge $\mathcal{Q}$ and the power $q$ affect both the oscillation frequencies (real part) and damping rates (imaginary part) of QNM spectra.
\end{enumerate}

This paper is organized as follows:
In Section $\mathrm{\uppercase\expandafter{\romannumeral2}}$, we introduce EPYMGB gravity and its corresponding high-dimensional static spherically symmetric black hole solution.
In Section $\mathrm{\uppercase\expandafter{\romannumeral3}}$, we develop a rigorous derivation of the general formula for the shadow radius of a high-dimensional static spherically symmetric black hole, while constructing a universal and standardized framework for applying shadow radius constraints.
In Section $\mathrm{\uppercase\expandafter{\romannumeral4}}$, we elaborate on the effective potential equations for various perturbations in high-dimensional static spherically symmetric flat space-time, while providing a concise overview of three prevalent computational methodologies for determining QNMs.
In Section $\mathrm{\uppercase\expandafter{\romannumeral5}}$, we present a series of numerical results derived from rigorous computational analysis.
In Section $\mathrm{\uppercase\expandafter{\romannumeral6}}$, we provide a comprehensive systematic summary of the complete manuscript.

Throughout this paper, we adopt the metric signature $\left(-,+,+,\cdots,+\right)$ and employ geometrized units such that $\mathscr{G}=c=1$ in each space-time dimension.

\section{Einstein-power-Yang-Mills-Gauss-Bonnet gravity and its solutions}

\subsection{Action and field equations}
The action of Einstein-Gauss-Bonnet gravity with the power-Yang-Mills field and a cosmological constant $\Lambda$ in the high-dimensional space-time background is as follows:
\begin{equation}
\mathcal{S}=\frac{1}{2} \int_{\mathcal{M}} \mathrm{d}^n x \sqrt{-g}\left[\mathcal{R}-\frac{(n-2)(n-1)}{3} \Lambda + {\alpha}_2 \mathcal{L}_{\text{GB}} + \mathcal{L}_{\text{PYM}}\right],
\end{equation}
where $8 \pi \mathscr{G}=1$, the parameter ${\alpha}_2$ is the Gauss-Bonnet coupling constant \cite{Cai:2001dz}, and the Gauss-Bonnet Lagrangian $\mathcal{L}_{\text{GB}}$ is given as
\begin{equation}
\mathcal{L}_{\text{GB}}=\mathcal{R}_{\gamma \zeta \varpi \delta} \mathcal{R}^{\gamma \zeta \varpi \delta}-4 \mathcal{R}_{\gamma \zeta} \mathcal{R}^{\gamma \zeta}+\mathcal{R}^{2}.
\end{equation}
Here, the Lagrangian density of the power-Yang-Mills field \cite{Mazharimousavi:2009mb, Ali:2020omz} is given by
\begin{equation}
\mathcal{L}_{\text{PYM}}=- \left(\mathcal{F}\right)^q,
\end{equation}
where the associated Yang-Mills invariant $\mathcal{F}$ is represented as
\begin{equation}
\mathcal{F}=\sum_{a=1}^{\frac{(n-1)(n-2)}{2}} \operatorname{Tr}\Big(F_{\gamma \zeta}{}{}^{(a)} F^{\gamma \zeta (a)}\Big).
\end{equation}
$\mathcal{R}_{\gamma \zeta \varpi \delta}$, $\mathcal{R}_{\gamma \zeta}$ and $\mathcal{R}$ are the Riemann tensor, Ricci tensor and Ricci scalar, respectively.
The power $q$ is chosen to be an arbitrary positive real number because the energy conditions and the causality conditions will be violated when $q<0$ \cite{Mazharimousavi:2009mb}.
The expression of the Yang-Mills field is defined as
\begin{equation}
\mathbf{F}^{(a)}=\mathbf{d} \mathbf{A}^{(a)}+\frac{1}{2 \sigma} C_{(b)(c)}^{(a)} \mathbf{A}^{(b)} \wedge \mathbf{A}^{(c)},
\end{equation} 
or
\begin{equation}
F_{\gamma \zeta}{}{}^{(a)}=\partial_{\gamma} A_{\zeta}{}^{(a)}-\partial_{\zeta} A_{\gamma}{}^{(a)}+\frac{1}{2 \sigma} C_{(b)(c)}^{(a)} A_{\gamma}{}^{(b)} A_{\zeta}{}^{(c)},
\end{equation}
where $\mathbf{A}^{(a)}$ or $A_{\zeta}{}^{(a)}$ are the Yang-Mills potentials of the gauge group $SO(n-1)$, $\sigma$ is the coupling constant and $C_{(b)(c)}^{(a)}$ are interpreted as the structure constants of $\frac{(n-1)(n-2)}{2}$-parameter Lie group.
Note that in the construction of high-dimensional spherically symmetric analytical solutions, the gauge group is generally chosen as $SO(n-1)$ to ensure compatibility with the space-time symmetry.
The field equation of the space-time metric $g_{\mu \nu}$ and the gauge potential $A_{\zeta}{}^{(a)}$ after the variation of action can be explicitly expressed as
\begin{equation}\label{eq:fe}
\begin{aligned}
\mathbf{d} \Big({ }^{\mathbf{\star}} \mathbf{F}^{(a)} \mathcal{F}^{q-1}\Big)+\frac{1}{\sigma} C_{(b)(c)}^{(a)} \mathcal{F}^{q-1} \mathbf{A}^{(b)} \wedge { }^{\mathbf{\star}} \mathbf{F}^{(c)}  &= 0, \\
\mathcal{G}_{\mu \nu}+\alpha_{2} \mathcal{H}_{\mu \nu}+\frac{(n-2)(n-1)}{6} \Lambda g_{\mu \nu} &= \mathcal{T}_{\mu \nu},
\end{aligned}
\end{equation}
where the Hodge star operator $\star$ denotes duality and the symbol $\wedge$ stands for wedge product.
$\mathcal{G}_{\mu \nu}$, $\mathcal{H}_{\mu \nu}$ and $\mathcal{T}_{\mu \nu}$ are Einstein, Lanczos and stress-energy tensors, respectively. 
And their specific expressions are given as follows:
\begin{equation}
\begin{aligned}
\mathcal{G_{\mu \nu}} &= \mathcal{R}_{\mu \nu}-\frac{1}{2} g_{\mu \nu} \mathcal{R},  \\
\mathcal{H_{\mu \nu}} &= 2\left( -\mathcal{R}_{\mu \sigma \kappa \tau} \mathcal{R}^{\kappa \tau \sigma}{ }{ }{ }_{\nu} - 2 \mathcal{R}_{\mu \rho \nu \sigma} \mathcal{R}^{\rho \sigma} -2 \mathcal{R}_{\mu \sigma} \mathcal{R}^{\sigma}{ }_{\nu} + \mathcal{R} \mathcal{R}_{\mu \nu} \right)-\frac{1}{2} g_{\mu \nu} \mathcal{L}_{\text{GB}},  \\
\end{aligned}
\end{equation}
and
\begin{equation}\label{eq:tmunu}
\mathcal{T}^{\mu}{ }_{\nu} = -\frac{1}{2} \delta^{\mu}{ }_{\nu} \mathcal{F}^{q} + 2q \operatorname{Tr}\Big(F_{\nu \lambda}{}{}^{(a)} F^{\mu \lambda (a)}\Big) \mathcal{F}^{q-1}.
\end{equation}

Here the higher-dimensional version of the Wu-Yang ansatz \cite{Mazharimousavi:2008ap, HabibMazharimousavi:2008ib} will be utilized so that the Yang-Mills magnetic gauge potential is written as
\begin{equation}
\mathbf{A}^{(a)}=\frac{\mathcal{Q}}{r^{2}} C_{(i)(j)}^{(a)} \chi^{i} \mathrm{d} \chi^{j},
\end{equation}
where 
\begin{equation}
\begin{array}{ll}
\chi_1=r \cos \theta_{n-3} \sin \theta_{n-4} \cdots \sin \theta_1, \\ \chi_2=r \sin \theta_{n-3} \sin \theta_{n-4} \cdots \sin \theta_1, \\
\chi_3=r \cos \theta_{n-4} \sin \theta_{n-5} \cdots \sin \theta_1, \\ \chi_4=r \sin \theta_{n-4} \sin \theta_{n-5} \cdots \sin \theta_1, \\
\vdots & \\
\chi_{n-2}=r \cos \theta_1,
\end{array}
\end{equation}
the parameter $\mathcal{Q}$ represents the Yang-Mills magnetic charge, and $r^{2}=\sum_{i=1}^{n-1} {\chi_{i}}^{2}$.
Moreover, $2 \leq j+1 \leq i \leq n-1$ and $1\leq a \leq \frac{(n-1)(n-2)}{2}$ are the ranges of indices $i$, $j$ and $a$.
Due to the normalization freedom of the gauge field, the Yang-Mills coupling constant $\sigma$ can be consistently absorbed into the magnetic charge through the redefinition $\mathcal{Q}\rightarrow{\mathcal{Q}}/{\sigma}$, such that the subsequent metric functions depend solely on the rescaled charge parameter.

These hypotheses are proved to satisfy the Yang-Mills equation, and the energy-momentum tensor \eqref{eq:tmunu} can be derived by utilizing
\begin{equation}
\mathcal{F}=\frac{(n-2)(n-3)\mathcal{Q}^2}{r^{4}},
\end{equation}
\begin{equation}
\operatorname{Tr}\Big(F_{\theta_{i} \lambda}{}{}^{(a)} F^{\theta_{i} \lambda (a)}\Big)=\frac{1}{(n-2)}\mathcal{F}=\frac{(n-3)\mathcal{Q}^2}{r^{4}}.
\end{equation}
Therefore, the energy-momentum tensor is given by
\begin{equation}
\mathcal{T}^{a}{}_{b}  =-\frac{1}{2} \mathcal{F}^{q} \; \operatorname{diag} \Bigg[1, 1, \left( 1- \frac{4q}{n-2} \right), \left( 1- \frac{4q}{n-2} \right), \cdots , \left( 1- \frac{4q}{n-2} \right) \Bigg],
\end{equation}
and its trace is $-\frac{1}{2} \mathcal{F}^{q} (n-4q)$.

\subsection{Black hole space-time solution}
The metric for high-dimensional ($n\geq5$) static spherically symmetric is ansatz \cite{Mazharimousavi:2009mb, Ali:2020omz} 
\begin{equation}\label{eq:metric1}
g_{\mu \nu} \mathrm{d}x^{\mu} \mathrm{d}x^{\nu} = -f(r) \mathrm{d}t^2 + \frac{1}{f(r)}\mathrm{d}r^2 + r^2 \sum_{i=2}^{n-2} \mathfrak{h}_{ij} \mathrm{d}x^i \mathrm{d}x^j,
\end{equation}
where $\mathfrak{h}_{ij} \mathrm{d}x^i \mathrm{d}x^j$ represents the line element of a $(n-2)$-dimensional hypersurface with the volume $\Sigma_{(n-2)}$.
The space-time is higher dimensional and isotropic, characterized by a spherical horizon topology $\mathbb{S}^{n-2}$.
The solution for a EPYMGB black hole in asymptotically flat space-time can be written as follows:
\begin{equation}\label{eq:lapsefunc1}
f_{\pm}(r) = \left\{\begin{array}{ll}
1+\frac{r^2}{2 \tilde{\alpha}_2}\left(1 \pm \sqrt{1+\frac{16 \mathfrak{M} \tilde{\alpha}_2}{(n-2) r^{n-1}}+\frac{4 \tilde{\alpha}_2 Q_1}{r^{4q}}}\right),  & q\neq\frac{n-1}{4}; \\
1+\frac{r^2}{2 \tilde{\alpha}_2}\left(1 \pm \sqrt{1+\frac{16 \mathfrak{M} \tilde{\alpha}_2}{(n-2) r^{n-1}}+\frac{4 \tilde{\alpha}_2 Q_2 \ln (r)}{r^{n-1}}}\right),  & q=\frac{n-1}{4}, \\
\end{array}\right. 
\end{equation}
where $\tilde{\alpha}_2=(n-3)(n-4) {\alpha}_2$,
\begin{equation}\label{eq:q1q2}
\begin{aligned}
Q_1 &= \frac{\big[\left(n-3\right)\left(n-2\right) \mathcal{Q}^2\big]^q}{\left(n-2\right)\left(n-1-4q\right)}, \\
Q_2 &= \frac{\big[\left(n-3\right) \left(n-2\right) \mathcal{Q}^2\big]^{\frac{\left(n-1\right)}{4}}}{\left(n-2\right)},
\end{aligned}
\end{equation}
and $\mathfrak{M}=\frac{\mathsf{M}}{2\Sigma_{(n-2)}}$.
The symbol $\mathfrak{M}$ is the integration constant expressed in terms of mass-dependent parameter, the parameter $\mathsf{M}$ represents the Arnowitt-Deser-Misner (ADM) mass of the black hole, and the formula for $\Sigma_{(n-2)}$ is defined as $\Sigma_{(n-2)}=\frac{2 \pi^{(n-1)/2}}{\Gamma\left(\frac{n-1}{2}\right)}$.

Since both the energy conditions and the causality conditions will be violated when $q<0$, the value of $q$ is determined to be $q>0$,
Furthermore, when $q$ is selected within the range $\frac{n-1}{4} \leqslant q < \frac{n-1}{2}$, the Weak Energy Condition (WEC), Strong Energy Condition (SEC), Dominant Energy Condition (DEC) and Causality Condition (CC) will all be satisfied.
Moreover, if $q$ is restricted to integers, specific $q$ values corresponding to different dimensions will be obtained, as shown in Table \ref{tb1:dq}.

\begin{table}[tbh]\centering
\caption{The allowed integer values for $q$ corresponding to different dimensions $n$.
\vspace{0.2cm}} \label{tb1:dq}
\begin{tabular*}{16cm}{*{8}{c @{\extracolsep\fill}}}
\hline
$n$ & \textbf{5} & 6 & 7 & 8 & \textbf{9} & 10 & 11 \\
\hline
$q$ & \textbf{1} & 2 & 2 & 2, 3 & \textbf{2}, 3 & 3, 4 & 3, 4 \\
\hline
\end{tabular*}
\end{table}

The branch ``$+$'' is unstable for small perturbations and generates a graviton ghost, so only one branch is physical \cite{Charmousis:2002rc, Boulware:1985wk}.
When considering the branch ``$-$'' within the effective range of $q$ and recovering $8 \pi \mathscr{G}$, the solution can be rewritten as 
\begin{equation}\label{eq:lapsefunc2}
f_{-}(r) = \left\{\begin{array}{ll}
1+\frac{r^2}{2 \tilde{\alpha}_2}\left(1 - \sqrt{1+\frac{64 \pi \mathscr{G} \mathsf{M} \tilde{\alpha}_2}{(n-2) \Sigma_{(n-2)} r^{n-1}}+\frac{4 \tilde{\alpha}_2 Q_1}{r^{4q}}}\right),  & \frac{n-1}{4} < q < \frac{n-1}{2}; \\
1+\frac{r^2}{2 \tilde{\alpha}_2}\left(1 - \sqrt{1+\frac{64 \pi \mathscr{G} \mathsf{M} \tilde{\alpha}_2}{(n-2) \Sigma_{(n-2)} r^{n-1}}+\frac{4 \tilde{\alpha}_2 Q_2 \ln (r)}{r^{n-1}}}\right),  & q=\frac{n-1}{4}. \\
\end{array}\right.
\end{equation}
Here, we name the solution corresponding to the condition $q=\frac{n-1}{4}$ as the special configuration and the solution corresponding to the condition $\frac{n-1}{4} < q < \frac{n-1}{2}$ as the general configuration.
Boldface in Table \ref{tb1:dq} highlights the integer case of the special configuration.

The Schwarzschild-Tangherlini solution \cite{Emparan:2008eg} is a static, spherically symmetric, strict vacuum solution in high-dimensional space-time as follows:
\begin{equation}\label{eq:sch-tang}
\mathsf{f}(r)=1-\frac{16 \pi \mathscr{G} \mathsf{M}}{(n-2) \Sigma_{(n-2)} r^{n-3}}.
\end{equation}
This metric is the generalization of Schwarzschild space-time in higher dimensions cases, which is proved to be Ricci-flat.
When considering an uncharged black hole, it can be observed that the EPYMGB solution \eqref{eq:lapsefunc2} can be transformed into the Schwarzschild-Tangherlini solution \eqref{eq:sch-tang} by applying the limit condition ${\alpha}_2\rightarrow0$.
When no limit condition is imposed on ${\alpha}_2$, the uncharged EPYMGB solution \eqref{eq:lapsefunc2} is the EGB solution \cite{Cai:2001dz} in asymptotically flat space-time, which is
\begin{equation}\label{eq:egb}
f(r)_{\mathrm{EGB}} = 1+\frac{r^2}{2 \tilde{\alpha}_2} \left(1 - \sqrt{1+\frac{64 \pi \mathscr{G} \mathsf{M} \tilde{\alpha}_2}{(n-2) \Sigma_{(n-2)} r^{n-1}} }\right).
\end{equation}
When the limit ${\alpha}_2\rightarrow0$ is set and $q=1$ is selected, the EPYMGB solution \eqref{eq:lapsefunc2} will be reduced to the EYM solution \cite{Guo:2020caw} as follows:
\begin{equation}\label{eq:eym2}
f(r)_{\mathrm{EYM}} = \left\{\begin{array}{ll}
1 - \frac{8 \mathscr{G} \mathsf{M}}{3 \pi r^{2}}-\frac{2\mathcal{Q}^{2} \ln(r)}{r^{2}},  & (n=5); \\
1 - \frac{16 \pi \mathscr{G} \mathsf{M}}{(n-2) \Sigma_{(n-2)} r^{n-3}} - \frac{(n-3) \mathcal{Q}^{2}}{(n-5) r^{2}},  & (n>5). \\
\end{array}\right.
\end{equation}

In order for the EPYMGB black hole solution to have at least one event horizon, that is, to ensure that the black hole does not expose a naked singularity, the values of the parameters ${\alpha}_2$ and $\mathcal{Q}$ are restricted to a fixed region, as shown in Figures \ref{figure2} and \ref{figure3}.
It is worth noting that in the process of numerical calculation of this fixed region, the specific conditions of the metric \eqref{eq:lapsefunc1} should also be considered as follows:
\begin{equation}\label{eq:sconditions}
\begin{aligned}
&1+\frac{16 \mathfrak{M} \tilde{\alpha}_2}{(n-2) r^{n-1}}+\frac{4 \tilde{\alpha}_2 Q_1}{r^{4q}} \geqslant 0, \\
&1+\frac{16 \mathfrak{M} \tilde{\alpha}_2}{(n-2) r^{n-1}}+\frac{4 \tilde{\alpha}_2 Q_2 \ln (r)}{r^{n-1}} \geqslant 0. \\
\end{aligned}
\end{equation}
According to Eq. \eqref{eq:q1q2}, we can know $Q_1 < 0$ and $Q_2 > 0$.
Since the parameters $\mathfrak{M}$, $n$, $q$ and $\tilde{\alpha}_2$ are all positive real numbers, the second condition in Eq. \eqref{eq:sconditions} must be true when $r > 1$, and we only need to pay attention to satisfying the first condition while performing numerical calculations.

\section{Black hole shadow formulas and applications}

\subsection{Geodesics and shadows}

\subsubsection{Null geodesics}
For the purpose of derivation in general, the expression of metric \eqref{eq:metric1} is rewritten as
\begin{equation}\label{eq:le}
\mathrm{d}s^2=-h(r)\mathrm{d}t^2+\frac{\mathrm{d}r^2}{f(r)}+k(r)\mathrm{d}\Omega_{n-2}^2,
\end{equation}
where the line element of the $(n-2)$-dimensional unit sphere is
\begin{equation}
\begin{aligned}
\mathrm{d}\Omega_{n-2}^2 = \mathrm{d}\theta_1^2 &+\sin^2\theta_1\mathrm{d}\theta_2^2+\cdots+\left(\sin^2\theta_1\cdots\sin^2\theta_{n-3}\right)\mathrm{d}\theta_{n-2}^2 \\
= \mathrm{d}\theta_1^2 &+\sin^2\theta_1\mathrm{d}\theta_2^2+\cdots+\left( \prod_{i=1}^{n-3}\sin^{2}\theta_{i} \right) \mathrm{d}\theta_{n-2}^2, \\
\end{aligned}
\end{equation}
thus, the non-zero component of the metric tensor can be written as
\begin{equation}\label{eq:gab}
\begin{aligned}
g_{a\:b} = \mathrm{diag} \bigg(& g_{0\:0}, g_{1\:1}, g_{2\:2}, g_{3\:3}, \cdots, g_{\chi+1\:\chi+1}, \cdots, g_{n-1\:n-1} \bigg) \\
= \mathrm{diag} \bigg(& -h(r), f^{-1}(r), k(r), k(r)\sin^2\theta_{1}, \\
&\cdots, k(r) \prod_{i=1}^{\chi-1}\sin^{2}\theta_{i}, \cdots, k(r) \prod_{i=1}^{n-3}\sin^{2}\theta_{i} \bigg).  \\
\end{aligned}
\end{equation}

The Lagrangian equation for a test particle with mass $m$ is 
\begin{equation}
\widetilde{\mathcal{L}}=\frac{1}{2}\left(g_{a\:b}\right){\dot{x}}^a{\dot{x}}^b =  \frac{1}{2} \left(g_{a\:b}\right) \frac{\mathrm{d}x^{a}}{\mathrm{d}\tau}\frac{\mathrm{d}x^{b}}{\mathrm{d}\tau},
\end{equation}
and its explicit expression can be obtained by substituting \eqref{eq:gab} into the Lagrangian equation as follows:
\begin{equation}\label{eq:l2}
\begin{aligned}
\widetilde{\mathcal{L}} = \frac{1}{2} \Bigg[& \left(-h(r)\right) \dot{t}^{2} +  \left(\frac{1}{f(r)}\right) {\dot{r}}^{2} + \left(k(r)\right) {\dot{\theta}_{1}}^{2} + \left(k(r)\sin^2\theta_{1} \right) {\dot{\theta}_{2}}^{2}  \\
&  + \cdots + \left( k(r) \prod_{i=1}^{\chi-1}\sin^{2}\theta_{i} \right) {\dot{\theta}_{\chi}}^{2} + \cdots + \left( k(r) \prod_{i=1}^{n-3}\sin^{2}\theta_{i} \right) {\dot{\theta}_{n-2}}^{2} \Bigg]. \\
\end{aligned}
\end{equation}
The component of canonically conjugate momentum for metric \eqref{eq:le} is defined as 
\begin{equation}\label{eq:paxa}
P_a=\frac{\partial\widetilde{\mathcal{L}}}{\partial\dot{x}^a}=\left( g_{a\:b} \right)\dot{x}^b=\dot{x}_a,
\end{equation}
and its components are specifically expressed as
\begin{equation}\label{eq:p1}
\begin{aligned}
&P_{0}=P_{t}=\frac{\partial\widetilde{\mathcal{L}}}{\partial\dot{x}^0}=\frac{\partial\widetilde{\mathcal{L}}}{\partial\dot{t}}= \left(-h(r)\right) \dot{t} \equiv -\mathrm{E}, \\
&P_{1}=P_{r}=\frac{\partial\widetilde{\mathcal{L}}}{\partial\dot{x}^1}=\frac{\partial\widetilde{\mathcal{L}}}{\partial\dot{r}}=\left(\frac{1}{f(r)}\right) \dot{r}, \\
&P_{2}=P_{{\theta}_{1}}=\frac{\partial\widetilde{\mathcal{L}}}{\partial\dot{x}^2}=\frac{\partial\widetilde{\mathcal{L}}}{\partial\dot{\theta}_{1}}=\left(k(r)\right) \dot{\theta}_{1}, \\
&P_{3}=P_{{\theta}_{2}}=\frac{\partial\widetilde{\mathcal{L}}}{\partial\dot{x}^3}=\frac{\partial\widetilde{\mathcal{L}}}{\partial\dot{\theta}_{2}}=\left(k(r)\sin^2\theta_{1} \right) \dot{\theta}_{2}, \\
&\vdots\\
&P_{\chi+1}=P_{{\theta}_{\chi}}=\frac{\partial\widetilde{\mathcal{L}}}{\partial\dot{x}^{\chi+1}}=\frac{\partial\widetilde{\mathcal{L}}}{\partial\dot{\theta}_{\chi}}=\left( k(r) \prod_{i=1}^{\chi-1}\sin^{2}\theta_{i} \right) \dot{\theta}_{\chi}, \\
&\vdots\\
&P_{n-1}=P_{\theta_{n-2}}=\frac{\partial\widetilde{\mathcal{L}}}{\partial\dot{x}^{n-1}}=\frac{\partial\widetilde{\mathcal{L}}}{\partial\dot{\theta}_{n-2}}=\left( k(r) \prod_{i=1}^{n-3}\sin^{2}\theta_{i} \right) \dot{\theta}_{n-2} \equiv {\L}, \\ 
\end{aligned}
\end{equation}
where $\mathrm{E} \equiv -P_{t}$ and ${\L} \equiv P_{\theta_{n-2}}$ are conserved quantities due to the Killing symmetries, which are interpreted as the energy and angular momentum of the test particle relative to a static observer at infinity, respectively.

The Hamilton-Jacobi method and the Carter constant separable method will be employed to obtain the complete geodesic equation of the test particle.
The Hamilton-Jacobi equation in high-dimensions is expressed as
\begin{equation}\label{eq:HJ}
\frac{\partial S}{\partial\tau}=-\mathscr{H}=-\frac{1}{2} \left( g^{a\:b} \right) \frac{\partial S}{\partial x^\mu}\frac{\partial S}{\partial x^\nu},
\end{equation}
where $S$ is the Jacobi action and the Hamiltonian $\mathscr{H}$ is given by
\begin{equation}
\mathscr{H}=P_a\dot{x}^a-\widetilde{\mathcal{L}}=\frac{1}{2}\left( g_{a\:b} \right)\dot{x}^a\dot{x}^b=-\frac{1}{2}{m}^2.
\end{equation}
Substituting Eq. \eqref{eq:gab} into Eq. \eqref{eq:HJ} yields the following expression
\begin{equation}\label{eq:ss}
\begin{aligned}
-2\frac{\partial S}{\partial\tau}=&-\frac1{h(r)}\left(\frac{\partial S_t}{\partial t}\right)^2+f(r)\left(\frac{\partial S_r}{\partial r}\right)^2 \\
&+\sum_{\chi=1}^{n-3}\frac{1}{\Big(k(r)\prod_{i=1}^{\chi-1}\sin^2\theta_i\Big)}\left(\frac{\partial S_{\theta_\chi}}{\partial\theta_\chi}\right)^2 \\
&+\frac{1}{\Big(k(r)\prod_{i=1}^{n-3}\sin^2\theta_i\Big)}\left(\frac{\partial S_{\theta_{n-2}}}{\partial\theta_{n-2}}\right)^2.
\end{aligned}
\end{equation}
The Jacobi ansatz action can be written as a separable solution
\begin{equation}\label{eq:ja}
S=\frac{1}{2}m^2\tau-\mathrm{E}t+S_r(r)+\sum_{\chi=1}^{n-3}S_{\theta_\chi}(\theta_\chi)+{\L}\theta_{n-2},
\end{equation}
the expression of the components in the Jacobi action satisfies
\begin{equation}\label{eq:partials}
\begin{aligned}
&\frac{\partial S_\tau}{\partial \tau} = -\frac{1}{2}\left( g^{a\:b} \right) P_a P_b = \frac{1}{2} m^2, \\
&\frac{\partial S_t}{\partial t}=P_{t} = -h(r)\dot t=-\mathrm{E}, \\
&\frac{\partial S_r}{\partial r}=P_{r} = \frac{1}{f(r)}\dot{r}, \\
&\frac{\partial S_{\theta_\chi}}{\partial\theta_\chi}=P_{{\theta}_{\chi}}=k(r) \left( \prod_{i=1}^{\chi-1}\sin^{2}\theta_{i} \right) \dot{\theta}_{\chi}, \\
&\frac{\partial S_{\theta_{n-2}}}{\partial\theta_{n-2}}=P_{\theta_{n-2}}  =k(r) \left( \prod_{i=1}^{n-3}\sin^{2}\theta_{i} \right) \dot{\theta}_{n-2} = {\L}.
\end{aligned}
\end{equation}
By inserting Eq. \eqref{eq:partials} into Eq. \eqref{eq:ss}, we can find the following equation
\begin{equation}\label{eq:krm}
\begin{aligned}
k(r) m^2 = &\frac{k(r)}{h(r)}\left(-\mathrm{E}\right)^2 - \frac{k(r)}{f(r)}\left(\dot{r}\right)^2 \\
& -k^{2}(r)\sum_{\chi=1}^{n-3} \Bigg[\left(\prod_{i=1}^{\chi-1}\sin^{2}\theta_{i}\right) \left(\dot{\theta}_{\chi}\right)^2\Bigg] - \Bigg[\prod_{i=1}^{n-3}\left(1+\cot^2\theta_i\right)\Bigg] \left({\L}\right)^2, \\
\end{aligned}
\end{equation}
the last term of which can be expanded to
\begin{equation}\label{eq:otheta}
\prod_{i=1}^{n-3}\left(1+\cot^2\theta_i\right) = 1+\prod_{i=1}^{n-3}\cot^2\theta_i+\mathfrak{O}\left(\theta_i\right),
\end{equation}
where $\mathfrak{O}$ is a function of $\theta_i$ as the independent variable.
Then rewrite Eq. \eqref{eq:krm} to be expressed in the following form:
\begin{equation}
\begin{aligned}
0 = & \Bigg\{ \frac{k(r)}{h(r)} \mathrm{E}^2 - \frac{k(r)}{f(r)}\left(\dot{r}\right)^2 - {\L}^2 - k(r) m^2 \Bigg\} \\
& - \Bigg\{ {k^{2}(r)}\sum_{\chi=1}^{n-3} \Bigg[\left(\prod_{i=1}^{\chi-1}\sin^{2}\theta_{i}\right) \left(\dot{\theta}_{\chi}\right)^2\Bigg] + \Bigg[\prod_{i=1}^{n-3}\cot^2\theta_i+\mathfrak{O}\left(\theta_i\right)\Bigg] {\L}^2 \Bigg\}.
\end{aligned}
\end{equation}
Introducing the generalized Carter constant $\mathscr{K}$ \cite{Carter:1968rr}, the above equation can be separated into
\begin{equation}
\begin{aligned}
\mathscr{K}&=\Bigg\{ \frac{k(r)}{h(r)} \mathrm{E}^2 - \frac{k(r)}{f(r)}\left(\dot{r}\right)^2 - {\L}^2 - k(r) m^2 \Bigg\}, \\
\mathscr{K}&=\Bigg\{ k^{2}(r)\sum_{\chi=1}^{n-3} \Bigg[\left(\prod_{i=1}^{\chi-1}\sin^{2}\theta_{i}\right) \left(\dot{\theta}_{\chi}\right)^2\Bigg] + \Bigg[\prod_{i=1}^{n-3}\cot^2\theta_i+\mathfrak{O}\left(\theta_i\right)\Bigg] {\L}^2 \Bigg\}. \\
\end{aligned}
\end{equation}
Subsequently, the complete geodesic equation for the test particle in high-dimensional space-time is obtained, which yields
\begin{equation}
\begin{aligned}
\dot{t}&=\frac{1}{h(r)} \mathrm{E}, \\
\dot{r}&=\pm\sqrt{\frac{f(r)}{h(r)} {\mathrm{E}}^2 - \frac{f(r)}{k(r)} \left(\mathscr{K} + {\L}^2\right) - f(r){m}^2}, \\
\sqrt{\sum_{\chi=1}^{n-3} \prod_{i=1}^{\chi-1}\sin^{2}\theta_{i} \left(\dot{\theta}_{\chi}\right)^2}&=\pm\sqrt{\frac{\mathscr{K}-\Big[\prod_{i=1}^{n-3}\cot^2\theta_i+\mathfrak{O}\left(\theta_i\right)\Big]{\L}^2}{k^{2}(r)}}, \\ 
\dot{\theta}_{n-2}&=\frac{1}{k(r) \left( \prod_{i=1}^{n-3}\sin^{2}\theta_{i} \right)} {\L},
\end{aligned}
\end{equation}
in which the ``$+$'' and ``$-$'' signs indicate that the motion of the test particle is in the outgoing and the ingoing radial direction, respectively.
Moreover, it should be noted that the ranges of integers $\chi$ and $i$ are $\left\{1, 2, \cdots, (n-2)\right\}$ and $\left\{0, 1, \cdots, (n-3)\right\}$, respectively.
It can also be written as
\begin{equation}\label{eq:cge}
\begin{aligned}
\dot{t}&=\frac{1}{h(r)} \mathrm{E}, \\
\dot{r}&=\pm\frac{1}{k(r)}\sqrt{\mathfrak{R}}, \\
\sqrt{\sum_{\chi=1}^{n-3} \prod_{i=1}^{\chi-1}\sin^{2}\theta_{i} \left(\dot{\theta}_{\chi}\right)^2}&=\pm\frac{1}{k(r)}\sqrt{\Theta_{\chi}}, \\ 
\dot{\theta}_{n-2}&=\frac{1}{k(r) \left( \prod_{i=1}^{n-3}\sin^{2}\theta_{i} \right)} {\L},
\end{aligned}
\end{equation}
where
\begin{equation}
\begin{aligned}
\mathfrak{R}&=k^{2}(r)\frac{f(r)}{h(r)} {\mathrm{E}}^2 - k(r)f(r) \left(\mathscr{K} + {\L}^2\right) - k^{2}(r)f(r){m}^2, \\
\Theta_{\chi}&=\mathscr{K}-\left[\prod_{i=1}^{n-3}\cot^2\theta_i+\mathfrak{O}\left(\theta_i\right)\right]{\L}^2.
\end{aligned}
\end{equation}

The shadow boundary of a black hole can be determined by an effective potential around it, which depends on the unstable circular orbits of the photons. 
To construct this effective potential, we will rewrite the radial equation as
\begin{equation}
\left(\frac{\mathrm{d}r}{\mathrm{d}\tau}\right)^2+V_{\mathrm{eff}}=0.
\end{equation}
When we consider the null geodesic equation, that is, the photon ($m=0$) as the test particle, the effective potential for the radial motion is
\begin{equation}
V_{\mathrm{eff}} = \frac{f(r)}{k(r)} \left(\mathscr{K} + {\L}^2\right) -\frac{f(r)}{h(r)} {\mathrm{E}}^2,
\end{equation}
and the function $\mathfrak{R}$ is given by
\begin{equation}
\mathfrak{R}=k^{2}(r)\frac{f(r)}{h(r)} {\mathrm{E}}^2 - k(r)f(r) \left(\mathscr{K} + {\L}^2\right).
\end{equation}
By defining two independent characteristic impact parameters
\begin{equation}
\begin{aligned}
&\xi=\frac{\L}{{\mathrm{E}}}, \\
&\eta=\frac{\mathscr{K}}{{\mathrm{E}}^2}, \\
\end{aligned}
\end{equation}
that describe the properties of photons near the black hole, the functions $V_{\mathrm{eff}}$ and $\mathfrak{R}$ will be rewritten as
\begin{equation}
V_{\mathrm{eff}}={\mathrm{E}}^2\Bigg\{\frac{f(r)}{k(r)}\Big(\eta+\xi^2\Big)-\frac{f(r)}{h(r)}\Bigg\},
\end{equation}
and
\begin{equation}
\mathfrak{R}={\mathrm{E}}^2\Bigg\{k^{2}(r)\frac{f(r)}{h(r)} {\mathrm{E}}^2 - k(r)f(r)\Big(\eta+\xi^2\Big)\Bigg\}.
\end{equation}
The unstable circular orbit of photons is also called the photon sphere radius, which corresponds to the maximum value of the effective potential.
Therefore, in order to solve this photon sphere radius, the effective potential needs to satisfy the following conditions
\begin{equation}
\begin{aligned}
&V_{\text{eff}}=0,  \\
&\frac{\partial V_{\text{eff}}}{\partial r}=0,  \\
&\frac{\partial^2 V_{\text{eff}}}{\partial r^2}<0, \\
\end{aligned}
\quad\quad \text{or} \quad\quad
\begin{aligned}
&\mathfrak{R}=0, \\
&\frac{\partial \mathfrak{R}}{\partial r}=0, \\
&\frac{\partial^2 \mathfrak{R}}{\partial r^2}>0. \\
\end{aligned}
\end{equation}
Solving the above equations, we will get
\begin{equation}
k\left(r_p\right) h^{\prime}\left(r_p\right) - h\left(r_p\right) k^{\prime}\left(r_p\right)=0,
\end{equation}
where $r_p$ is the photon sphere radius, and its value will be determined by this equation.
Moreover, during the solving process, after utilizing the condition ${\partial \mathfrak{R}}/{\partial r}=0$, we can also provide the general solution for the parameter $\eta+\xi^2$ as follows:
\begin{equation}
\Big(\eta+\xi^2\Big)=\frac{h(r) k^2(r) f'(r)-f(r) k^2(r) h'(r)+2 f(r) h(r) k(r) k'(r)}{h^2(r) \Big[k(r) f'(r)+f(r) k'(r)\Big]},
\end{equation}
substituting $r_p$ into the above formula, we will obtain the definite value of $\eta+\xi^2$ as
\begin{equation}\label{eq:etaxi2}
\Big(\eta+\xi^2\Big) = \frac{k^2\left(r_p\right)}{h\left(r_p\right) \big[k\left(r_p\right) f^{\prime}\left(r_p\right)+f\left(r_p\right) k^{\prime}\left(r_p\right)\big]} \Bigg\{ f^{\prime}\left(r_p\right) - f\left(r_p\right) \bigg[ \frac{h^{\prime}\left(r_p\right)}{h\left(r_p\right)} - \frac{2k^{\prime}\left(r_p\right)}{k\left(r_p\right)} \bigg] \Bigg\}.
\end{equation}
The parameter $\eta+\xi^2$ directly corresponds to the shadow radius, and this relationship will be discussed in the next subsection.

\subsubsection{Geometric shape of shadow}
First, let us define the inertial reference frame of an observer infinitely far away from the black hole in high-dimensional space-time, whose basis vector \cite{Johannsen:2013vgc, Li:2020drn} is
\begin{equation}
e_{\hat{\alpha}} = \left\{ e_{\hat{t}}, e_{\hat{r}}, \cdots, e_{{\hat{\theta}}_{\chi}}, \cdots, e_{{\hat{\theta}}_{n-2}} \right\}.
\end{equation}
The coordinate basis vector of the metric is
\begin{equation}
e_{\mu} = \left\{ e_{t}, e_{r}, \cdots, e_{{\theta}_{\chi}}, \cdots, e_{{\theta}_{n-2}} \right\},
\end{equation}
and its relation to the basis vector of the observer is given by
\begin{equation}
e_{\hat{\alpha}}=e_{\hat{\alpha}}^\mu e_\mu.
\end{equation}
Note that the coefficients $e_{\hat{\alpha}}^\mu$ are the transformation matrices that satisfy the following relation:
\begin{equation}
e_{\hat{\alpha}}^{\mu} e_{\hat{\beta}}^{\nu} g_{\mu\nu} = \eta_{\hat{\alpha} \hat{\beta}},
\end{equation}
where $\eta_{\hat{\alpha} \hat{\beta}}$ is the Minkowski metric.
Since the observer's basis vector has a Minkowski normalization relation
\begin{equation}
-1 = e_{\hat{t}} \cdot e_{\hat{t}}, \quad 1 = e_{\hat{r}} \cdot e_{\hat{r}} = e_{{\hat{\theta}}_{\chi}} \cdot e_{{\hat{\theta}}_{\chi}} = e_{{\hat{\theta}}_{n-2}} \cdot e_{{\hat{\theta}}_{n-2}}, 
\end{equation}
and needs to satisfy $0=e_{\hat{t}} \cdot e_{{\hat{\theta}}_{n-2}}$, we choose it as
\begin{equation}
e_{\hat{t}}= \zeta e_{t} + \gamma e_{{\theta}_{n-2}}, \quad e_{\hat{r}}= \iota e_{r}, \quad e_{{\hat{\theta}}_{\chi}}= \varsigma e_{{\theta}_{\chi}}, \quad e_{{\hat{\theta}}_{n-2}}= \varrho e_{{\theta}_{n-2}}, 
\end{equation}
where the coefficients are defined as
\begin{equation}
\begin{aligned}
\zeta&=e_{\hat{t}}^t=\sqrt{\frac{g_{n-1\:n-1}}{\left({g_{0\:n-1}}\right)^2 - g_{0\:0}g_{n-1\:n-1}}}, \\
\gamma&=e_{\hat{t}}^{{\theta}_{n-2}}=-\frac{g_{0\:n-1}}{g_{n-1\:n-1}} \sqrt{\frac{g_{n-1\:n-1}}{\left({g_{0\:n-1}}\right)^2 - g_{0\:0}g_{n-1\:n-1}}}, \\
\iota&=e_{\hat{r}}^r=\frac{1}{\sqrt{g_{1\:1}}}, \\
\varsigma&=e_{{\hat{\theta}}_{\chi}}^{{\theta}_{\chi}}=\frac{1}{\sqrt{g_{\chi+1\:\chi+1}}}, \\
\varrho&=e_{{\hat{\theta}}_{n-2}}^{{\theta}_{n-2}}=\frac{1}{\sqrt{g_{n-1\:n-1}}}. \\
\end{aligned}
\end{equation}

\begin{figure}[htbp]
\centering
\includegraphics[width=1\textwidth]{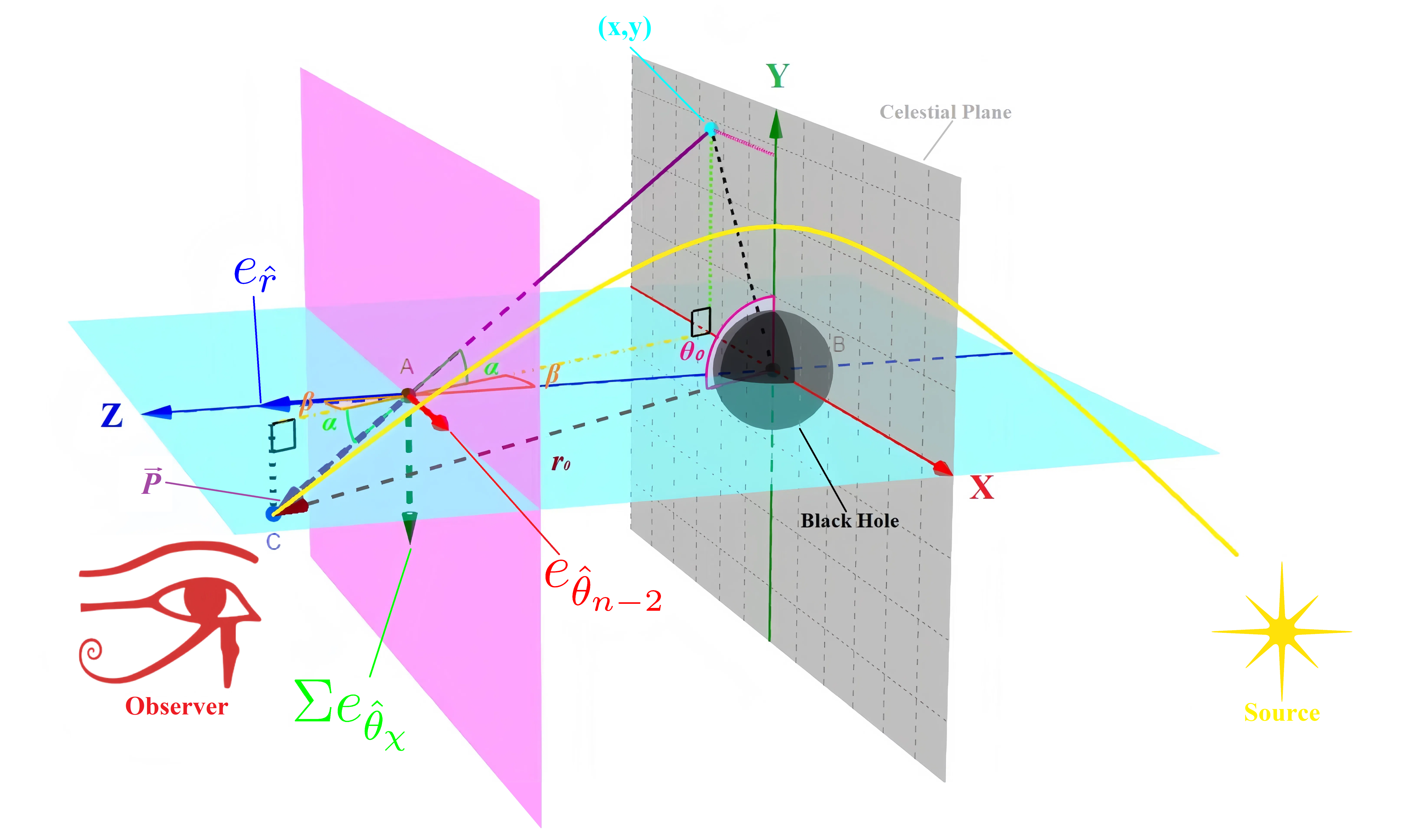}
\caption{
The schematic of the geometric projection of the photon's linear momentum $\boldsymbol{P}$ in the observer reference frame $\left\{ e_{\hat{r}}, \cdots, e_{{\hat{\theta}}_{\chi}}, \cdots, e_{{\hat{\theta}}_{n-2}} \right\}$.
The vector $\vec{P}$ is represented as the $(n-1)$-vector $\boldsymbol{P}$.
The symbol $r_{o}$ is the distance between the observer and the black hole, and $\theta_{o}$ is the inclination angle of the observer relative to the celestial plane.
}
\label{figure1}
\end{figure}

The photon's locally measured energy $P^{\hat{t}}$ is given by the projection of its $n$-momentum $P^{a}$ onto the $e_{\hat{t}}$, where the choice for the affine parameter is $P^{a}={\dot{x}}^{a}$, i.e., Eq. \eqref{eq:paxa}.
After adding the photon's locally measured momentum in $(n-1)$-dimensional space, the contravariant component of the linear momentum can be written as
\begin{equation}
\begin{aligned}
&P^{\hat{t}} = -\left( \zeta P_{t} + \gamma P_{{\theta}_{n-2}} \right), \\
&P^{\hat{r}} = \iota P_{r}, \\
&P^{\hat{\theta}_{\chi}} = \varsigma P_{{\theta}_{\chi}}, \\
&P^{\hat{\theta}_{n-2}} = \varrho P_{{\theta}_{n-2}}, \\
\end{aligned}
\end{equation}
according to Eq. \eqref{eq:p1}, the results will be explicitly expressed as
\begin{equation}\label{eq:pthat}
\begin{aligned}
&P^{\hat{t}} = \zeta \mathrm{E} - \gamma {\L} = \left(\frac{1}{h(r)}\right)^{\frac{1}{2}} \mathrm{E}, \\
&P^{\hat{r}} = \frac{1}{\sqrt{g_{1\:1}}} \left(\frac{1}{f(r)}\right) \dot{r} = \left(\frac{1}{f(r)}\right)^{\frac{1}{2}} \left|\dot{r}\right|, \\
&P^{\hat{\theta}_{\chi}} = \frac{1}{\sqrt{g_{\chi+1\:\chi+1}}} \left( k(r) \prod_{i=1}^{\chi-1}\sin^{2}\theta_{i} \right) \dot{\theta}_{\chi}  = \left(k(r) \prod_{i=1}^{\chi-1}\sin^{2}\theta_{i}\right)^{\frac{1}{2}} \dot{\theta}_{\chi}, \\
&P^{\hat{\theta}_{n-2}} = \frac{1}{\sqrt{g_{n-1\:n-1}}} {\L} = \left( \frac{1}{k(r) \prod_{i=1}^{n-3}\sin^{2}\theta_{i}} \right)^{\frac{1}{2}} {\L}. \\
\end{aligned}
\end{equation}
For the stationary observer towards the black hole, the photon from the edge of the shadow has $P^{\hat{r}} \geq 0$, so $\dot{r} \geq 0$ should be selected.
The $(n-1)$-vector $\boldsymbol{P}$ is defined as the linear momentum of the photon, whose components $\left\{ P^{\hat{r}}, \cdots, P^{\hat{\theta}_{\chi}}, \cdots, P^{\hat{\theta}_{n-2}} \right\}$ are in the orthonormal basis $\left\{ e_{\hat{r}}, \cdots, e_{{\hat{\theta}}_{\chi}}, \cdots, e_{{\hat{\theta}}_{n-2}} \right\}$ and are expressed as follows:
\begin{equation}
\boldsymbol{P} = P^{\hat{r}}e_{\hat{r}} + \sum_{\chi=1}^{n-3} P^{\hat{\theta}_{\chi}}e_{{\hat{\theta}}_{\chi}} + P^{\hat{\theta}_{n-2}}e_{{\hat{\theta}}_{n-2}}.
\end{equation}
In addition, it is worth noting that there is $\left|\boldsymbol{P}\right| = P^{\hat{t}}$ in the observer's reference frame, therefore,
\begin{equation}
\Big(P^{\hat{t}}\Big)^2 = \boldsymbol{P}^2 = \Big(P^{\hat{r}}\Big)^2 + \sum_{\chi=1}^{n-3}\Big(P^{\hat{\theta}_{\chi}}\Big)^2 + \Big(P^{\hat{\theta}_{n-2}}\Big)^2.
\end{equation}
The projection of momentum $\boldsymbol{P}$ in the observer coordinate system is shown in Figure \ref{figure1}.
It can be seen that in the observer's reference frame, the angles $(\alpha, \beta)$ formed by the photon projected to the observer are defined as the observation angles, which can be used to describe the coordinates $(\mathrm{X}, \mathrm{Y})$ of the photon on the two-dimensional celestial plane.
When the observer is on the $\mathrm{Z}$ axis, the observation angles $\alpha$ and $\beta$ are both zero.
Based on the geometric structure constructed in Figure \ref{figure1}, we can obtain
\begin{equation}
P^{\hat{r}} = \left|\boldsymbol{P}\right| \cos\alpha\cos\beta, \quad  P^{\hat{\theta}_{\chi}} = \left|\boldsymbol{P}\right| \sin\alpha, \quad  P^{\hat{\theta}_{n-2}} = \left|{\boldsymbol{P}}\right| \cos\alpha\sin\beta, 
\end{equation}
where $\alpha \in \left[-\frac{\pi}{2}, \frac{\pi}{2}\right]$ and $\cos\beta \geq 0$ should be satisfied due to $P^{\hat{r}} \geq 0$.
Then, we have
\begin{equation}
\tan\beta = \frac{P^{\hat{\theta}_{n-2}}}{P^{\hat{r}}}, \quad \sin\alpha = \frac{P^{\hat{\theta}_{\chi}}}{P^{\hat{t}}}.
\end{equation}
For a stationary observer at infinity, the celestial coordinates $(\mathrm{X}, \mathrm{Y})$ can be defined as
\begin{equation}
\begin{aligned}
&\mathrm{X} \equiv -r_{o} \beta, \\
&\mathrm{Y} \equiv r_{o} \alpha, \\
\end{aligned}
\end{equation}
and since the observation angles $(\alpha, \beta)$ are very small, $\beta \simeq \tan\beta$ and $\alpha \simeq \sin\alpha$ will be allowed.
Therefore, the precise formula for celestial coordinates in high-dimensional space-time should be represented as
\begin{equation}\label{eq:xy}
\begin{aligned}
\mathrm{X} &\equiv \lim_{r_o\to\infty} \left[ -r_{o} \left. \left(\frac{P^{\hat{\theta}_{n-2}}}{P^{\hat{r}}}\right) \right|_{\left(r=r_{o}\right)} \right], \\
\mathrm{Y} &\equiv \lim_{r_o\to\infty} \left[ r_{o} \left. \left(\frac{\left\|\boldsymbol{P^{\hat{\theta}_{\chi}}}\right\|}{P^{\hat{t}}}\right) \right|_{\left(r=r_{o}\right)} \right], \\
\end{aligned}
\end{equation}
where
\begin{equation}
\left\| \boldsymbol{P^{\hat{\theta}_{\chi}}} \right\| \equiv \sqrt{\sum_{\chi=1}^{n-3} \left|P^{\hat{\theta}_{\chi}}\right|^2 } = \sqrt{\sum_{\chi=1}^{n-3} \left(k(r) \prod_{i=1}^{\chi-1}\sin^{2}\theta_{i}\right)\left(\dot{\theta}_{\chi}\right)^{2}}.
\end{equation}

Substituting Eqs. \eqref{eq:pthat} and \eqref{eq:cge} into Eq. \eqref{eq:xy} and selecting the function $k(r)$ as $k(r)=\pounds r^2$, then the expression of celestial coordinates $(\mathrm{X}, \mathrm{Y})$ can be deduced as
\begin{equation}\label{eq:x1}
\mathrm{X} = -\frac{1}{\sqrt{\pounds}} \left( \frac{\xi}{\prod_{i=1}^{n-3}\sin\theta_{i}} \right), 
\end{equation}
and
\begin{equation}\label{eq:y1}
\mathrm{Y} = \pm\frac{1}{\sqrt{\pounds}} \sqrt{\eta-{\xi}^2\left[\prod_{i=1}^{n-3}\cot^2\theta_i+\mathfrak{O}\left(\theta_i\right)\right]}.
\end{equation}
When we choose the observer's position on the equatorial hyperplane $\theta_\chi=\frac{\pi}{2}$ (i.e., $\theta_i=\frac{\pi}{2}$ by the same reasoning.) in space-time, and $\mathfrak{O}\left(\frac{\pi}{2}\right)\equiv0$ must hold by considering Eq. \eqref{eq:otheta}, the celestial coordinates $(\mathrm{X}, \mathrm{Y})$ will be simplified to
\begin{equation}
\begin{aligned}
\mathrm{X} &= -\frac{\xi}{\sqrt{\pounds}}, \\
\mathrm{Y} &= \pm\sqrt{\frac{\eta}{\pounds}}. \\
\end{aligned}
\end{equation}
Consequently, the radius of the shadow on the celestial plane is
\begin{equation}
{\left(R_{s}\right)}^2 \equiv \mathrm{X}^2 + \mathrm{Y}^2 = \frac{\Big(\xi^2+\eta\Big)}{\pounds}.
\end{equation}
Inserting Eq. \eqref{eq:etaxi2} into the equation above, we will ultimately obtain the general formula for the shadow radius taking into account the metric \eqref{eq:le}:
\begin{equation}\label{eq:highdimrseq}
\begin{aligned}
R_{s} &= \sqrt{ \frac{\left(r_p\right)^4 \pounds^2}{\pounds h\left(r_p\right) \big[\left(r_p\right)^2 \pounds f^{\prime}\left(r_p\right)+ 2 r_p \pounds f\left(r_p\right)\big]} \Bigg\{ f^{\prime}\left(r_p\right) - f\left(r_p\right) \bigg[ \frac{h^{\prime}\left(r_p\right)}{h\left(r_p\right)} - \frac{4 r_p \pounds}{\left(r_p\right)^2 \pounds} \bigg] \Bigg\} } \\
&= \sqrt{ \frac{\left(r_p\right)^3}{h\left(r_p\right) \big[r_p f^{\prime}\left(r_p\right)+ 2 f\left(r_p\right)\big]} \Bigg\{ f^{\prime}\left(r_p\right) - f\left(r_p\right) \bigg[ \frac{h^{\prime}\left(r_p\right)}{h\left(r_p\right)} - \frac{4}{r_p} \bigg] \Bigg\} }, \\
\end{aligned}
\end{equation}
from which we can see that this result is independent of $\pounds$. Therefore, parameter $\pounds$ can be any function or constant that has no effect on the shadow radius $R_{s}$.

In addition, the coordinate $\mathrm{X}$ can also be defined as another expression that ultimately yields the same result as Eq. \eqref{eq:x1}, and its derivation is as follows:
\begin{equation}
\begin{aligned}
\mathrm{X} &= \lim_{r_o\to\infty} \left[ -r_{o} \left. \left(\frac{P^{\hat{\theta}_{n-2}}}{P^{\hat{t}}}\right) \right|_{\left(r=r_{o}\right)} \right] \\
&= -\frac{1}{\sqrt{\pounds}} \left(\frac{\xi}{\prod_{i=1}^{n-3}\sin\theta_{i}}\right).
\end{aligned}
\end{equation}

\subsection{Constraints on the shadow radius}
Previously, the stellar-dynamics measurements of supermassive black holes were used to produce a posterior distribution function of the angular gravitational radius $\theta_{\mathrm{g}}$, which was determined by the EHT in 2017 through observations of the $\mathrm{M87}^{\star}$ and $\mathrm{Sgr~A}^{\star}$ black holes.
The EHT team used the prior information on the angular gravitational radius $\theta_{\mathrm{g}}$ of the $\mathrm{M87}^{\star}$ and $\mathrm{Sgr~A}^{\star}$ black holes to calculate the predicted size of their shadows. 
Afterwards, the posterior of $\theta_{\mathrm{g}}$ was analyzed by comparing its prior results with the size inferred from the EHT images and visibility-domain model fitting.

The angular gravitational radius $\theta_{\mathrm{g}}$ is defined as
\begin{equation}\label{eq:agr}
\theta_{\mathrm{g}} = \frac{G M}{c^2 D},
\end{equation}
where $G$ represents the gravitational constant term, $M$ is the black hole mass inferred from the EHT and $D$ is the observed distance between the black hole and the observer.
Next, the fractional deviation of the angular gravitational radius is introduced as follows:
\begin{equation}\label{eq:agrfd}
\delta \equiv \frac{\theta_{\mathrm{dyn}}}{\theta_{\mathrm{g}}} - 1,
\end{equation}
where $\theta_{\mathrm{g}}$ and $\theta_{\mathrm{dyn}}$ are used to represent the real measurement and the inference of the stellar-dynamics of the angular gravitational radius, respectively.
Although the definition $\delta = {\theta_{\mathrm{g}}} / \theta_{\mathrm{dyn}} - 1$ is referenced in some related papers, the two definitions are numerically similar; To maintain the clarity of the derivation, we adopt the expression $\delta = \theta_{\mathrm{dyn}} / {\theta_{\mathrm{g}}} - 1$.
Here, parameters $\theta_{\mathrm{g}}$ and $\theta_{\mathrm{dyn}}$ correspond to the measured value and estimated value, respectively. 
Thus, $\theta_{\mathrm{dyn}}$ can be written as 
\begin{equation}
\theta_{\mathrm{dyn}} = \frac{G M_{\mathrm{dyn}}}{c^2 D_{\mathrm{dyn}}}.
\end{equation}

In addition, the angular shadow radius can be defined as
\begin{equation}\label{eq:asr}
\theta_{\mathrm{sh}} = \frac{r_{\mathrm{sh}}}{D},
\end{equation}
then, with the known value of the shadow radius $r_{\mathrm{sh}}$, the predicted value of the angular shadow radius $\theta_{\mathrm{sh}}$ can be determined using $\theta_{\mathrm{dyn}}$.

Since the stellar-dynamics measurements are only sensitive to the monopole of the metric (i.e., the mass) and spin-dependent effects are negligible at the distances involved in that analysis, it is reasonable to model the $\mathrm{M87}^{\star}$ and $\mathrm{Sgr~A}^{\star}$ using the Schwarzschild solution.
The EHT has given several limits on the fractional deviations from different observation instruments \cite{EventHorizonTelescope:2019ggy, EventHorizonTelescope:2022xqj}, as shown in Table \ref{tb2:delta}.
And then, reasonable deviations in the shadow radius size can be obtained.
Since these deviations can still be consistent with the imaging data, an effective constraint range can be imposed on certain underlying parameters in the black hole metric \cite{EventHorizonTelescope:2021dqv, EventHorizonTelescope:2020qrl}.

\begin{table}[tbh]\centering
\caption{
Reasonable boundaries of the fractional deviation $\delta$ are determined based on different observational instruments.
The confidence level of these intervals was adopted as $68\%$ for a conservative evaluation.
\vspace{0.2cm}} \label{tb2:delta}
\begin{tabular*}{12cm}{*{3}{c @{\extracolsep\fill}}}
\hline
Black Hole & Observation Instrument & $\delta$ \\
\hline
$\mathrm{M87}^{\star}$ & \text{EHT} & $-0.01_{-0.17}^{+0.17}$ \\
\multirow{2}{*}{$\mathrm{Sgr~A}^{\star}$} & \text{VLTI} & $-0.08_{-0.09}^{+0.09}$ \\
                                          & \text{Keck} & $-0.04_{-0.10}^{+0.09}$ \\ 
\hline
\end{tabular*}
\end{table}

\subsubsection{Four-dimensional space-time}
When we choose the Schwarzschild solution in four-dimensional space-time as the predicted value of the shadow radius $r_{\mathrm{sh}}$, it can be expressed as
\begin{equation}
r_{\mathrm{sh}} = \frac{3\sqrt{3} G M_{\mathrm{dyn}}}{c^2},
\end{equation}
where $G$ is the four-dimensional gravitational constant, and then $\theta_{\mathrm{sh}}$ can be determined as
\begin{equation}
\theta_{\mathrm{sh}} = \frac{r_{\mathrm{sh}}}{D_{\mathrm{dyn}}} = 3\sqrt{3} \left( \frac{G M_{\mathrm{dyn}}}{c^2 D_{\mathrm{dyn}}} \right),
\end{equation}
thus,
\begin{equation}
\theta_{\mathrm{sh}} = 3\sqrt{3} \theta_{\mathrm{dyn}}.
\end{equation}
Substituting the above formula into Eq. \eqref{eq:agrfd}, we can get
\begin{equation}
\delta = \frac{\left( \frac{\theta_{\mathrm{sh}}}{3\sqrt{3}} \right)}{\theta_{\mathrm{g}}} - 1,
\end{equation}
that is,
\begin{equation}\label{eq:thetashdelta}
\theta_{\mathrm{sh}} = 3\sqrt{3} \left(\delta + 1\right) \theta_{\mathrm{g}}.
\end{equation}
Utilizing Eqs. \eqref{eq:agr} and \eqref{eq:thetashdelta}, it is natural to get the following relation:
\begin{equation}
\theta_{\mathrm{sh}} = \frac{r_{\mathrm{sh}}}{D} = \frac{r_{\mathrm{sh}}}{\left( \frac{G M}{c^2 \theta_{\mathrm{g}}} \right)} = 3\sqrt{3} \left(\delta + 1\right) \theta_{\mathrm{g}},
\end{equation}
when we choose $c=1$, the shadow radius is denoted by
\begin{equation}\label{eq:4dimrange}
r_{\mathrm{sh}} = 3\sqrt{3} \left(\delta + 1\right) G M.
\end{equation}

Therefore, utilizing this effective domain of shadow radius can constrain the unique parameters in the majority of four-dimensional black hole metrics.

\subsubsection{High-dimensional space-time}
In the previous subsection, we used the Schwarzschild space-time metric to construct a reasonable constraint on the shadow radius of a four-dimensional black hole.
However, it is incorrect for some papers \cite{Nozari:2024jiz, Nozari:2023flq} to apply Eq. \eqref{eq:4dimrange} to high-dimensional black hole solutions.
Firstly, Eq. \eqref{eq:4dimrange} is derived by modeling the theoretical value of the shadow radius of the four-dimensional Schwarzschild black hole.
Consequently, it is evident that the shadow radius of a high-dimensional black hole cannot logically be constrained within this four-dimensional framework.
Secondly, if Eq. \eqref{eq:4dimrange} is forcibly applied to constrain the characteristic parameters in high-dimensional black hole solutions, unphysical scenarios may arise. 
For instance, certain characteristic parameters could erroneously appear as constants or fixed functions as the dimension grows, which is clearly impossible.

To ensure the constraint range is valid in high-dimensional black hole solutions, we will derive the constraint equation for the shadow radius of black holes in high-dimensional space-time.
Here, we will use the shadow radius of the Schwarzschild-Tangherlini metric \eqref{eq:sch-tang} as the predicted value for modeling, and its radius formula ($c=1$) can be expressed as
\begin{equation}
r_{\mathrm{sh}} = \sqrt{ \frac{n-1}{n-3} \left[ \frac{16 \pi \mathscr{G}M_{\mathrm{dyn}} (n-1)}{2 (n-2) \Sigma_{(n-2)}} \right]^{\frac{2}{n-3}} },
\end{equation}
and then,
\begin{equation}\label{eq:dthsh}
\begin{aligned}
\theta_{\mathrm{sh}} &= \frac{r_{\mathrm{sh}}}{D_{\mathrm{dyn}}} \\
&= \left\{ \left(\frac{n-1}{n-3}\right)^{\frac{1}{2}} \left[ \frac{16 \pi (n-1)}{2 (n-2) \Sigma_{(n-2)}} \right]^{\frac{1}{n-3}} \cdot \left(\mathscr{G} M_{\mathrm{dyn}}\right)^{\frac{1}{n-3}} \right\} \cdot \frac{1}{D_{\mathrm{dyn}}}.
\end{aligned}
\end{equation}
Next, we define the angular gravitational radius in higher-dimensional space-time, denoted by
\begin{equation}\label{eq:nagr}
\theta_{\mathrm{g}}=\frac{1}{D} \left(\frac{\mathscr{G} M}{c^2}\right)^{\frac{1}{n-3}},
\end{equation}
which is constructed to ensure dimensional consistency in arbitrary space-time dimensions.
We further introduce the function $\mathfrak{C}(n)$, defined as follows:
\begin{equation}
\mathfrak{C}(n) \equiv \left(\frac{n-1}{n-3}\right)^{\frac{1}{2}} \left[ \frac{16 \pi (n-1)}{2 (n-2) \Sigma_{(n-2)}} \right]^{\frac{1}{n-3}},
\end{equation}
so that Eq. \eqref{eq:dthsh} can be rewritten as
\begin{equation}\label{eq:dthsh2}
\theta_{\mathrm{sh}} = \mathfrak{C}(n) \cdot \left[\frac{1}{D_{\mathrm{dyn}}} \left(\mathscr{G} M_{\mathrm{dyn}}\right)^{\frac{1}{n-3}}\right] = \mathfrak{C}(n) \cdot \theta_{\mathrm{dyn}}. \\
\end{equation}
Substituting Eq. \eqref{eq:dthsh2} into Eq. \eqref{eq:agrfd}, we can get
\begin{equation}
\delta = \frac{\left[ \frac{\theta_{\mathrm{sh}}}{\mathfrak{C}(n)} \right]}{\theta_{\mathrm{g}}} - 1,
\end{equation}
that is,
\begin{equation}\label{eq:dthetashdelta}
\theta_{\mathrm{sh}} = \mathfrak{C}(n) \left(\delta + 1\right) \theta_{\mathrm{g}}.
\end{equation}
Combining Eqs. \eqref{eq:nagr} and \eqref{eq:dthetashdelta}, we have
\begin{equation}
\theta_{\mathrm{sh}} = \frac{r_{\mathrm{sh}}}{D} = \frac{r_{\mathrm{sh}}}{\left[ \frac{\left(\mathscr{G} M\right)^{\frac{1}{n-3}}}{\theta_{\mathrm{g}}} \right]} = \mathfrak{C}(n) \left(\delta + 1\right) \theta_{\mathrm{g}},
\end{equation}
therefore, the shadow radius is given by
\begin{equation}\label{eq:highdimrange}
r_{\mathrm{sh}} = \mathfrak{C}(n) \left(\delta + 1\right) \left(\mathscr{G} M\right)^{\frac{1}{n-3}}.
\end{equation}

As defined above, it follows clearly that $M_{\mathrm{dyn}} \equiv \mathsf{M}$.
In practical calculations, we adopt a system of units in which the ADM mass $\mathsf{M}$, the mass ${M}$, and the high-dimensional gravitational constant $\mathscr{G}$ are set to unity, i.e., $\mathsf{M}=M=\mathscr{G}=1$.
It should be emphasized that this choice does not imply that the physical value of $\mathscr{G}$ is identical across different space-time dimensions; rather, the gravitational constant in each dimension is simply used as a reference unit.
Consequently, as long as one is not concerned with physical comparisons between different dimensions or with dimensional effects associated with the gravitational constant, this choice of units is entirely justified.

The shadow radius constraint boundaries $r_{\mathrm{sh}}$ in high-dimensional space-time exhibit explicit dimensional dependence, with distinct parameter bounds emerging for each space-time dimension $n$.
Additionally, an intriguing phenomenon is that for the high-dimensional constraint equation, the contribution of $\mathscr{G}$ to $r_{\mathrm{sh}}$ is affected by the dimension $n$ when $\mathscr{G}\neq1$.
It can be seen that Eqs. \eqref{eq:highdimrange} and \eqref{eq:4dimrange} are equivalent when the space-time dimension is set to four.

The confidence interval for the fractional deviation parameter $\delta$ is obtained here using the following measurement method.
For $\mathrm{Sgr~A}^{\star}$, $\delta$ is computed relative to the expected shadow size from monitoring stellar orbits \cite{EventHorizonTelescope:2022wkp};
For $\mathrm{M87}^{\star}$, $\delta$ is computed for the expected shadow both from stellar dynamics and from gas dynamics \cite{EventHorizonTelescope:2019ggy}.
Although the methods used to obtain the $\delta$ indirectly imply a gravitational framework governed by general relativity, the physical definition of the parameter $\delta$ does not mandate that the observed black hole space-time must be described and interpreted using general relativity.
Therefore, it can be considered that the parameter $\delta$ itself does not depend on a specific theoretical model.
This is reflected in the fact that the EHT employs certain parameterized test metrics as well as modified metrics to examine whether the black hole geometry deviates from the Schwarzschild (or Kerr) metric \cite{EventHorizonTelescope:2022xqj, EventHorizonTelescope:2020qrl, EventHorizonTelescope:2021dqv}.
If $\delta\simeq0$ is theoretically satisfied, then it can be definitively assumed that the space-time geometry of the black hole is described by the corresponding Schwarzschild (or Kerr) solution of general relativity.
Therefore, the effective range of $\delta$ obtained with the current observational accuracy still allows for the possibility that other metrics may exist within reasonable ranges of their characteristic parameters \cite{Khodadi:2020gns, Uniyal:2022vdu, Xavier:2023exm, Pantig:2022gih, Pantig:2022qak}.

Furthermore, when considering black hole space-times with extra dimensions, if the Schwarzschild (or Kerr) metric adopted by the EHT is used as the default benchmark for modeling, the corresponding parameter $\delta$ should systematically deviate from $\delta\simeq0$.
Here we speculate that the confidence interval of $\delta$ may already include the deviation induced by the presence of extra dimensions, and that the higher-dimensional deviation to the shadow remains smaller than the current observational uncertainty.
Based on this conjecture, we proceed to investigate two classes of high-dimensional black hole space-times: black string space-times and general high-dimensional spherically symmetric black hole space-times.
For black string space-times, their geometric structure can be effectively regarded as a four-dimensional Schwarzschild space-time once the extra dimensions are compactified. Accordingly, we still adopt the Schwarzschild (or Kerr) metric as the benchmark for modeling \cite{Yan:2024rsx}.
For the space-time of a general spherically symmetric high-dimensional black hole, since the high-dimensional shadow is projected onto a two-dimensional $(\mathrm{X}, \mathrm{Y})$ plane, the influence of higher dimensions on the shadow scale is encoded in the space-time metric.
In other words, the metric that serves as the baseline for constraint modeling should be highly correlated with the metric used to compute the high-dimensional shadow radius.
Therefore, when discussing the EPYMGB metric, which deviates slightly from the Schwarzschild-Tangherlini metric, we adopt the Schwarzschild-Tangherlini metric modeling as the benchmark.

\section{Perturbation equations and various calculation methods}
In this section, we will introduce the equations of motion governing a static spherically symmetric black hole in asymptotically flat high-dimensional space-time for different background fields.
This paper will follow the results given by covariantly performing the $2 + (n - 2)$ decomposition \cite{Hui:2020xxx} adopted in the paper \cite{Charalambous:2024tdj}.

The symbol $t^{a}$ involved in this paper is defined as time-like Killing vector, the 2-vector $r^{a}$ are orthogonal to each other and $t_{a}r^{a}=0$.
The functions $f_{t}\left(r\right)$ and $f_{r}\left(r\right)$ will be permitted to be covariantly defined as $f_{t}\left(r\right)=-t_{a}t^{a}$ and $f_{r}\left(r\right)=r_{a}r^{a}$, respectively.
The symbols $D_a$ and $D_A$ denote the covariant derivatives with respect to $g_{ab}$ and the unit-sphere metric $\Omega_{AB}$, respectively.

\subsection{Equations of motion and master variables}

\subsubsection{Spin-0 perturbations}
The action of the free scalar field minimally coupled to gravity is
\begin{equation}
\mathcal{S}^{(0)}=\int \mathrm{d}^n x \sqrt{-g}\left[-\frac{1}{2}(\nabla \Phi)^2-\frac{1}{2} m^2 \Phi^2\right],
\end{equation}
and the scalar field is decomposed into spherical harmonic modes according to $2 + (n - 2)$ as follows:
\begin{equation}
\Phi(x)=\sum_{\ell, \mathbf{m}} \frac{\Psi_{\ell, \mathbf{m}}^{(0)}(t, r)}{r^{(n-2) / 2}} Y_{\ell, \mathbf{m}}(\theta).
\end{equation}
Furthermore, the reduced action describes the scalar field minimally coupled to 2-$n$ gravity,
\begin{equation}
\begin{aligned}
\mathcal{S}^{(0)} & = \sum_{\ell, \mathbf{m}} \mathcal{S}_{\ell, \mathbf{m}}^{(0)}, \\
\mathcal{S}_{\ell, \mathbf{m}}^{(0)} & = \int \mathrm{d}^2 x \sqrt{-g^{(2)}}\left[-\frac{1}{2} D_a \bar{\Psi}_{\ell, \mathbf{m}}^{(0)} D^a \Psi_{\ell, \mathbf{m}}^{(0)}-\frac{1}{2 f_{t}} V_{\ell}^{(0)}(r)\left|\Psi_{\ell, \mathbf{m}}^{(0)}\right|^2\right],
\end{aligned}
\end{equation}
and the potential is represent as
\begin{equation}\label{eq:vl0}
V_{\ell}^{(0)}\left(r\right) \equiv f_{t}\Bigg\{ \frac{\ell\left(\ell+n-3\right)}{r^{2}}+\frac{\left(n-2\right)\left(n-4\right)}{4r^{2}}f_{r}+\frac{n-2}{2r} \left[\frac{\left(f_{t}f_{r}\right)^{\prime}}{2f_{t}}\right]+m^{2} \Bigg\}.
\end{equation}
In the tortoise coordinate $\mathrm{d}r_\ast=\mathrm{d}r / \sqrt{f_{t}f_{r}}$, this simplifies to the action of a scalar field propagating in 2-$n$ flat space-time under the influence of potential,
\begin{equation}
\mathcal{S}_{\ell, \mathbf{m}}^{(0)} = \int \mathrm{d} t \mathrm{d} r_\ast\left[\frac{1}{2}\left|\partial_t \Psi_{\ell, \mathbf{m}}^{(0)}\right|^2-\frac{1}{2}\left|\partial_{r_\ast} \Psi_{\ell, \mathbf{m}}^{(0)}\right|^2-\frac{1}{2} V_{\ell}^{(0)}(r)\left|\Psi_{\ell, \mathbf{m}}^{(0)}\right|^2\right],
\end{equation}
and the equation of motion for the scalar field perturbation is expressed as a Schr\"{o}dinger-like equation as follows:
\begin{equation}
\left[\partial_{r_\ast}^2-\partial_t^2-V_{\ell}^{(0)}(r)\right]\Psi_{\ell, \mathbf{m}}^{(0)}\left(r_\ast\right)=0.
\end{equation}

\subsubsection{Spin-1 perturbations}
The general Maxwell action used to describe the massless vector field of electromagnetic perturbations is
\begin{equation}
\begin{aligned}
\mathcal{S}^{(1)}&= \int \mathrm{d}^n x \sqrt{-g}\left[-\frac{1}{4} F_{\mu \nu} F^{\mu \nu}\right], \\
F_{\mu \nu}&= \partial_{\mu} A_{\nu} - \partial_{\nu} A_{\mu}, \\
\end{aligned}
\end{equation}
the components of the gauge field are separated into irreducible representations of $SO(n-1)$ by $2 + (n - 2)$ decomposition
\begin{equation}
A_{\mu}\left(x\right)=
\begin{pmatrix}
A_{a}\left(x\right) \\
D_{A}A^{\left(\mathrm{L}\right)}\left(x\right)+A_{A}^{\left(\mathrm{T}\right)}\left(x\right)
\end{pmatrix}.
\end{equation}
The quantities $A_{a}$ and the longitudinal component $A^{\left(\mathrm{L}\right)}$ transform as scalars under $SO(n-1)$, while the transverse component $A_{A}^{\left(\mathrm{T}\right)}$ transforms as a vector and satisfies $D^{A}A_{A}^{\left(\mathrm{T}\right)}=0$.
Under the gauge transformation $\delta_\Lambda A_\mu=\partial_\mu\Lambda$, the transformations of these individual components are
\begin{equation}\label{eq:gt}
\begin{aligned}
\delta_\Lambda A_a &= \partial_a\Lambda, \\
\delta_\Lambda A^{\left(\mathrm{L}\right)} &= \Lambda, \\
\delta_\Lambda A_A^{\left(\mathrm{T}\right)} &= 0. \\
\end{aligned}
\end{equation}
Note that the transverse vectors $A_A^{\left(\mathrm{T}\right)}$ are gauge invariant, while the longitudinal modes $A^{\left(\mathrm{L}\right)}$ are redundant degrees of freedom, being pure gauge.
Another gauge invariant quantity can be constructed as
\begin{equation}
\mathcal{A}_a=A_a-D_aA^{\left(\mathrm{L}\right)},
\end{equation}
then, the $(2 + (n - 2))$-decomposed field strength tensor can be written as
\begin{equation}
\begin{aligned}
F_{ab} &= D_a \mathcal{A}_b - D_b \mathcal{A}_a, \\
F_{aA} &= D_a A_A^{\left(\mathrm{T}\right)} - D_A \mathcal{A}_a, \\
F_{AB} &= D_{A} A_{B}^{\left(\mathrm{T}\right)} - D_{B} A_{A}^{\left(\mathrm{T}\right)}. \\
\end{aligned}
\end{equation}
Next, the scalar $\mathcal{A}^{a}$ is expanded into the scalar spherical harmonic modes $\mathcal{Y}_{\ell,\mathbf{m}}\left(\theta\right)$, and the transverse vector $A_{A}^{\left(\mathrm{T}\right)}$ is expanded into the transverse vector spherical harmonic mode $\mathcal{Y}_{A;\ell,\mathbf{m}}^{\left(\mathrm{T}\right)}\left(\theta\right)$,
\begin{equation}
\begin{aligned}
\mathcal{A}^{a}\left(x\right) &= \sum_{\ell,\mathbf{m}} \mathcal{A}_{\ell,\mathbf{m}}^{a}\left(t,r\right) \mathcal{Y}_{\ell,\mathbf{m}}\left(\theta\right), \\
A_{A}^{\left(\mathrm{T}\right)}\left(x\right) &= \sum_{\ell,\mathbf{m}} A_{\ell,\mathbf{m}}^{\left(\mathrm{V}\right)}\left(t,r\right) \mathcal{Y}_{A;\ell,\mathbf{m}}^{\left(\mathrm{T}\right)}\left(\theta\right). \\
\end{aligned}
\end{equation}
Due to the spherical symmetry of the background, the scalar modes $\mathcal{A}_{\ell,\mathbf{m}}^{a}$ and the vector modes $A_{\ell,\mathbf{m}}^{\left(\mathrm{T}\right)}$ will be completely decouple from each other.
Therefore, the Maxwell action is expanded into the following expression:
\begin{equation}
\begin{aligned}
&\mathcal{S}^{(1)}=\sum_{\ell,\mathbf{m}}\left(\mathcal{S}_{\ell,\mathbf{m}}^{\left(\mathrm{V}\right)}+\mathcal{S}_{\ell,\mathbf{m}}^{\left(\mathrm{S}\right)}\right), \\
&\mathcal{S}_{\ell,\mathbf{m}}^{\left(\mathrm{V}\right)}=\int \mathrm{d}^{2}x\sqrt{-g^{(2)}}r^{n-4}\left[-\frac{1}{2}D_{a}\bar{A}_{\ell,\mathbf{m}}^{\left(\mathrm{V}\right)}D^{a}A_{\ell,\mathbf{m}}^{\left(\mathrm{V}\right)}-\frac{1}{2}\frac{(\ell+1)\left(\ell+n-4\right)}{r^{2}}\left|A_{\ell,\mathbf{m}}^{\left(\mathrm{V}\right)}\right|^{2}\right], \\
&\mathcal{S}_{\ell,\mathbf{m}}^{\left(\mathrm{S}\right)}=\int \mathrm{d}^{2}x\sqrt{-g^{(2)}}r^{n-2}\left[-\frac{1}{4}\bar{\digamma}_{ab;\ell,\mathbf{m}}\digamma_{\ell,\mathbf{m}}^{ab}-\frac{1}{2}\frac{\ell\left(\ell+n-3\right)}{r^{2}}\bar{\mathcal{A}}_{a;\ell,\mathbf{m}}\mathcal{A}_{\ell,\mathbf{m}}^{a}\right], \\
\end{aligned}\end{equation}
where
\begin{equation}
\digamma_{\ell,\mathbf{m}}^{ab}\equiv D^a\mathcal{A}_{\ell,\mathbf{m}}^b-D^b\mathcal{A}_{\ell,\mathbf{m}}^a.
\end{equation}

For the vector modes $A_{\ell,\mathbf{m}}^{\left(\mathrm{V}\right)}$, they are odd under parity transformations and gauge invariant. 
Here we define the following variable:
\begin{equation}
A_{\ell,\mathbf{m}}^{\left(\mathrm{V}\right)}\left(t,r\right) = \frac{\Psi_{\ell,\mathbf{m}}^{\left(\mathrm{V}\right)}\left(t,r\right)}{r^{(n-4)/2}},
\end{equation}
and then the reduced action of the vector modes can be written as
\begin{equation}
\mathcal{S}_{\ell,\mathbf{m}}^{\left(\mathrm{V}\right)}=\int \mathrm{d}^2x\sqrt{-g^{(2)}}\left[-\frac{1}{2}D_a\bar{\Psi}_{\ell,\mathbf{m}}^{\left(\mathrm{V}\right)}D^a\Psi_{\ell,\mathbf{m}}^{\left(\mathrm{V}\right)}-\frac{1}{2f_{t}}V_{\ell}^{\left(\mathrm{V}\right)}\left(r\right)\left|\Psi_{\ell,\mathbf{m}}^{\left(\mathrm{V}\right)}\right|^2\right],
\end{equation}
where the potential is given by
\begin{equation}\label{eq:vlv}
V_{\ell}^{\left(\mathrm{V}\right)}\left(r\right) \equiv f_{t}\Bigg\{ \frac{(\ell+1)\left(\ell+n-4\right)}{r^{2}}+\frac{\left(n-4\right)\left(n-6\right)}{4r^{2}}f_{r}+\frac{n-4}{2r} \left[\frac{\left(f_{t}f_{r}\right)^{\prime}}{2f_{t}}\right] \Bigg\}.
\end{equation}
Using the tortoise coordinate, the action in 2-$n$ flat space-time will become
\begin{equation}
\mathcal{S}_{\ell,\mathbf{m}}^{\left(\mathrm{V}\right)} = \int \mathrm{d}t\mathrm{d}{r_\ast}\left[\frac{1}{2}\left|\partial_t\Psi_{\ell,\mathbf{m}}^{\left(\mathrm{V}\right)}\right|^2-\frac{1}{2}\left|\partial_{r_\ast}\Psi_{\ell,\mathbf{m}}^{\left(\mathrm{V}\right)}\right|^2-\frac{1}{2}V_\ell^{\left(\mathrm{V}\right)}\left(r\right)\left|\Psi_{\ell,\mathbf{m}}^{\left(\mathrm{V}\right)}\right|^2\right],
\end{equation}
and the equation of motion for $\Psi_{\ell,\mathbf{m}}^{\left(\mathrm{V}\right)}\left(r_\ast\right)$ is the Schr\"{o}dinger-like equation
\begin{equation}
\left[\partial_{r_\ast}^2-\partial_t^2-V_\ell^{\left(\mathrm{V}\right)}\left(r\right)\right]\Psi_{\ell,\mathbf{m}}^{\left(\mathrm{V}\right)}\left(r_\ast\right)=0.
\end{equation}

For the analysis of the scalar modes, we will introduce an auxiliary field $\Psi_{\ell,\mathbf{m}}^{\left(\mathrm{S}\right)}(t,r)$, whose action is
\begin{equation}
\begin{aligned}
\tilde{\mathcal{S}}_{\ell,\mathbf{m}}^{\left(\mathrm{S}\right)} =& \int \mathrm{d}^2x\sqrt{-g^{(2)}}\Bigg[ \frac{1}{2}\sqrt{\ell\left(\ell+n-3\right)} r^{\frac{(n-4)}{2}} \mathrm{Re}\left\{\bar{\Psi}_{\ell,\mathbf{m}}^{\left(\mathrm{S}\right)}\varepsilon_{ab}\digamma_{\ell,\mathbf{m}}^{ab}\right\} \\
 & -\frac{1}{2}\frac{\ell\left(\ell+n-3\right)}{r^{2}}\left(\left|\Psi_{\ell,\mathbf{m}}^{\left(\mathrm{S}\right)}\right|^{2}+r^{n-2}\bar{\mathcal{A}}_{a;\ell,\mathbf{m}}\mathcal{A}_{\ell,\mathbf{m}}^{a}\right) \Bigg],
\end{aligned}
\end{equation}
where $\varepsilon_{ab}$ is the Levi-Civita tensor.
The auxiliary field $\Psi_{\ell,\mathbf{m}}^{\left(\mathrm{S}\right)}$ can be written as
\begin{equation}
\Psi_{\ell,\mathbf{m}}^{\left(\mathrm{S}\right)}=\frac{1}{2}\frac{r^{\frac{n}{2}}}{\sqrt{\ell\left(\ell+n-3\right)}}\varepsilon_{ab}\digamma_{\ell,\mathbf{m}}^{ab},
\end{equation}
we can know that $\Psi_{\ell,\mathbf{m}}^{\left(\mathrm{S}\right)}$ is gauge invariant by using the gauge transformation \eqref{eq:gt}.
Utilizing this new action $\tilde{\mathcal{S}}_{\ell,\mathbf{m}}^{\left(\mathrm{S}\right)}$ as an alternative, we will be able to reformulate it in the form of the scalar field modes and the gauge field vector modes.
$\mathcal{A}_{\ell,\mathbf{m}}^a$ can be obtained as
\begin{equation}
\mathcal{A}_{\ell,\mathbf{m}}^a=\frac{r^{-\frac{(n-4)}{2}}}{\sqrt{\ell\left(\ell+n-3\right)}}\left\{\varepsilon^{ab}D_b-t^a\frac{n-4}{2r}\sqrt{\frac{f_r}{f_t}}\right\}\Psi_{\ell,\mathbf{m}}^{\left(\mathrm{S}\right)}.
\end{equation}
The action for $\Psi_{\ell,\mathbf{m}}^{\left(\mathrm{S}\right)}$ can be expressed as
\begin{equation}
\begin{aligned}
\tilde{\mathcal{S}}_{\ell,\mathbf{m}}^{\left(\mathrm{S}\right)} &= \int \mathrm{d}^2x\sqrt{-g^{(2)}}\left[-\frac{1}{2}D_a\bar{\Psi}_{\ell,\mathbf{m}}^{\left(\mathrm{S}\right)}D^a\Psi_{\ell,\mathbf{m}}^{\left(\mathrm{S}\right)}-\frac{1}{2f_{t}}V_\ell^{\left(\mathrm{S}\right)}\left(r\right)\left|\Psi_{\ell,\mathbf{m}}^{\left(\mathrm{S}\right)}\right|^2\right] \\
&= \int \mathrm{d}t\mathrm{d}r_{\ast}\left[\frac{1}{2}\left|\partial_{t}\Psi_{\ell,\mathbf{m}}^{\left(\mathrm{S}\right)}\right|^{2}-\frac{1}{2}\left|\partial_{r_{\ast}}\Psi_{\ell,\mathbf{m}}^{\left(\mathrm{S}\right)}\right|^{2}-\frac{1}{2}V_{\ell}^{\left(\mathrm{S}\right)}\left(r\right)\left|\Psi_{\ell,\mathbf{m}}^{\left(\mathrm{S}\right)}\right|^{2}\right],
\end{aligned}
\end{equation}
which can also be written in the form of the Schr\"{o}dinger-like as
\begin{equation}
\left[\partial_{r_\ast}^2-\partial_t^2-V_\ell^{\left(\mathrm{S}\right)}\left(r\right)\right]\Psi_{\ell,\mathbf{m}}^{\left(\mathrm{S}\right)}\left(r_\ast\right)=0,
\end{equation}
where the potential is
\begin{equation}\label{eq:vls}
V_{\ell}^{\left(\mathrm{S}\right)}\left(r\right) \equiv f_{t}\Bigg\{ \frac{\ell\left(\ell+n-3\right)}{r^2}+\frac{\left(n-2\right)\left(n-4\right)}{4r^2}f_{r}-\frac{n-4}{2r} \left[\frac{\left(f_{t}f_{r}\right)^{\prime}}{2f_{t}}\right] \Bigg\}.
\end{equation}

\subsubsection{$p$-form perturbations}
A completely antisymmetric gauge field tensor $A_{\mu_1\mu_2\cdots\mu_p}$ will generate the so-called $p$-form perturbations in high-dimensional space-time, We defined the $p$-form field $\mathbb{A}^{(p)}$ of rank $1 \leqslant p \leqslant n-3$ as follows:
\begin{equation}
\mathbb{A}^{(p)}=\frac{1}{p!}A_{\mu_1\mu_2\cdots\mu_p}\mathrm{d}x^{\mu_1}\wedge\mathrm{d}x^{\mu_2}\wedge\cdots\wedge\mathrm{d}x^{\mu_p}.
\end{equation}
The $p$-form generalization of Maxwell's action is defined as
\begin{equation}
\mathcal{S}^{(p)}=-\frac{1}{2} \int \mathbb{F}^{(p+1)}\wedge{ }^{\mathbf{\star}}\mathbb{F}^{(p+1)},
\end{equation}
where $\mathbb{F}^{(p+1)}=\mathrm{d}\mathbb{A}^{(p)}$ is the $(p+1)$-form field strength tensor, or written as
\begin{equation}
F_{\mu_1\mu_2\cdots\mu_{p+1}}=(p+1)\partial_{[\mu_1}A_{\mu_2\cdots\mu_{p+1}]},
\end{equation}
and the $p$-form action can be rewritten as
\begin{equation}
\mathcal{S}^{(p)}=-\frac{1}{2\left(p+1\right)!} \int \mathrm{d}^{n}x\sqrt{-g}F_{\mu_1\mu_2\cdots\mu_{p+1}}F^{\mu_1\mu_2\cdots\mu_{p+1}}.
\end{equation}
The operator $\mathrm{d}$ is defined as the exterior derivative satisfying $\mathrm{d}^2 = 0$.
Moreover, we can define the co-derivative operator $\delta$ of $p$-form as $\delta\equiv(-1)^{n(p+1)+1}\,\mathbf{\star}\,\mathrm{d}\, \mathbf{\star}$, and it also satisfies $\delta^2 = 0$.

The $2 + (n - 2)$ decomposition of $p$-form gauge fields into irreducible representations of $SO(n-1)$ is accomplished via the Hodge decomposition on $\mathbb{S}^{n-2}$ \cite{Yoshida:2019tvk}.
This analysis has been extended to general spherically symmetric background geometry.
Here, we follow the analysis and notation in the paper \cite{Charalambous:2024tdj}.
A $p$-form on $\mathbb{S}^{n-2}$ is denoted by
\begin{equation}
\hat{\mathbb{A}}^{(p)} = \frac{1}{p!} A_{A_1 A_2 \cdots A_p} \mathscr{D}^{A_1 A_2 \cdots A_p},
\end{equation}
where
\begin{equation}
\mathscr{D}^{A_1 A_2 \cdots A_p} \equiv \mathrm{d}x^{A_1}\wedge\mathrm{d}x^{A_2}\wedge\cdots\wedge\mathrm{d}x^{A_p},
\end{equation}
and the spherical coordinate is expressed by $x^A = \left( \theta^1, \theta^2, \cdots, \theta^{n-2} \right)$.
Subsequently, the components of $\mathbb{A}^{(p)}$ can be expressed as the following equation:
\begin{equation}
\begin{aligned}
\mathbb{A}^{(p)} =& \frac{1}{2} \frac{1}{(p-2)!} A_{ab A_1 \cdots A_{p-2}} \mathrm{d}x^a\wedge\mathrm{d}x^b\wedge\mathscr{D}^{A_1 \cdots A_{p-2}} \\
& +\frac{1}{(p-1)!} A_{a A_1 \cdots A_{p-1}} \mathrm{d}x^a\wedge\mathscr{D}^{A_1 \cdots A_{p-1}}  \\
& +\frac{1}{p!} A_{A_1 \cdots A_p} \mathscr{D}^{A_1 \cdots A_{p}}.
\end{aligned}
\end{equation}
We then define the components of the space-time $p$-form gauge field $\mathbb{A}^{(p)}$ as
\begin{equation}
\begin{aligned}
\left(T_{ab}\right)_{A_1 \cdots A_{p-2}} &\equiv A_{ab A_{1} \cdots A_{p-2}}, \\
\left(V_a\right)_{A_1 \cdots A_{p-1}} &\equiv A_{a A_{1} \cdots A_{p-1}}, \\
X_{A_1 \cdots A_p} &\equiv A_{A_{1} \cdots A_p}. \\
\end{aligned}
\end{equation}
Thus, $\mathbb{A}^{(p)}$ can be written as
\begin{equation}
\mathbb{A}^{(p)}=\frac{1}{2}\mathrm{d}x^a\wedge\mathrm{d}x^b\wedge\hat{\mathbb{T}}_{ab}^{(p-2)}+\mathrm{d}x^a\wedge\hat{\mathbb{V}}_a^{(p-1)}+\hat{\mathbb{X}}^{(p)},
\end{equation}
where the components of $\hat{\mathbb{T}}_{ab}^{(p-2)}$, $\hat{\mathbb{V}}_a^{(p-1)}$ and $\hat{\mathbb{X}}^{(p)}$ on the sphere are represented as $\left(T_{ab}\right)_{A_1 \cdots A_{p-2}}$, $\left(V_a\right)_{A_1 \cdots A_{p-1}}$ and $X_{A_1 \cdots A_p}$, respectively.
The Hodge decomposition on $\mathbb{S}^{n-2}$ is operated as a general $p$-form $\hat{\mathbb{A}}^{(p)}$ on the sphere can be decomposed into a ``longitudinal'' $(p-1)$-form $\hat{\mathbb{A}}^{(p-1)}$ and a ``transverse'' $p$-form $\hat{\mathcal{A}}^{(p)}$, that is,
\begin{equation}\label{eq:hatap}
\begin{aligned}
\hat{\mathbb{A}}^{(p)} &= \hat{\mathrm{d}}\hat{\mathbb{A}}^{(p-1)}+\hat{\mathcal{A}}^{(p)} \\
&= \hat{\mathrm{d}}\left(\hat{\mathrm{d}}\hat{\mathbb{A}}^{(p-2)}+\hat{\mathcal{A}}^{(p-1)}\right)+\hat{\mathcal{A}}^{(p)} \\
&= \hat{\mathrm{d}}\hat{\mathcal{A}}^{(p-1)}+\hat{\mathcal{A}}^{(p)},
\end{aligned}
\end{equation}
where $\hat{\mathcal{A}}^{(p)}$ is the co-exact $p$-form on the sphere $\mathbb{S}^{n-2}$.
This naturally satisfies $\hat{\delta}\hat{\mathcal{A}}^{(p)}=0$, with its component form expressed as
\begin{equation}
D^{A_1} \mathcal{A}_{A_1 A_2 \cdots A_{p+1}}=0.
\end{equation}
The mathematical formula \eqref{eq:hatap} reveals that the co-exact form fields can comprehensively replace all the general form fields on $\mathbb{S}^{n-2}$.
Therefore, the $(2 + (n - 2))$-decomposition of the $p$-form gauge field into irreducible representation of $SO(n-1)$ can be written as:
\begin{equation}
\begin{aligned}
\mathbb{A}^{(p)}=&\frac{1}{2} \mathrm{d}x^a\wedge\mathrm{d}x^b\wedge\left(\hat{\mathrm{d}}\hat{\mathcal{T}}_{ab}^{(p-3)}+\hat{\mathcal{T}}_{ab}^{(p-2)}\right) \\
&+\mathrm{d}x^a\wedge\left(\hat{\mathrm{d}}\hat{\mathcal{V}}_a^{(p-2)}+\hat{\mathcal{V}}_a^{(p-1)}\right) \\
&+\left(\hat{\mathrm{d}}\hat{\mathcal{X}}^{(p-1)}+\hat{\mathcal{X}}^{(p)}\right),
\end{aligned}
\end{equation}
where $\hat{\mathcal{T}}_{ab}$, $\hat{\mathcal{V}}_a$ and $\hat{\mathcal{X}}$ appear as co-exact forms on $\mathbb{S}^{n-2}$.
Since the $p$-form is invariant under the gauge transformations,
\begin{equation}
\begin{aligned}
\mathrm{d}\Lambda^{(p-1)}  =& \delta_{\Lambda}\mathbb{A}^{(p)} \\
=& \frac{1}{2} \mathrm{d}x^a\wedge\mathrm{d}x^b\wedge \left(\hat{\mathrm{d}}\hat{\varLambda}_{ab}^{(p-3)} + 2D_{[a} \hat{\varLambda}_{b]}^{(p-2)}\right) \\
& +\mathrm{d}x^a\wedge\left(-\hat{\mathrm{d}}\hat{\varLambda}_a^{(p-2)}+D_a\hat{\varLambda}^{(p-1)}\right) \\
& +\hat{\mathrm{d}}\hat{\varLambda}^{(p-1)}.
\end{aligned}
\end{equation}
The aforementioned outcomes were achieved by utilizing the following transformation methods:
\begin{equation}
\begin{aligned}
&\delta_{\Lambda}\hat{\mathcal{T}}_{ab}^{(p-3)} = \hat{\varLambda}_{ab}^{(p-3)}, \quad\quad \delta_{\Lambda}\hat{\mathcal{T}}_{ab}^{(p-2)} = 2 D_{[a} \hat{\varLambda}_{b]}^{(p-2)}; \\
&\delta_{\Lambda}\hat{\mathcal{V}}_a^{(p-2)} = -\hat{\varLambda}_a^{(p-2)}, \quad\quad \delta_{\Lambda}\hat{\mathcal{V}}_a^{(p-1)} = D_{a}\hat{\varLambda}^{(p-1)}; \\
&\delta_{\Lambda}\hat{\mathcal{X}}^{(p-1)} = \hat{\varLambda}^{(p-1)}, \quad\quad \delta_{\Lambda}\hat{\mathcal{X}}^{(p)} = 0. \\
\end{aligned}
\end{equation}
By constructing gauge invariant combinations
\begin{equation}
\begin{aligned}
\hat{\mathcal{H}}_{ab}^{(p-2)} &= \hat{\mathcal{T}}_{ab}^{(p-2)} + 2D_{[a} \hat{\mathcal{V}}_{b]}^{(p-2)}, \\
\hat{\mathcal{A}}_a^{(p-1)} &= \hat{\mathcal{V}}_a^{(p-1)} - D_{a}\hat{\mathcal{X}}^{(p-1)},
\end{aligned}
\end{equation}
the $(p+1)$-form field strength can be written as
\begin{equation}
\begin{aligned}
\mathbb{F}^{(p+1)} =& \frac{1}{2} \mathrm{d}x^a\wedge\mathrm{d}x^b\wedge\left(2D_{[a}\hat{\mathcal{A}}_{b]}^{(p-1)}+(p-2)!\,\hat{\mathrm{d}}\hat{\mathcal{H}}_{ab}^{(p-2)}\right) \\
& +\mathrm{d}x^a\wedge\left(D_a\hat{\mathcal{X}}^{(p)}-\hat{\mathrm{d}}\hat{\mathcal{A}}_a^{(p-1)}\right) \\
& +\hat{\mathrm{d}}\hat{\mathcal{X}}^{(p)}.
\end{aligned}
\end{equation}
We can now expand it into the spherical harmonics of the co-exact $p$-form field on the sphere,
\begin{equation}
\begin{aligned}
\left(\mathcal{H}^{ab}\right)^{A_{1} \cdots A_{p-2}}(x) &= \sum_{\ell,\mathbf{m}}\mathcal{H}_{\ell,\mathbf{m}}^{ab}\left(t,r\right) \mathfrak{Y}_{\ell,\mathbf{m}}^{\left(\mathrm{T}\right)) A_{1} \cdots A_{p-2}}\left(\theta\right), \\
\left(\mathcal{A}^a\right)^{A_1 \cdots A_{p-1}}\left(x\right) &= \sum_{\ell,\mathbf{m}}\mathcal{A}_{\ell,\mathbf{m}}^a\left(t,r\right) \mathfrak{Y}_{\ell,\mathbf{m}}^{\left(\mathrm{T}\right)) A_1 \cdots A_{p-1}}\left(\theta\right), \\
\mathcal{X}^{A_1 \cdots A_p}\left(x\right) &= \sum_{\ell,\mathbf{m}}\mathcal{X}_{\ell,\mathbf{m}}\left(t,r\right) \mathfrak{Y}_{\ell,\mathbf{m}}^{\left(\mathrm{T}\right))A_1 \cdots A_p}\left(\theta\right),
\end{aligned}
\end{equation}
where $\mathfrak{Y}_{A_1 \cdots A_p; \ell,\mathbf{m}}^{\left(\mathrm{T}\right)}$ is defined as
\begin{equation}
\hat{\mathfrak{Y}}_{\ell,\mathbf{m}}^{\left(\mathrm{T}\right)\,(p)} = \frac{1}{p!} \mathfrak{Y}_{A_1 \cdots A_p; \ell,\mathbf{m}}^{\left(\mathrm{T}\right)} \mathscr{D}^{A_1 \cdots A_p},
\end{equation}
which satisfies 
\begin{equation}
\begin{aligned}
\hat{\Delta} \hat{\mathfrak{Y}}_{\ell,\mathbf{m}}^{\left(\mathrm{T}\right)\,(p)} &= \lambda_{p} \hat{\mathfrak{Y}}_{\ell,\mathbf{m}}^{\left(\mathrm{T}\right)\,(p)}, \\
\hat{\delta}_{\Lambda} \hat{\mathfrak{Y}}_{\ell,\mathbf{m}}^{\left(\mathrm{T}\right)\,(p)} &= 0. \\
\end{aligned}
\end{equation}
Here, the symbol $\hat{\Delta}$ is represented as the Laplace-Beltrami operator $\hat{\Delta} \equiv \hat{\delta}\hat{\mathrm{d}}+\hat{\mathrm{d}}\hat{\delta}$, and $\lambda_{p} \equiv \left(\ell+p\right)\left(\ell+n-p-3\right)$.
Thus, we can get the following equation
\begin{equation}
\hat{\delta}\hat{\mathrm{d}} \hat{\mathfrak{Y}}_{\ell,\mathbf{m}}^{\left(\mathrm{T}\right)\,(p)} = \lambda_{p} \hat{\mathfrak{Y}}_{\ell,\mathbf{m}}^{\left(\mathrm{T}\right)\,(p)}. \\
\end{equation}
The above equation can be further rewritten as
\begin{equation}
\tilde{\Delta} \mathfrak{Y}_{A_1 \cdots A_p; \ell,\mathbf{m}}^{\left(\mathrm{T}\right)} = -\gamma_{p} \mathfrak{Y}_{A_1 \cdots A_p; \ell,\mathbf{m}}^{\left(\mathrm{T}\right)},
\end{equation}
where $\tilde{\Delta}$ is defined as $\tilde{\Delta} \mathfrak{Y}_{A_1 \cdots A_p; \ell,\mathbf{m}}^{\left(\mathrm{T}\right)} = {{\mathfrak{Y}_{A_1 \cdots A_p; \ell,\mathbf{m}}}^{:A}}{ }_{:A}$, and $\gamma_{p} = \ell\left(\ell+n-3\right)-p$.

The action of the $p$-form can be written as
\begin{equation}
\begin{aligned}
\mathcal{S}^{(p)}=& \sum_{\ell,\mathbf{m}}\left(\mathcal{S}_{\ell,\mathbf{m}}^{(p)}+\mathcal{S}_{\ell,\mathbf{m}}^{(p-1)}+\mathcal{S}_{\ell,\mathbf{m}}^{(p-2)}\right), \\
\mathcal{S}_{\ell,\mathbf{m}}^{(p)} =& \int \mathrm{d}^2x \sqrt{-g^{(2)}}r^{n-2p-2} \left[-\frac{1}{2p!}D_{a}\bar{\mathcal{X}}_{\ell,\mathbf{m}}D^{a}\mathcal{X}_{\ell,\mathbf{m}}-\frac{1}{2p!}\frac{\left(\ell+p\right)\left(\ell+n-p-3\right)}{r^2}\left|\mathcal{X}_{\ell,\mathbf{m}}\right|^2\right], \\
\mathcal{S}_{\ell,\mathbf{m}}^{(p-1)} =& \int \mathrm{d}^2x\sqrt{-g^{(2)}}r^{n-2p} \left[-\frac{1}{4\left(p-1\right)!}\bar{\mathfrak{F}}_{ab;\ell,\mathbf{m}}\mathfrak{F}_{\ell,\mathbf{m}}^{ab}\right. \\
& \left. -\frac{1}{2\left(p-1\right)!}\frac{\left(\ell+p-1\right)\left(\ell+n-p-2\right)}{r^{2}}\bar{\mathcal{A}}_{a;\ell,\mathbf{m}}\mathcal{A}_{\ell,\mathbf{m}}^{a}\right], \\
\mathcal{S}_{\ell,\mathbf{m}}^{(p-2)} =& \int \mathrm{d}^2x\sqrt{-g^{(2)}}r^{n-2p+2} \left[-\frac{(p-2)!}{4}\frac{\left(\ell+p-2\right)\left(\ell+n-p-1\right)}{r^2}\bar{\mathcal{H}}_{ab;\ell,\mathbf{m}}\mathcal{H}_{\ell,\mathbf{m}}^{ab}\right],
\end{aligned}
\end{equation}
where
\begin{equation}
\mathfrak{F}_{\ell,\mathbf{m}}^{ab}\equiv D^a\mathcal{A}_{\ell,\mathbf{m}}^b-D^b\mathcal{A}_{\ell,\mathbf{m}}^a.
\end{equation}
Notably, the $(p-2)$-form sector generated by spherical harmonic mode $\mathcal{H}_{\ell,\mathbf{m}}^{ab}$ is trivial, namely $\mathcal{H}_{\ell,\mathbf{m}}^{ab}=0$.

The master variable for the $p$-form component is
\begin{equation}
\mathcal{X}_{\ell,\mathbf{m}}\left(t,r\right)=\sqrt{p!}\frac{\Psi_{\ell,\mathbf{m}}^{(p)}\left(t,r\right)}{r^{\left(n-2p-2\right)/2}},
\end{equation}
the reduced action for the $p$-form modes can be written as
\begin{equation}
\mathcal{S}_{\ell,\mathbf{m}}^{(p)}=\int \mathrm{d}^2x\sqrt{-g^{(2)}}\left[-\frac{1}{2}D_a\bar{\Psi}_{\ell,\mathbf{m}}^{(p)}D^a\Psi_{\ell,\mathbf{m}}^{(p)}-\frac{1}{2f_{t}}V_{\ell}^{(p)}\left(r\right)\left|\Psi_{\ell,\mathbf{m}}^{(p)}\right|^2\right],
\end{equation}
where the potential is
\begin{equation}\label{eq:vlp}
\begin{aligned}
V_{\ell}^{(p)}\left(r\right) \equiv & f_{t}\Bigg\{ \frac{(\ell+p)\left(\ell+n-p-3\right)}{r^{2}}+\frac{\left(n-2p-2\right)\left(n-2p-4\right)}{4r^{2}}f_{r} \\
& + \frac{n-2p-2}{2r} \left[\frac{\left(f_{t}f_{r}\right)^{\prime}}{2f_{t}}\right] \Bigg\}.
\end{aligned}
\end{equation}

For the analysis of the $(p-1)$-form, we will introduce an auxiliary 2-$n$ scalar field $\Psi_{\ell,\mathbf{m}}^{(\tilde{p})}\left(t,r\right)$ and the action is denoted by
\begin{equation}
\begin{aligned}
\tilde{\mathcal{S}}_{\ell,\mathbf{m}}^{(p-1)} =& \int \mathrm{d}^2x\sqrt{-g^{(2)}} \Bigg[\frac{1}{2}\sqrt{\frac{\left(\ell+p-1\right)\left(\ell+n-p-2\right)}{(p-1)!}}r^{\frac{\left(n-2p-2\right)}{2}}\mathrm{Re}\left\{\bar{\Psi}_{\ell,\mathbf{m}}^{\left(\tilde{p}\right)}\varepsilon_{ab}\mathfrak{F}_{\ell,\mathbf{m}}^{ab}\right\} \\
& -\frac{1}{2}\frac{\left(\ell+p-1\right)\left(\ell+n-p-2\right)}{r^2}\left(\left|\Psi_{\ell,\mathbf{m}}^{\left(\tilde{p}\right)}\right|^2+\frac{r^{n-2p}}{(p-1)!}\bar{\mathcal{A}}_{a;\ell,\mathbf{m}}\mathcal{A}_{\ell,\mathbf{m}}^a\right) \Bigg],
\end{aligned}
\end{equation}
which is in principle equivalent to the original action $\mathcal{S}_{\ell,\mathbf{m}}^{(p-1)}$.
The auxiliary field $\Psi_{\ell,\mathbf{m}}^{\left(\tilde{p}\right)}$ can be written as
\begin{equation}
\Psi_{\ell,\mathbf{m}}^{\left(\tilde{p}\right)} = \frac{1}{2}\frac{r^{\frac{\left(n-2p+2\right)}{2}}}{\sqrt{(p-1)!\left(\ell+p-1\right)\left(\ell+n-p-2\right)}}\varepsilon_{ab}\mathfrak{F}_{\ell,\mathbf{m}}^{ab},
\end{equation}
and $\mathcal{A}_{\ell,\mathbf{m}}^{a}$ is given by
\begin{equation}
\mathcal{A}_{\ell,\mathbf{m}}^{a} = \sqrt{\frac{(p-1)!}{\left(\ell+p-1\right)\left(\ell+n-p-2\right)}} r^{-\frac{\left(n-2p-2\right)}{2}}\left\{\varepsilon^{ab}D_b-t^a\frac{n-2p-2}{2r}\sqrt{\frac{f_r}{f_t}}\right\}\Psi_{\ell,\mathbf{m}}^{\left(\tilde{p}\right)}.
\end{equation}
Then the action $\tilde{\mathcal{S}}_{\ell,\mathbf{m}}^{(p-1)}$ is reduced to
\begin{equation}
\tilde{\mathcal{S}}_{\ell,\mathbf{m}}^{(p-1)} = \int \mathrm{d}^2x\sqrt{-g^{(2)}}\left[-\frac{1}{2}D_a\bar{\Psi}_{\ell,\mathbf{m}}^{(\tilde{p})}D^a\Psi_{\ell,\mathbf{m}}^{\left(\tilde{p}\right)}-\frac{1}{2f_{t}}V_\ell^{\left(\tilde{p}\right)}\left(r\right)\left|\Psi_{\ell,\mathbf{m}}^{\left(\tilde{p}\right)}\right|^2\right],
\end{equation}
where the potential is
\begin{equation}\label{eq:vltildep}
\begin{aligned}
V_{\ell}^{\left(\tilde{p}\right)}\left(r\right) \equiv& f_{t}\Bigg\{ \frac{\left(\ell+p-1\right)\left(\ell+n-p-2\right)}{r^{2}}+\frac{\left(n-2p\right)\left(n-2p-2\right)}{4r^{2}}f_{r} \\
& -\frac{\left(n-2p-2\right)}{2r} \left[\frac{\left(f_{t}f_{r}\right)^{\prime}}{2f_{t}}\right] \Bigg\}.
\end{aligned}
\end{equation}

We hereby present a comprehensive formulation of the effective potential governing the perturbations of spin-0, spin-1, and $p$-form as follows:
\begin{equation}
\begin{aligned}
V_{\ell}^{(\varkappa)}\left(r\right)\equiv& f_{t}\Bigg\{ \frac{(\ell+\varkappa)\left(\ell+n-\varkappa-3\right)}{r^{2}}+\frac{\left(n-2\varkappa-2\right)\left(n-2\varkappa-4\right)}{4r^{2}}f_{r} \\
& +\frac{n-2\varkappa-2}{2r} \left[\frac{\left(f_{t}f_{r}\right)^{\prime}}{2f_{t}}\right] \Bigg\}.
\end{aligned}
\end{equation}\
This general formulation can be reduced to Eqs. \eqref{eq:vl0}, \eqref{eq:vlv}, \eqref{eq:vls}, \eqref{eq:vlp} and \eqref{eq:vltildep} through the specific assignment of the parameter $\varkappa$, that is,
\begin{equation}\label{eq:v0vvvsvp1vp2}
V_{\ell}^{(\varkappa)}\left(r\right) \Rightarrow \left\{\begin{array}{ll}
V_{\ell}^{(0)}\left(r\right), \quad & \varkappa=0; \\
V_{\ell}^{\left(\mathrm{V}\right)}\left(r\right), \quad & \varkappa=1; \\
V_{\ell}^{\left(\mathrm{S}\right)}\left(r\right), \quad & \varkappa=n-3; \\
V_{\ell}^{(p)}\left(r\right), \quad & \varkappa=p; \\
V_{\ell}^{(\tilde{p})}\left(r\right), \quad & \varkappa=\tilde{p}=n-p-2.
\end{array}\right.
\end{equation}

\subsubsection{Spin-2 perturbations}
We will eventually investigate gravitational perturbations, which are more complicated and require the Kodama-Ishibashi classification \cite{Kodama:2003jz, Ishibashi:2011ws} of high-dimensional perturbations (scalar-type, vector-type and tensor-type).
When analyzing the action that describes massless spin-2 field perturbations, we first decompose the metric perturbation $h_{\mu\nu}$ under the irreducible $SO(n-1)$ representation.
The $\frac{n(n+1)}{2}$ components of the $h_{\mu\nu}$ are expressed as follows:
\begin{equation}
\begin{aligned}
h_{ab}\left(x\right) &= H_{ab}\left(x\right), \\
h_{aA}\left(x\right) &= D_AH_a^{\left(\mathrm{S}\right)}\left(x\right)+h_{aA}^{\left(\mathrm{V}\right)}\left(x\right), \\
h_{AB}\left(x\right) &= r^{2} \left[K\left(x\right)\Omega_{AB}+D_{\langle A}D_{B \rangle}G\left(x\right)+D_{( A}h_{B )}^{\left(\mathrm{V}\right)}\left(x\right)+h_{AB}^{\left(\mathrm{TT}\right)}\left(x\right)\right],
\end{aligned}
\end{equation}
where the notation $\langle\cdots\rangle$ designates the trace-free symmetrized tensor portion associated with the bracketed indices.
In order to decompose the spin-2 field $h_{\mu\nu}$, it is essential to employ three specialized spherical harmonics: tensor harmonics $\mathscr{Y}_{AB;\ell,\mathbf{m}}^{\left(\mathrm{TT}\right)}$; (transverse) vector harmonics $\mathscr{Y}_{A;\ell,\mathbf{m}}^{\left(\mathrm{T}\right)}$; and scalar spherical harmonics $\mathscr{Y}_{\ell,\mathbf{m}}$.
One is composed of the perturbations proportional to seven scalar harmonics:
\begin{equation}
H_{ab}\left(x\right), \quad H_{a}^{\left(\mathrm{L}\right)}\left(x\right), \quad K\left(x\right),\quad G\left(x\right).
\end{equation}
Another is composed of the perturbations proportional to three vector harmonics:
\begin{equation}
h_{aA}^{\left(\mathrm{V}\right)}\left(x\right),\quad h_{A}^{\left(\mathrm{V}\right)}\left(x\right);
\end{equation}
with
\begin{equation}
\begin{aligned}
D^{A} h_{aA}^{\left(\mathrm{V}\right)}\left(x\right) &= 0, \\
D^{A} h_{A}^{\left(\mathrm{V}\right)}\left(x\right) &= 0.
\end{aligned}
\end{equation}
The last one is a single perturbation proportional to the tensor harmonic:
\begin{equation}
h_{AB}^{\left(\mathrm{TT}\right)}\left(x\right);
\end{equation}
with
\begin{equation}
\begin{aligned}
D^{A} h_{AB}^{\left(\mathrm{TT}\right)}\left(x\right) &= 0, \\
\Omega^{AB} h_{AB}^{\left(\mathrm{TT}\right)}\left(x\right) &= 0.
\end{aligned}
\end{equation}

Under infinitesimal diffeomorphisms $x^\mu\to x^\mu+\xi^\mu\left(x\right)$, the perturbation of a massless spin-2 field with gauge invariance is
\begin{equation}
\delta_{\xi} h_{\mu\nu} = \nabla_{\mu}\xi_{\nu}+\nabla_{\nu}\xi_{\mu}.
\end{equation}
Decomposing the diffeomorphism parameter $\xi$ into irreducible representations yields three scalars $\left(\xi_{a},\,\xi^{\left(\mathrm{S}\right)}\right)$ and one transverse vector $\left(\xi_{A}^{\left(\mathrm{V}\right)}\right)$ with $D^{A}\xi_{A}^{\left(\mathrm{V}\right)}=0$,
\begin{equation}
\begin{aligned}
&\xi_{a}\left(x\right), \\
&\xi_A\left(x\right) = D_A\xi^{\left(\mathrm{S}\right)}\left(x\right)+\xi_{A}^{\left(\mathrm{V}\right)}\left(x\right).
\end{aligned}
\end{equation}
The gauge transformation properties of the various $SO(n-1)$-decomposed components of $h_{\mu\nu}$ can be explicitly expressed as
\begin{equation}
\begin{aligned}
\delta_{\xi} H_{ab} &= D_{a}\xi_{b}+D_{b}\xi_{a}, \\
\delta_{\xi} H_{a}^{\left(\mathrm{S}\right)} &= \xi_{a}+D_{a}\xi^{\left(\mathrm{S}\right)}-\frac{2}{r}r_{a}\xi^{\left(\mathrm{S}\right)}, \\
\delta_{\xi} K &= \frac{2}{r} r^{a}\xi_{a}+\frac{2}{n-2}\frac{1}{r^2}D_{A}D^{A}\xi^{\left(\mathrm{S}\right)}, \\
\delta_{\xi} G &= \frac{2}{r^2}\xi^{\left(\mathrm{S}\right)}; \\
\delta_\xi h_{aA}^{\left(\mathrm{V}\right)} &= D_{a}\xi_A^{\left(\mathrm{V}\right)}-\frac{2}{r}r_{a}\xi_A^{\left(\mathrm{V}\right)}, \\
\delta_{\xi} h_A^{\left(\mathrm{V}\right)} &= \frac{2}{r^2}\xi_A^{\left(\mathrm{V}\right)}; \\
\delta_{\xi} h_{AB}^{\left(\mathrm{TT}\right)} &= 0.
\end{aligned}
\end{equation}
The transverse symmetric trace-free tensor is gauge invariant, and $H_{a}^{\left(\mathrm{S}\right)}$, $G$ and $h_a^{\left(\mathrm{V}\right)}$ are redundant degrees of freedom.
Here, instead of fixing the gauge, the gauge invariant quantities is adopted for research.
For tensor modes, there is the transverse symmetric trace-free tensor:
\begin{equation}
h_{AB}^{\left(\mathrm{TT}\right)}.
\end{equation}
For vector modes, we consider the gauge invariant combination:
\begin{equation}
\mathcal{H}_{aA}^{\left(\mathrm{V}\right)}=h_{aA}^{\left(\mathrm{V}\right)}-\frac{1}{2}r^2D_ah_A^{\left(\mathrm{V}\right)}.
\end{equation}
For scalar modes, the following two sets of gauge invariant combinations are included:
\begin{equation}
\begin{aligned}
\mathcal{H}_{ab} &= H_{ab}-2D_{(a}H_{b)}^{\left(\mathrm{S}\right)} + D_{(a}\left(r^2 D_{b)}G\right), \\
\mathcal{K} &= K-\frac{1}{n-2}D_{A}D^{A}G + r r^{a}D_{a}G - \frac{2}{r}r^{a}H_{a}^{\left(\mathrm{S}\right)}.
\end{aligned}
\end{equation}
After applying the $2 + (n - 2)$ decomposition of the field, we expand it into spherical harmonics with scalar, transverse vector and transverse symmetric trace-free tensor:
\begin{equation}\label{eq:sphericalharmonics}
\begin{aligned}
\mathcal{H}_{ab}\left(x\right) &= \sum_{\ell,\mathbf{m}}\mathcal{H}_{ab;\ell,\mathbf{m}}\left(t,r\right) \mathscr{Y}_{\ell,\mathbf{m}}\left(\theta\right), \\
\mathcal{K}\left(x\right) &= \sum_{\ell,\mathbf{m}}\mathcal{K}_{\ell,\mathbf{m}}\left(t,r\right) \mathscr{Y}_{\ell,\mathbf{m}}\left(\theta\right), \\
\mathcal{H}_{aA}^{\left(\mathrm{V}\right)}\left(x\right) &= \sum_{\ell,\mathbf{m}}\mathcal{H}_{a;\ell,\mathbf{m}}\left(t,r\right) \mathscr{Y}_{A;\ell,\mathbf{m}}^{\left(\mathrm{T}\right)}\left(\theta\right), \\
h_{AB}^{\left(\mathrm{TT}\right)}\left(x\right) &= \sum_{\ell,\mathbf{m}}h_{\ell,\mathbf{m}}^{\left(\mathrm{T}\right)}\left(t,r\right) \mathscr{Y}_{AB;\ell,\mathbf{m}}^{\left(\mathrm{TT}\right)}\left(\theta\right).
\end{aligned}
\end{equation}

When considering the gravitational field framework described by the Einstein-Hilbert action, perturbations propagating on an asymptotically flat vacuum are represented by the massless Fierz-Pauli action:
\begin{equation}
\mathcal{S}^{\left(\mathrm{gr}\right)} = \int \mathrm{d}^{n}x\sqrt{-g}\left[-\frac{1}{2}\nabla_\rho h_{\mu\nu}\nabla^\rho h^{\mu\nu}+\nabla_\rho h_{\mu\nu}\nabla^\nu h^{\mu\rho}-\nabla_\mu h\nabla_\nu h^{\mu\nu}+\frac{1}{2}\nabla_\mu h\nabla^\mu h\right],
\end{equation}
where canonical variables are used, i.e., the perturbed metric around a background $g_{\mu\nu}$ is $g_{\mu\nu}^{\text{full}}=g_{\mu\nu}+ \sqrt{32\pi\mathscr{G}}h_{\mu\nu}$.
Substituting the spherical harmonic expansion \eqref{eq:sphericalharmonics} of the metric perturbations, we will be able to derive the decoupling of the tensor ``$\left(\mathrm{T}\right)$'', vector ``$\left(\mathrm{RW}\right)$'' and scalar ``$\left(\mathrm{Z}\right)$'' modes as follows:
\begin{equation}
\begin{aligned}
\mathcal{S}^{\left(\mathrm{gr}\right)} =& \sum_{\ell,\mathbf{m}} \left( \mathcal{S}_{\ell,\mathbf{m}}^{\left(\mathrm{T}\right)} + \mathcal{S}_{\ell,\mathbf{m}}^{\left(\mathrm{RW}\right)} + \mathcal{S}_{\ell,\mathbf{m}}^{\left(\mathrm{Z}\right)} \right), \\
\mathcal{S}_{\ell,\mathbf{m}}^{\left(\mathrm{T}\right)} =& \int \mathrm{d}^2x\sqrt{-g^{(2)}}r^{n-2}\left[-\frac{1}{2}D_a\bar{h}_{\ell,\mathbf{m}}^{\left(\mathrm{T}\right)}D^ah_{\ell,\mathbf{m}}^{\left(\mathrm{T}\right)}-\frac{1}{r}r^aD_a\left|h_{\ell,\mathbf{m}}^{\left(\mathrm{T}\right)}\right|^2 \right. \\
& \left. -\frac{1}{2}\frac{\ell\left(\ell+n-3\right)+2\left(n-3\right)}{r^2}\left|h_{\ell,\mathbf{m}}^{\left(\mathrm{T}\right)}\right|^2 \right], \\
\mathcal{S}_{\ell,\mathbf{m}}^{\left(\mathrm{RW}\right)} =& \int \mathrm{d}^2x\sqrt{-g^{(2)}}2r^{n-4}\left[-\frac{1}{4}\bar{\mathscr{F}}_{ab;\ell,\mathbf{m}}\mathscr{F}_{\ell,\mathbf{m}}^{ab}-\frac{2}{r}r_a\mathrm{Re}\left\{\bar{\mathcal{H}}_{\ell,\mathbf{m}}^bD_b\mathcal{H}_{\ell,\mathbf{m}}^a\right\}\right. \\
& \left. -\frac{1}{2}\left(\frac{\left(\ell+1\right)\left(\ell+n-4\right)}{r^2}\bar{\mathcal{H}}_{a;\ell,\mathbf{m}}\mathcal{H}_{\ell,\mathbf{m}}^a-4\left|r_a\mathcal{H}_{\ell,\mathbf{m}}^a\right|^2\right)\right], \\
\mathcal{S}_{\ell,\mathbf{m}}^{\left(\mathrm{Z}\right)} =& \int \mathrm{d}^2x\sqrt{-g^{(2)}}r^{n-2}\bigg[-\frac{1}{2}D_c\bar{\mathcal{H}}_{ab;\ell,\mathbf{m}}D^c\mathcal{H}_{\ell,\mathbf{m}}^{ab}+D_c\bar{\mathcal{H}}_{ab;\ell,\mathbf{m}}D^b\mathcal{H}_{\ell,\mathbf{m}}^{ac} \\
& -\mathrm{Re}\left\{D_a\bar{\mathcal{H}}_{\ell,\mathbf{m}}D_b\mathcal{H}_{\ell,\mathbf{m}}^{ab}\right\}\\
& +\frac{1}{2}D_a\bar{\mathcal{H}}_{\ell,\mathbf{m}}D^a\mathcal{H}_{\ell,\mathbf{m}}+\frac{(n-2)(n-3)}{2}D_a\bar{\mathcal{K}}_{\ell,\mathbf{m}}D^a\mathcal{K}_{\ell,\mathbf{m}} \\
& -(n-2)\mathrm{Re}\left\{D_a\bar{\mathcal{K}}_{\ell,\mathbf{m}}\left(D_b\mathcal{H}_{\ell,\mathbf{m}}^{ab}-D^a\mathcal{H}_{\ell,\mathbf{m}}\right)\right\} \\
& -\frac{n-2}{r}\mathrm{Re}\left\{\left(D_{a}\bar{\mathcal{H}}_{\ell,\mathbf{m}}+(n-4)D_{a}\bar{\mathcal{K}}_{\ell,\mathbf{m}}\right)\left(r_{b}\mathcal{H}_{\ell,\mathbf{m}}^{ab}-r^{a}\mathcal{K}_{\ell,\mathbf{m}}\right)\right\} \\
& -\frac{1}{2}\frac{\ell(\ell+n-3)}{r^2}\Big(\bar{\mathcal{H}}_{ab;\ell,\mathbf{m}}\mathcal{H}_{\ell,\mathbf{m}}^{ab}-|\mathcal{H}_{\ell,\mathbf{m}}|^2-(n-3)(n-4)|\mathcal{K}_{\ell,\mathbf{m}}|^2 \\
& -2(n-3)\mathrm{Re}\left\{\bar{\mathcal{H}}_{\ell,\mathbf{m}}\mathcal{K}_{\ell,\mathbf{m}}\right\}\Big) \bigg], \\
\end{aligned}
\end{equation}
where
\begin{equation}
\begin{aligned}
\mathscr{F}_{\ell,\mathbf{m}}^{ab} &\equiv D^a\mathcal{H}_{\ell,\mathbf{m}}^b-D^b\mathcal{H}_{\ell,\mathbf{m}}^a, \\
\mathcal{H}_{\ell,\mathbf{m}} &\equiv g_{ab}\mathcal{H}_{\ell,\mathbf{m}}^{ab}.
\end{aligned}
\end{equation}

For the case of tensor modes, we can also define its field as follows:
\begin{equation}
h_{\ell,\mathbf{m}}^{\left(\mathrm{T}\right)} = \frac{\Psi_{\ell,\mathbf{m}}^{\left(\mathrm{T}\right)}}{r^{(n-2)/2}}.
\end{equation}
The action in terms of a canonically normalized variable can be expressed as
\begin{equation}
\begin{aligned}
\mathcal{S}_{\ell,\mathbf{m}}^{\left(\mathrm{T}\right)} &= \int \mathrm{d}^2x\sqrt{-g^{(2)}}\left[-\frac{1}{2}D_a\bar{\Psi}_{\ell,\mathbf{m}}^{\left(\mathrm{T}\right)}D^a\Psi_{\ell,\mathbf{m}}^{\left(\mathrm{T}\right)}-\frac{1}{2f_{t}}V_{\ell}^{\left(\mathrm{T}\right)}\left(r\right)\left|\Psi_{\ell,\mathbf{m}}^{\left(\mathrm{T}\right)}\right|^2\right] \\
&= \int \mathrm{d}t\mathrm{d}r_{\ast}\left[\frac{1}{2}\left|\partial_{t}\Psi_{\ell,\mathbf{m}}^{\left(\mathrm{T}\right)}\right|^{2}-\frac{1}{2}\left|\partial_{r_{\ast}}\Psi_{\ell,\mathbf{m}}^{\left(\mathrm{T}\right)}\right|^{2}-\frac{1}{2}V_{\ell}^{\left(\mathrm{T}\right)}\left(r\right)\left|\Psi_{\ell,\mathbf{m}}^{\left(\mathrm{T}\right)}\right|^{2}\right],
\end{aligned}
\end{equation}
where the tensor modes potential is
\begin{equation}\label{eq:gtmvt}
V_{\ell}^{\left(\mathrm{T}\right)}\left(r\right) \equiv f_{t}\Bigg\{ \frac{\ell\left(\ell+n-3\right)+2\left(n-3\right)}{r^{2}}+\frac{n^{2}-14n+32}{4r^{2}}f_{r}+\frac{n-6}{2r} \left[\frac{\left(f_{t}f_{r}\right)^{\prime}}{2f_{t}}\right] \Bigg\}.
\end{equation}

For the vector modes (odd parity), namely the Regge-Wheeler modes.
We introduce an auxiliary Regge-Wheeler variable $\Psi_{\ell,\mathbf{m}}^{\left(\mathrm{RW}\right)}$ and consider the action
\begin{equation}
\begin{aligned}
\tilde{\mathcal{S}}_{\ell,\mathbf{m}}^{\left(\mathrm{RW}\right)} =& \int \mathrm{d}^2x\sqrt{-g^{(2)}}\Bigg[\sqrt{\frac{\boldsymbol{\mathbbm{F}}_{\ell}\left(r\right)}{2}}r^{\frac{(n-6)}{2}}\mathrm{Re}\left\{\bar{\Psi}_{\ell,\mathbf{m}}^{\left(\mathrm{RW}\right)}\left(\varepsilon_{ab}\mathscr{F}_{\ell,\mathbf{m}}^{ab}-\frac{4}{r}\sqrt{\frac{f_r}{f_t}}t_a\mathcal{H}_{\ell,\mathbf{m}}^a\right)\right\} \\
& -\frac{1}{2}\frac{\boldsymbol{\mathbbm{F}}_{\ell}\left(r\right)}{r^{2}}\left(\left|\Psi_{\ell,\mathbf{m}}^{\left(\mathrm{RW}\right)}\right|^{2}+2r^{n-4}\bar{\mathcal{H}}_{a;\ell,\mathbf{m}}\mathcal{H}_{\ell,\mathbf{m}}^{a}\right)\Bigg],
\end{aligned}
\end{equation}
with
\begin{equation}
\boldsymbol{\mathbbm{F}}_{\ell}\left(r\right) \equiv \left(\ell+1\right)\left(\ell+n-4\right)-2\left(n-3\right)r_{a}r^{a}-2rD_{a}r^{a}.
\end{equation}
For the Schwarzschild-Tangherlini metric, $\boldsymbol{\mathbbm{F}}_{\ell}\left(r\right)$ will become a constant $\left(\ell+1\right)\left(\ell+n-2\right)$.
This alternative action $\tilde{\mathcal{S}}_{\ell,\mathbf{m}}^{\left(\mathrm{RW}\right)}$ can be restored to the original action $\mathcal{S}_{\ell,\mathbf{m}}^{\left(\mathrm{RW}\right)}$ after integrating the auxiliary field
\begin{equation}
\Psi_{\ell,\mathbf{m}}^{\left(\mathrm{RW}\right)}=\frac{r^{\frac{(n-2)}{2}}}{\sqrt{2\boldsymbol{\mathbbm{F}}_{\ell}\left(r\right)}}\left(\varepsilon_{ab}\mathscr{F}_{\ell,\mathbf{m}}^{ab}-\frac{4}{r}\sqrt{\frac{f_r}{f_t}}t_a\mathcal{H}_{\ell,\mathbf{m}}^a\right),
\end{equation}
and $\mathcal{H}_{\ell,\mathbf{m}}^{a}$ is given by
\begin{equation}
\mathcal{H}_{\ell,\mathbf{m}}^{a}=\frac{r^{-\frac{(n-6)}{2}}}{\sqrt{2\boldsymbol{\mathbbm{F}}_{\ell}\left(r\right)}}\left\{\varepsilon^{ab}D_{b}-t^{a}\left[\frac{n-2}{2r}+\frac{\boldsymbol{\mathbbm{F}}_{\ell}^{\prime}\left(r\right)}{2\boldsymbol{\mathbbm{F}}_{\ell}\left(r\right)}\right]\sqrt{\frac{f_{r}}{f_{t}}}\right\}\Psi_{\ell,\mathbf{m}}^{\left(\mathrm{RW}\right)}.
\end{equation}
We ultimately obtained a canonically normalized action for the field $\Psi_{\ell,\mathbf{m}}^{\left(\mathrm{RW}\right)}$,
\begin{equation}
\begin{aligned}
\tilde{\mathcal{S}}_{\ell,\mathbf{m}}^{\left(\mathrm{RW}\right)} &= \int \mathrm{d}^2x\sqrt{-g^{(2)}}\left[-\frac{1}{2}D_a\bar{\Psi}_{\ell,\mathbf{m}}^{\left(\mathrm{RW}\right)}D^a\Psi_{\ell,\mathbf{m}}^{\left(\mathrm{RW}\right)}-\frac{1}{2f_{t}}V_{\ell}^{\left(\mathrm{RW}\right)}\left(r\right)\left|\Psi_{\ell,\mathbf{m}}^{\left(\mathrm{RW}\right)}\right|^2\right] \\
&= \int \mathrm{d}t\mathrm{d}r_{\ast}\left[\frac{1}{2}\left|\partial_{t}\Psi_{\ell,\mathbf{m}}^{\left(\mathrm{RW}\right)}\right|^{2}-\frac{1}{2}\left|\partial_{r_{\ast}}\Psi_{\ell,\mathbf{m}}^{\left(\mathrm{RW}\right)}\right|^{2}-\frac{1}{2}V_{\ell}^{\left(\mathrm{RW}\right)}\left(r\right)\left|\Psi_{\ell,\mathbf{m}}^{\left(\mathrm{RW}\right)}\right|^{2}\right].
\end{aligned}
\end{equation}
The Regge-Wheeler potential is explicitly given as
\begin{equation}\label{eq:gvmvrw}
\begin{aligned}
V_{\ell}^{\left(\mathrm{RW}\right)}\left(r\right) \equiv& f_{t}\Bigg\{ \frac{\left(\ell+1\right)\left(\ell+n-4\right)}{r^{2}}+\frac{\left(n-4\right)\left(n-6\right)}{4r^{2}}f_{r}-\frac{n+2}{2r} \left[\frac{\left(f_{t}f_{r}\right)^{\prime}}{2f_{t}}\right] \\
& +\left[\frac{\boldsymbol{\mathbbm{F}}_{\ell}^{\prime}}{2\boldsymbol{\mathbbm{F}}_\ell}\left(\frac{n-4}{r}+\frac{\boldsymbol{\mathbbm{F}}_{\ell}^{\prime}}{2\boldsymbol{\mathbbm{F}}_\ell}\right)-\left(\frac{\boldsymbol{\mathbbm{F}}_{\ell}^{\prime}}{2\boldsymbol{\mathbbm{F}}_\ell}\right)^{\prime}\right]f_{r}-\frac{\boldsymbol{\mathbbm{F}}_{\ell}^{\prime}}{2\boldsymbol{\mathbbm{F}}_\ell} \left[\frac{\left(f_{t}f_{r}\right)^{\prime}}{2f_{t}}\right] \Bigg\},
\end{aligned}
\end{equation}
where
\begin{equation}
\boldsymbol{\mathbbm{F}}_{\ell}\left(r\right) \equiv \left(\ell+1\right)\left(\ell+n-4\right)-2\left(n-3\right)f_{r}-2r \left[\frac{\left(f_{t}f_{r}\right)^{\prime}}{2f_{t}}\right].
\end{equation}

For the scalar modes (even parity), namely the Zerilli modes.
While this perturbation mode does not provide the expression of the definite potential $V_{\ell}^{\left(\mathrm{Z}\right)}\left(r\right)$ in the general scenarios, instead, the formula of the Zerilli potential corresponding to the Schwarzschild-Tangherlini metric is given as follows:
\begin{equation}\label{eq:ZerilliST}
\begin{aligned}
V_{\ell}^{\left(\mathrm{Z}\right)}\left(r\right) \equiv& \Bigg\{ \frac{\ell\left(\ell+n-3\right)}{r^{2}}+\frac{\left(n-2\right)\left(n-4\right)}{4r^{2}}\mathsf{f}\left(r\right)+\frac{n-2}{2r}\mathsf{f}^{\prime}\left(r\right) \Bigg\}\mathsf{f}\left(r\right) \\
& -\Bigg\{ \frac{2\mathsf{f}^{\prime}\left(r\right)}{r}\frac{\left[2\lambda_{\ell}+\left(n-1\right)\left(n-2\right)\right]\left[\mathsf{H}_{\ell}\left(r\right)+2\left(n-3\right)\lambda_{\ell}\right]}{\mathsf{H}_{\ell}^2\left(r\right)} \Bigg\}\mathsf{f}\left(r\right),
\end{aligned}
\end{equation}
where the constant $\lambda_{\ell}$ is defined as $\lambda_{\ell} \equiv \left(\ell-1\right)\left(\ell+n-2\right)$ and
\begin{equation}
\mathsf{H}_{\ell}\left(r\right) = 2\ell\left(\ell+n-3\right)-2\left(n-2\right)\mathsf{f}\left(r\right)+\left(n-2\right)r\mathsf{f}^{\prime}\left(r\right).
\end{equation}
Here, the symbol $\mathsf{f}$ refers specifically to the Schwarzschild-Tangherlini metric, whose explicit form is given as follows:
\begin{equation}
\mathsf{f}\left(r\right)=f_{t}=f_{r}=1-\frac{2\boldsymbol{\mu}}{r^{n-3}},
\end{equation}
where $\boldsymbol{\mu}$ can be determined as $\boldsymbol{\mu}=\frac{8 \pi \mathscr{G} \mathsf{M}}{(n-2) \Sigma_{(n-2)}}$ according to Eq. \eqref{eq:sch-tang}.

Obviously, the potentials associated with various types of gravitational perturbations in higher-dimensional space-time deviate from the previously established expressions.
Within the framework of the Kodama-Ishibashi master equation in Schr\"{o}dinger-like form, the potentials for the three types of perturbation were derived by Kodama and Ishibashi \cite{Kodama:2003jz, Kodama:2003kk, Ishibashi:2003ap, Ishibashi:2011ws}.
These results have been extensively employed in subsequent research \cite{Konoplya:2003dd, Natario:2004jd, Cardoso:2005vb, Lopez-Ortega:2006aal, Kunduri:2006qa, Siopsis:2010zn, Takahashi:2010ye, Konoplya:2020bxa}.

In accordance with the conventions established in previous studies \cite{Kodama:2003jz, Natario:2004jd, Lopez-Ortega:2006aal, Siopsis:2010zn, Ishibashi:2011ws}, we present the expression for the scenario in which $f_{t}=f_{r}=f(r)$.

For tensor-type perturbations, the corresponding potential is given by
\begin{equation}
V_{\ell}^{\left(\mathrm{T}\right)}\left(r\right) = f(r) \left\{\frac{\ell\left(\ell+n-3\right)}{r^2}+\frac{\left(n-2\right)\left(n-4\right)}{4r^2}f(r)+\frac{\left(n-2\right)f^{\prime}(r)}{2r}\right\}.
\end{equation}
It is evident from Eq. \eqref{eq:vl0} that the tensor-type gravitational perturbation is governed by the same effective potential as the scalar perturbation.

When considering the case where there is no charge and the cosmological constant is zero, the potential associated with the vector-type perturbations is expressed as
\begin{equation}
\begin{aligned}
V_{\ell}^{\left(\mathrm{RW}\right)}\left(r\right) & = \mathsf{f}(r)\left\{\frac{\ell\left(\ell+n-3\right)}{r^2}+\frac{\left(n-2\right)\left(n-4\right)}{4 r^2} \mathsf{f}(r)-\frac{\left(n-1\right)\left(n-2\right) \boldsymbol{\mu}}{r^{n-1}}\right\} \\
& = f(r)\left\{\frac{\ell\left(\ell+n-3\right)}{r^{2}}+\frac{\left(n-2\right)\left(n-4\right)}{4r^{2}}f(r)-\frac{rf^{\prime\prime\prime}(r)}{2\left(n-3\right)}\right\}.
\end{aligned}
\end{equation}

For scalar-type perturbations, when considering the Schwarzschild-Tangherlini metric in an asymptotically flat space-time with vanishing charge and cosmological constant, the potential can be expressed as
\begin{equation}
V_{\ell}^{\left(\mathrm{Z}\right)}\left(r\right) = \frac{\mathsf{f}(r) \mathsf{U}(r)}{16 r^2 \mathsf{H}^2(r)},
\end{equation}
where
\begin{equation}
\mathsf{H}(r)=\ell\left(\ell+n-3\right)-\left(n-2\right)+\frac{\left(n-1\right)\left(n-2\right) \boldsymbol{\mu}}{r^{n-3}},
\end{equation}
and
\begin{equation}
\begin{aligned}
\mathsf{U}(r) =& +8(n-1)^2(n-2)^4 \frac{\boldsymbol{\mu}^3}{r^{3 n-9}} \\
&+4(n-1)(n-2)\Bigg\{4\left(2 n^2-11 n+18\right)\Big[\ell(\ell+n-3)-(n-2)\Big] \\
& +(n-1)(n-2)(n-4)(n-6)\Bigg\} \frac{\boldsymbol{\mu}^2}{r^{2 n-6}} \\
& -24(n-2) \Bigg\{(n-6) \Big[\ell(\ell+n-3)-(n-2)\Big] \\
&+(n-1)(n-2)(n-4)\Bigg\} \Big[\ell(\ell+n-3)-(n-2)\Big] \frac{\boldsymbol{\mu}}{r^{n-3}} \\
& +16 \Big[\ell(\ell+n-3)-(n-2)\Big]^3+4 n(n-2)\Big[\ell(\ell+n-3)-(n-2)\Big]^2.
\end{aligned}
\end{equation}

During the calculations, it is important to emphasize that gravitational perturbations in the EPYMGB space-time are extremely complex, rendering the derivation of the corresponding master equations technically challenging.
Moreover, the physically admissible ranges of the parameters ${\alpha}_{2}$ and $\mathcal{Q}$ are considerably restricted, which leads to only a slight deviation between the EPYMGB solution and the Schwarzschild-Tangherlini solution when considering cases with small parameter values.
Therefore, this approximation is well justified when employing the gravitational perturbation equations derived from the higher-dimensional vacuum Einstein field equations, although it may omit certain features of the full gravitational spectrum.

\subsection{Numerical methods}
The quasinormal modes (QNMs) of an asymptotically flat black hole are complex frequencies characterized by wave functions satisfying specific boundary conditions.
These frequencies can be expressed as $\omega = \text{Re}(\omega) + i \text{Im}(\omega)$, where the real part corresponds to the oscillation frequency and the imaginary part indicates the damping.
Various computational approaches are currently employed to determine QNMs, and this section provides an overview of several widely-used methodologies.

\subsubsection{Wentzel-Kramers-Brillouin method}
The semi-analytic Wentzel-Kramers-Brillouin (WKB) approach was first implemented by Schutz and Will \cite{Schutz:1985zz} to determine quasinormal frequencies.
Subsequently, Iyer and collaborators extended this formalism to third order \cite{Iyer:1986np}, a framework later advanced to sixth order by Konoplya \cite{Konoplya:2003ii}.
Currently, this method has been systematically enhanced to 13th order.
Furthermore, the predictive accuracy of this method is significantly enhanced through the implementation of Pad\'e approximants, which systematically characterize the asymptotic convergence properties of the WKB series \cite{Matyjasek:2017psv}.

The wavelike equation involving the effective potential $V(x)$ can be written as
\begin{equation}
\frac{\mathrm{d}^2\Psi}{\mathrm{d} x^2}+\left[\omega^2-V(x)\right]\Psi(x)=0,
\end{equation}
where $x$ is the tortoise coordinate.
The WKB approximation provides an effective framework for calculating dominant QNMs while inherently satisfying the boundary conditions:
\begin{equation}\label{eq:bcwkb}
\Psi\left(x \rightarrow\pm\infty\right) \propto e^{\pm i \omega x},
\end{equation}
which are purely ingoing wave at the horizon ($x\rightarrow -\infty$) and purely outgoing wave at spatial infinity ($x\rightarrow +\infty$) or cosmological horizon.
The WKB method relies on expanding the asymptotic solutions of the wave function at the two boundary conditions \eqref{eq:bcwkb} to a certain order and matching these WKB asymptotic solutions to the Taylor expansion near the peak of the effective potential.
Therefore, the two turning points and the single maximum of the effective potential will be effectively employed in the WKB method.
The general higher-order WKB formula has the form \cite{Konoplya:2019hlu}
\begin{equation}\label{eq:higherorderwkb}
\begin{aligned}
\omega^{2} = & V_0+A_2\left(\mathcal{K}^2\right)+A_4\left(\mathcal{K}^2\right)+A_6\left(\mathcal{K}^2\right)+\cdots \\
 & -i\mathcal{K}\sqrt{-2V_{2}} \Big[ 1+A_{3}\left(\mathcal{K}^{2}\right)+A_{5}\left(\mathcal{K}^{2}\right)+A_{7}\left(\mathcal{K}^{2}\right)+\cdots \Big],
\end{aligned}
\end{equation}
where $V_0$ is the value of the potential at its maximum, $V_{2}$ is the value of the second derivative of the potential at this point.
The corrections $A_{k}\left(\mathcal{K}^2\right)$ of order $k$ are polynomials of $\mathcal{K}^2$ with rational coefficients.
And these corrections depend on the values of the higher derivatives of the potential $V(r)$ at its maximum.
The complex eigenfrequencies correspond to the poles of $\Gamma \left(-\mathcal{K}+\frac{1}{2}\right)$ for $\mathrm{Re}(\omega)>0$ and $\Gamma \left(\mathcal{K}+\frac{1}{2}\right)$ for $\mathrm{Re}(\omega)<0$ \cite{Iyer:1986np}.
Thus, the QNMs can be determined by substituting the half-integer value of $\mathcal{K}$ into Eq. \eqref{eq:higherorderwkb},
\begin{equation}
\mathcal{K}=\left\{
\begin{array}{ll}
+\mathrm{n}+\frac{1}{2}, & \mathrm{Re}(\omega)>0; \\
-\mathrm{n}-\frac{1}{2}, & \mathrm{Re}(\omega)<0,
\end{array}\right.
\end{equation}
where $\mathrm{n}$ is the overtone number.
Since we only consider positive values of the real oscillation frequencies ($\mathrm{Re}(\omega)>0$) and the modes are decaying ($\mathrm{Im}(\omega)<0$), The branch $\mathcal{K}=\mathrm{n}+\frac{1}{2}$ will be chosen.

It is important to note that the WKB series is asymptotically convergent and does not guarantee improved accuracy at each order. 
Therefore, in order to improve the accuracy of the higher-order WKB formula, we will use Pad\'e approximants following the procedure of Matyjasek and Opala \cite{Matyjasek:2017psv}.
This method requires introducing the powers of the order parameter $\epsilon$ in the WKB formula \eqref{eq:higherorderwkb} to define a polynomial $P_{k}(\epsilon)$ as follows:
\begin{equation}
\begin{aligned}
P_{k}(\epsilon) = & V_0+A_2\left(\mathcal{K}^2\right)\epsilon^2+A_4\left(\mathcal{K}^2\right)\epsilon^4+A_6\left(\mathcal{K}^2\right)\epsilon^6+\cdots \\
 & -i\mathcal{K}\sqrt{-2V_2} \Big[ \epsilon+A_3\left(\mathcal{K}^2\right)\epsilon^3+A_5\left(\mathcal{K}^2\right)\epsilon^5+A_7\left(\mathcal{K}^2\right)\epsilon^7+\cdots \Big],
\end{aligned}
\end{equation}
where the polynomial $k$ order is consistent with the order of the WKB formula, and the squared frequency can be determined by selecting $\epsilon=1$,
\begin{equation}
\omega^{2} = P_{k}(1).
\end{equation}

For the polynomial $P_{k}\left(\epsilon\right)$, we consider a class of rational functions known as Pad\'e approximants, denoted by
\begin{equation}
P_{\tilde{n}/\tilde{m}}\left(\epsilon\right) = \frac{Q_{0}+Q_{1}\epsilon+\cdots+Q_{\tilde{n}}\epsilon^{\tilde{n}}}{R_{0}+R_{1}\epsilon+\cdots+R_{\tilde{m}}\epsilon^{\tilde{m}}},
\end{equation}
where $\tilde{n}+\tilde{m}=k$, and near $\epsilon=0$ which satisfies:
\begin{equation}
P_{\tilde{n}/\tilde{m}}\left(\epsilon\right) - P_{k}\left(\epsilon\right) = \mathcal{O}\left(\epsilon^{k+1}\right).
\end{equation}
The Pad\'e approximants significantly enhances the accuracy of the WKB method, facilitating inferences regarding the asymptotic behavior of the WKB series. 
Further details on the Pad\'e approximants can be found in Ref. \cite{Konoplya:2019hlu}.

Additionally, the outburst of overtone cannot be theoretically determined by the WKB formula when the multipole number $\ell$ is less than the overtone number $\mathrm{n}$.
Hence, throughout our analysis, we restrict ourselves to the case $\ell>\mathrm{n}$.

\subsubsection{Asymptotic iteration method}
The asymptotic iteration method (AIM) was initially developed in Ref. \cite{Ciftci:2005xn} as a systematic approach for solving second-order differential equations.
Subsequently, this innovative approach proved particularly effective in determining the quasinormal frequencies for field perturbations in the Schwarzschild black hole space-time \cite{Cho:2009cj}.
To implement this method, it is necessary to adopt a specific initial value for the massless scalar field while accounting for its asymptotic behavior both at the event horizon and spatial infinity.
Consequently, we give a homogeneous linear second-order differential equation of the form:
\begin{equation}\label{2nd Eq}
\chi''=\lambda_{0}(x)\chi'+s_{0}(x)\chi,
\end{equation}
where $\lambda_{0}(x)$ and $s_{0}(x)$ are well-defined and sufficiently smooth functions.
Differentiating the above equation with respect to $x$ yields
\begin{equation}
\chi'''=\lambda_{1}(x)\chi'+s_{1}(x)\chi,
\end{equation}
where the two coefficients are $\lambda_{1}(x) = \lambda'_{0}+s_{0}+\lambda_{0}^{2}$ and $s_{1}(x) = s'_{0}+s_{0}\lambda_{0}$.
By applying this differentiation process iteratively $\Xi$ times with respect to the independent variable, we obtain:
\begin{equation}
\chi^{(\Xi+2)}=\lambda_{\Xi}(x)\chi'+s_{\Xi}(x)\chi,
\end{equation}
where the new coefficients $\lambda_{\Xi}(x)$ and $s_{\Xi}(x)$ are associated with the older ones through the following recurrence relation:
\begin{equation}\label{eq:iteration}
\begin{aligned}
\lambda_{\Xi}(x)&=\lambda'_{\Xi-1}(x)+s_{\Xi-1}(x)+\lambda_{0}(x)\lambda_{\Xi-1}(x), \\
s_{\Xi}(x)&=s'_{\Xi-1}(x)+s_{0}(x)\lambda_{\Xi-1}(x).
\end{aligned}
\end{equation}
For sufficiently large $\Xi$, the asymptotic properties of the AIM are established \cite{Cho:2011sf} by
\begin{equation}\label{eq:Quantum condition}
\frac{s_{\Xi}(x)}{\lambda_{\Xi}(x)}=\frac{s_{\Xi-1}(x)}{\lambda_{\Xi-1}(x)} = \text{Constant}.
\end{equation}

The perturbation frequency can be derived from the aforementioned ``quantization condition.''
Subsequently, the functions $\lambda_{\Xi}(x)$ and $s_{\Xi}(x)$ are expanded as Taylor series around the point $\breve{x}$, at which the AIM is performed.
\begin{equation}
\begin{aligned}
\lambda_{\Xi}\left(\breve{x}\right)&=\sum_{i=0}^{\infty} c_{\Xi}^{i}\left(x-\breve{x}\right)^{i}, \\
s_{\Xi}\left(\breve{x}\right)&=\sum_{i=0}^{\infty} d_{\Xi}^{i}\left(x-\breve{x}\right)^{i},
\end{aligned}
\end{equation}
where $c_{\Xi}^{i}$ and $d_{\Xi}^{i}$ are defined as the $i$-th coefficients in the Taylor expansions of $\lambda_{\Xi}\left(\breve{x}\right)$ and $s_{\Xi}\left(\breve{x}\right)$, respectively.
Substituting the preceding equations into Eq. \eqref{eq:iteration} yields a set of recursion relations for the Taylor coefficients as
\begin{equation}\label{eq:recursion}
\begin{aligned}
c_{\Xi}^{i}&=\sum_{k=0}^{i} c_{0}^{k} c_{\Xi-1}^{i-k}+(i+1) c_{\Xi-1}^{i+1}+d_{\Xi-1}^{i}, \\
d_{\Xi}^{i}&=\sum_{k=0}^{i} d_{0}^{k} c_{\Xi-1}^{i-k}+(i+1) d_{\Xi-1}^{i+1}.
\end{aligned}
\end{equation}

After applying the recursive relation given in Eq. \eqref{eq:recursion} in Eq. \eqref{eq:Quantum condition}, the quantization condition \cite{Ciftci:2005xn} can be obtained
\begin{equation}
d_{\Xi}^{0}c_{\Xi-1}^{0}- c_{\Xi}^{0} d_{\Xi-1}^{0}=0,
\end{equation}
which can be used to calculate the QNMs of black holes.
The AIM demonstrates significantly improved accuracy and efficiency without requiring derivative computations.

\subsubsection{Time-domain integration method}
The time-domain integration method, commonly referred to as the finite difference approach, was originally developed by Gundlach, Price and Pullin \cite{Gundlach:1993tp}.
We rewrite the general wavelike equation using the following ansatz:
\begin{equation}\label{eq:wavelikeeq}
\frac{\partial^2\Psi}{\partial t^2}-\frac{\partial^2\Psi}{\partial x^2}+V(t,x)\Psi=0,
\end{equation}
where $x$ is the tortoise coordinate, which is defined as $\mathrm{d}x = \mathrm{d}r / \sqrt{h f}$.
We will reformulate the wavelike equation \eqref{eq:wavelikeeq} in terms of the so-called light-cone coordinates $\mathrm{d}u=\mathrm{d}t-\mathrm{d}x$ and $\mathrm{d}v=\mathrm{d}t+\mathrm{d}x$,
\begin{equation}
\left[ 4\frac{\partial^2}{\partial u\partial v}+V(u,v) \right]\Psi(u,v)=0.
\end{equation}
The initial data are specified on the two null surfaces $u=u_{0}$ and $v=v_{0}$.
Since the fundamental behaviour of field attenuation is independent of the initial conditions, it is assumed that the field $\Psi$ is initially in the form of Gaussian wave packets.
Accordingly, we impose the initial conditions as $\Psi\left(u=u_{0}, v\right)=\exp \left[-\frac{\left(v-v_{c}\right)^{2}}{2 \sigma^{2}}\right]$, $\Psi\left(u, v=v_{0}\right)=\Psi\left(u=u_{0}, 0\right)$, and choose the characteristic parameters $v_{c}$ and $\sigma$ for the initial Gaussian wave package in the practical computation.
The appropriate discretization scheme is
\begin{equation}\label{eq:nwes}
\Psi(N)=\Psi(W)+\Psi(E)-\Psi(S)- \frac{\Delta^{2}}{8}\Big[\Psi(W)+\Psi(E)\Big]V(S)+\mathcal{O}\left(\Delta^{4}\right),
\end{equation}
where we have used the following definitions for the points: $N=(u+\Delta, v+\Delta)$, $W=(u+\Delta, v)$, $E=(u, v+\Delta)$ and $S=(u, v)$.

After the integration is completed, extract the value of $\Psi\left(u_{\max}, v\right)$, where $u_{\max}$ represents the maximum value of $u$ on the numerical grid.
Provided $u_{\max}$ is sufficiently large, this approach yields an accurate approximation of the wave function on the event horizon.
In this way, we obtain the time-domain profile, which is a series of values of the perturbation field $\Psi\left(t=\frac{v+u}{2}, x=\frac{v-u}{2}\right)$ at a given position $x$ and discrete moments $\left\{ t_{0}, t_{0}+\Delta, t_{0}+2\Delta, \cdots, t_{0}+\infty\Delta \right\}$.
The asymptotic tails of the process can be constructed by integrating the wave-like equation in the time-domain at a fixed value of the radial coordinates.

The Prony method can be used to extract the quasinormal frequencies from the time-domain profiles.
While this technique yields accurate results for dominant frequencies, it typically cannot extract higher overtones.
In this paper, we develop a customized ``least square analysis'' methodology to extract the quasinormal frequencies from the time-domain profiles.
Specifically, this approach involves fitting an exponentially attenuated linear regression model to the waveform data, enabling robust determination of approximate QNMs in the time-domain profiles.

\section{Numerical results}
In this section, we systematically present a series of numerical results obtained through detailed computations.

\subsection{Reasonable ranges of parameters}
\begin{figure}[htbp]
\centering
\includegraphics[width=1\textwidth]{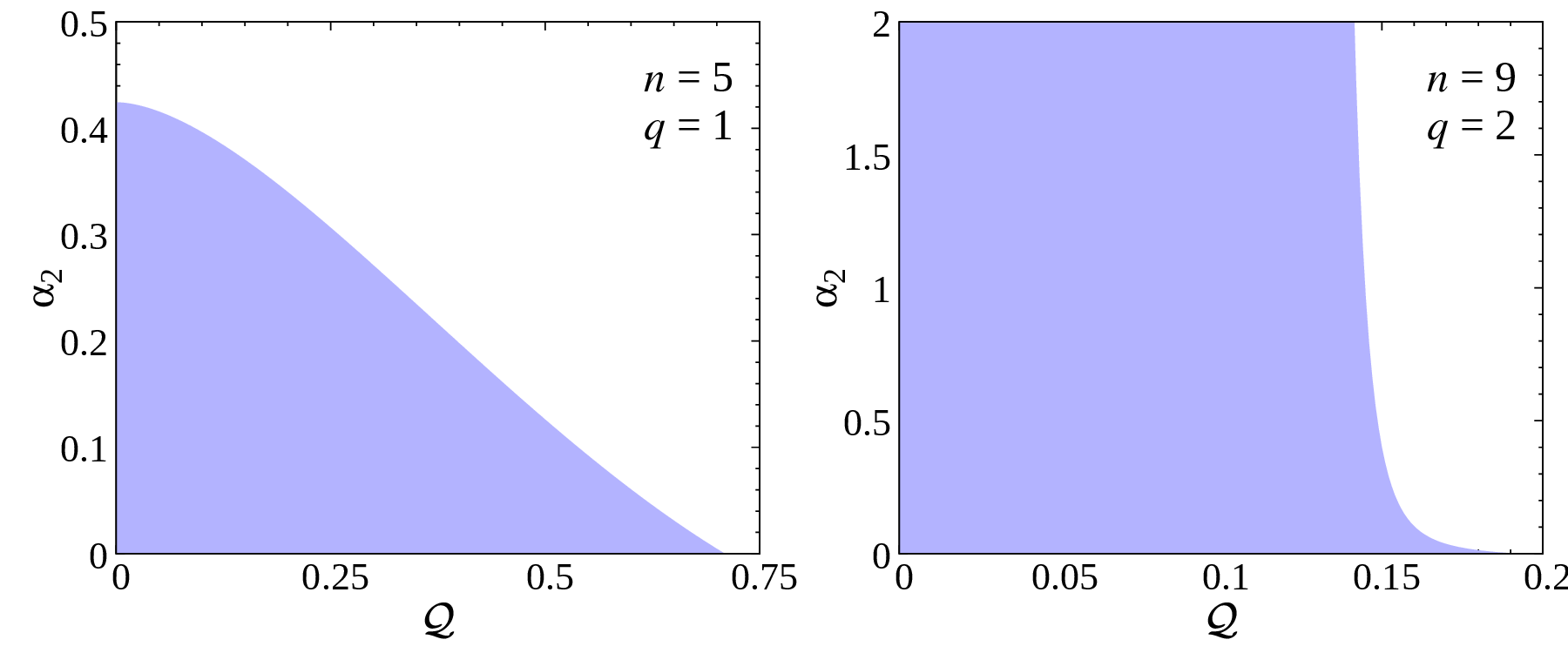}
\caption{
The blue-shaded region demarcates the admissible parameter space $\left({\alpha}_2, \mathcal{Q}\right)$ that ensures the existence of the event horizon in the special configuration of the EPYMGB black hole.
The parameters $\mathsf{M}=1$ and $\mathscr{G}=1$ are selected.
}
\label{figure2}
\end{figure}

\begin{figure}[htbp]
\centering
\includegraphics[width=1\textwidth]{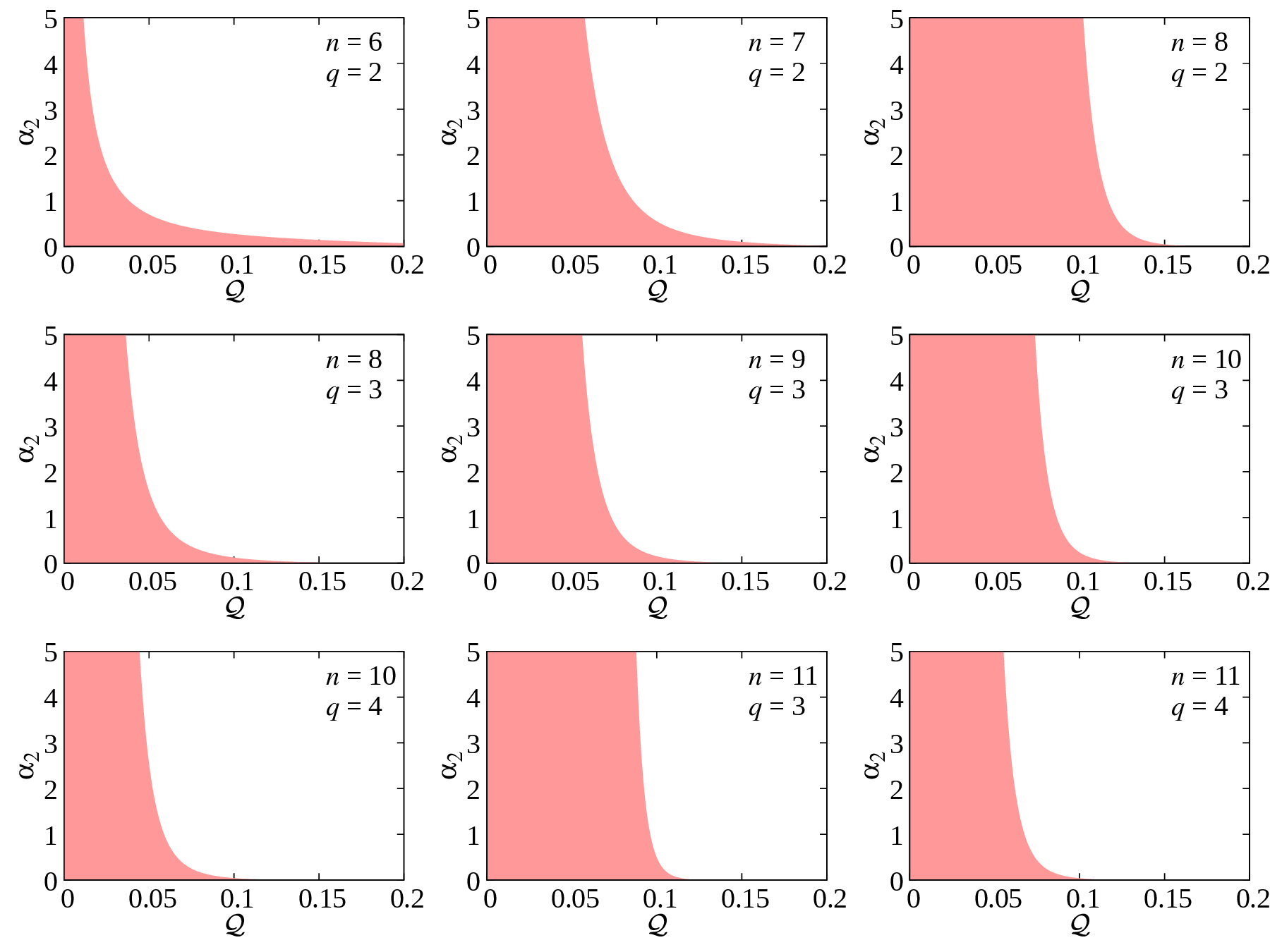}
\caption{
The red-shaded region demarcates the admissible parameter space $\left({\alpha}_2, \mathcal{Q}\right)$ that ensures the existence of the event horizon in the general configuration of the EPYMGB black hole.
The parameters $\mathsf{M}=1$ and $\mathscr{G}=1$ are selected.
}
\label{figure3}
\end{figure}

To guarantee the existence of at least one event horizon in the black hole solution \eqref{eq:lapsefunc2}, the parameters ${\alpha}_2$ and $\mathcal{Q}$ must be constrained to specific ranges within the allowable parameter space $\left({\alpha}_2, \mathcal{Q}\right)$, as illustrated in Figures \ref{figure2} and \ref{figure3}.
In this calculation we only consider the parameter conditions ${\alpha}_2>0$ and $\mathcal{Q}\geqslant0$ for demonstration.
It can be observed that, unlike in other dimensions, when $n=5$, the parameter space forms a closed region.
Determining this valid region will assist us in making a reasonable selection of parameters for the subsequent calculations.

\subsection{Constraint results for the shadow radius}
\begin{figure}[htbp]
\centering
\includegraphics[width=1\textwidth]{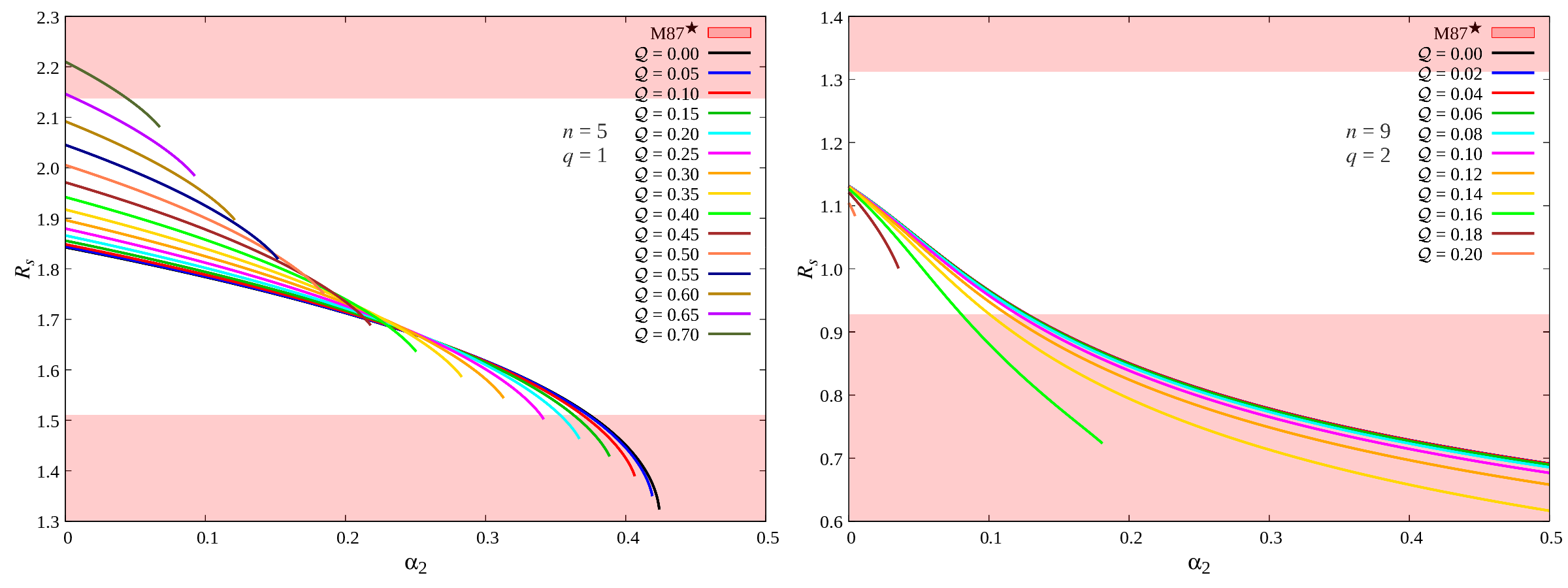}
\caption{
The relevant parameters of the special configuration of the EPYMGB black hole are constrained through the high-dimensional shadow constraint formula \eqref{eq:highdimrange}. The unfilled color region corresponds to the effective range $r_{\mathrm{sh}}$ of the shadow radius derived from the $\mathrm{M87}^{\star}$ observational data.
}
\label{figure4}
\end{figure}

\begin{figure}[htbp]
\centering
\includegraphics[width=1\textwidth]{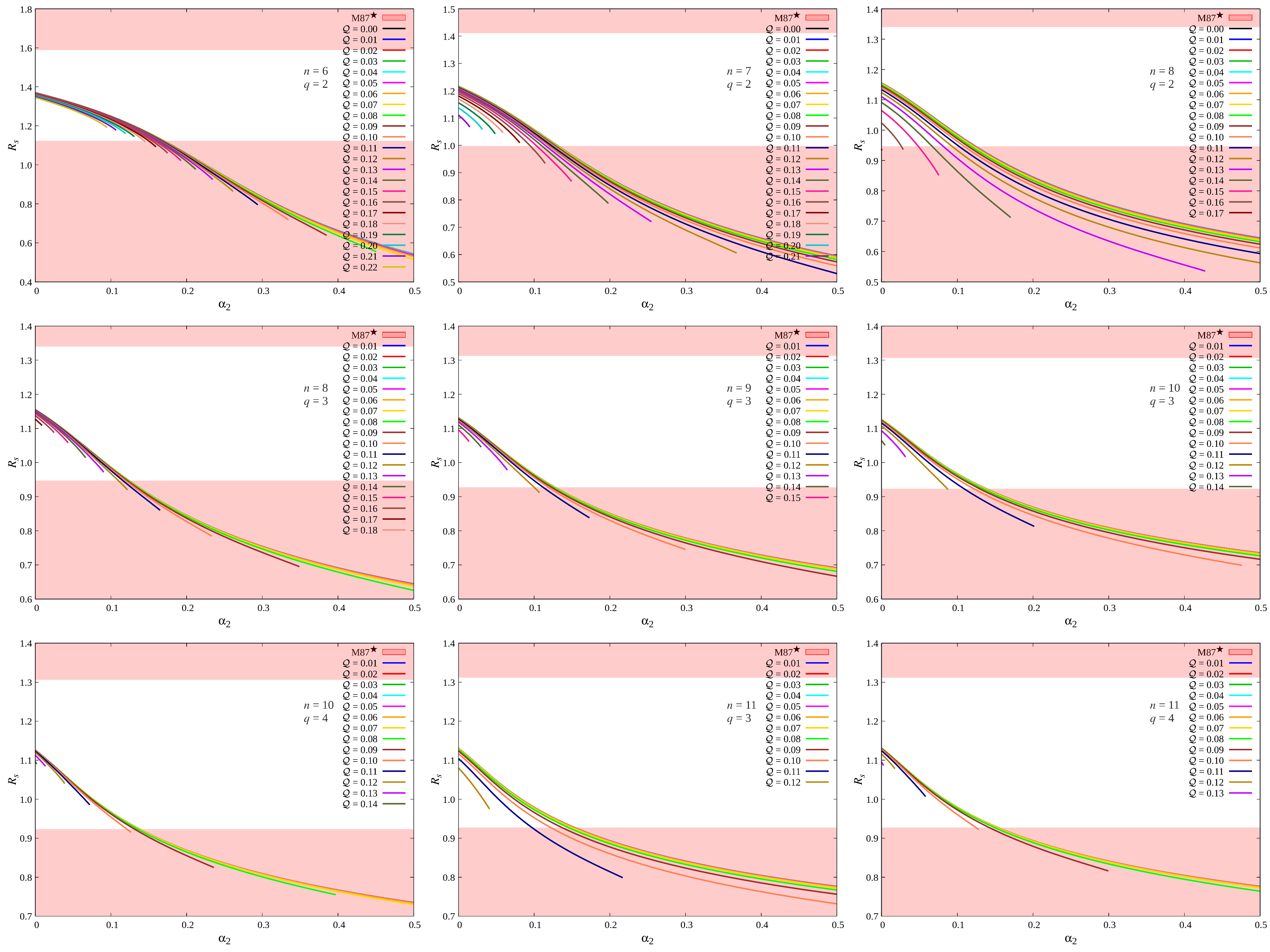}
\caption{
Similar to Figure \ref{figure4}, but with the metric corresponding to the general configuration of the EPYMGB black hole.
}
\label{figure5}
\end{figure}

\begin{figure}[htbp]
\centering
\includegraphics[width=1\textwidth]{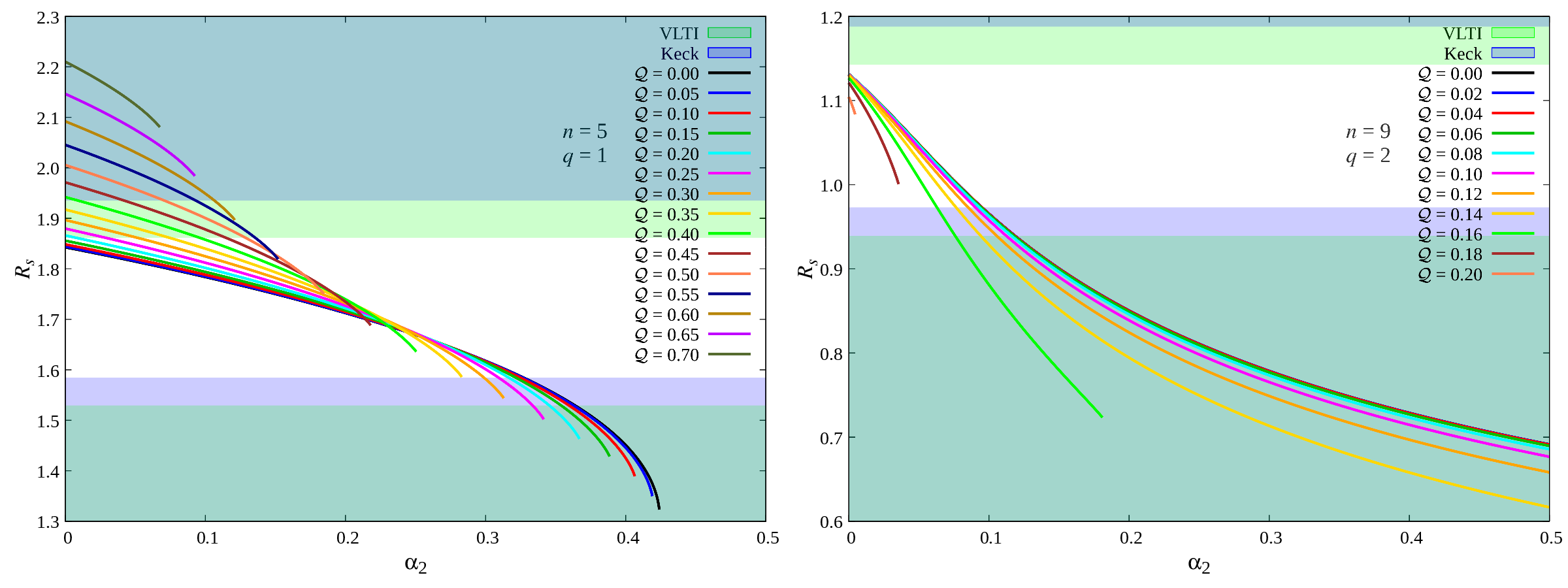}
\caption{
The relevant parameters of the special configuration of the EPYMGB black hole are constrained through the high-dimensional shadow constraint formula \eqref{eq:highdimrange}. The unfilled color region corresponds to the effective range $r_{\mathrm{sh}}$ of the shadow radius derived from the $\mathrm{Sgr~A}^{\star}$ observational data.
}
\label{figure6}
\end{figure}

\begin{figure}[htbp]
\centering
\includegraphics[width=1\textwidth]{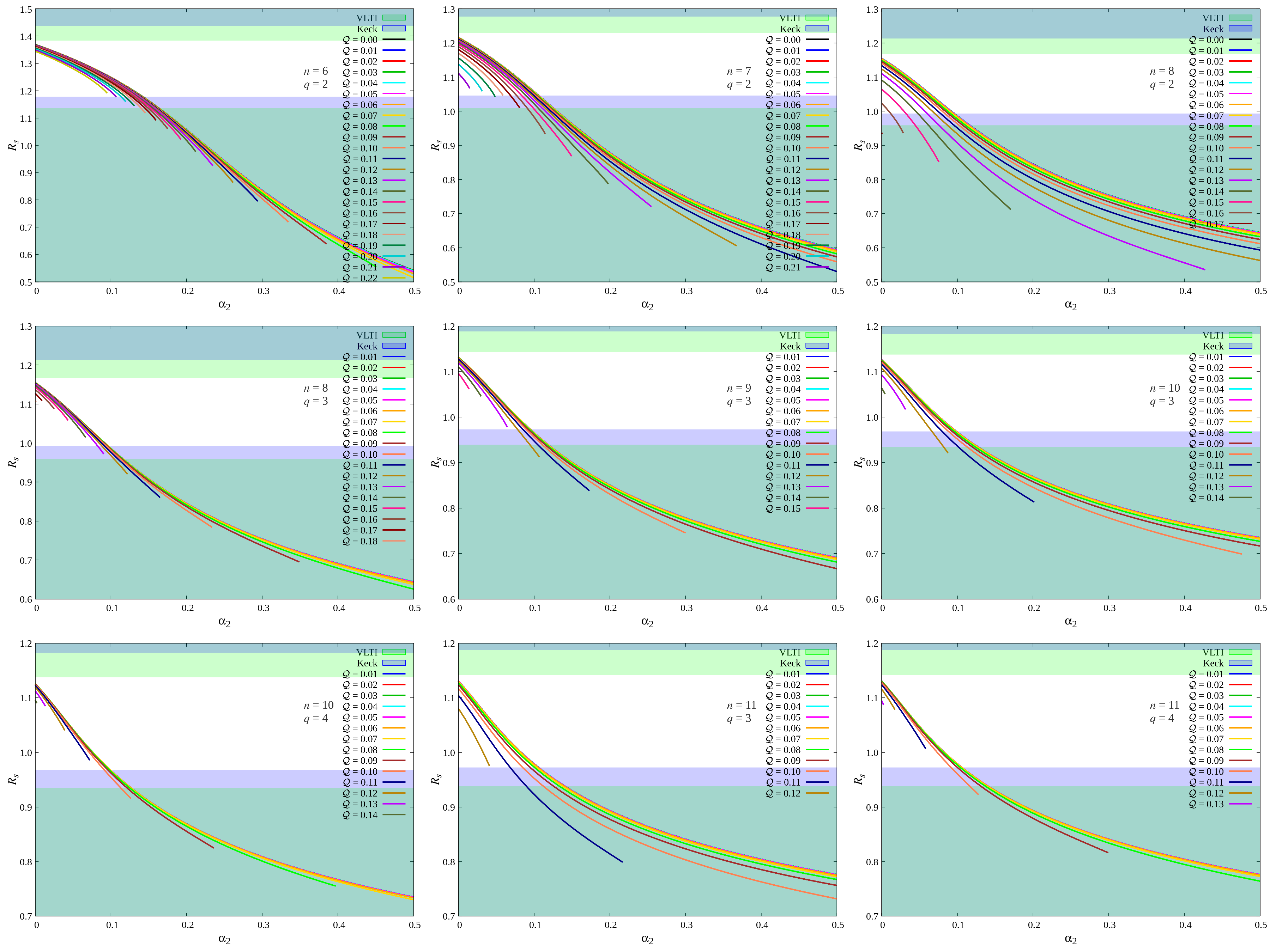}
\caption{
Similar to Figure \ref{figure6}, but with the metric corresponding to the general configuration of the EPYMGB black hole.
}
\label{figure7}
\end{figure}

Firstly, we apply the high-dimensional shadow radius formula \eqref{eq:highdimrseq} constructed in this paper to investigate the EPYMGB black hole solutions, and we present the corresponding numerical result of $R_{s}$ within the allowed parameter space $\left({\alpha}_2, \mathcal{Q}\right)$.
Subsequently, we utilize the high-dimensional shadow constraint formula \eqref{eq:highdimrange} to provide constraints on the previously obtained values of $R_{s}$ based on observational data from the $\mathrm{M87}^{\star}$ and $\mathrm{Sgr~A}^{\star}$, respectively.
The results are shown in Figures \ref{figure4}, \ref{figure5}, \ref{figure6} and \ref{figure7}.
Among these results, we set the ADM mass of the black hole and the four-dimensional gravitational constant as $\mathsf{M}=M=1$ and $\mathscr{G}=1$, respectively.

\subsection{Quasinormal frequencies and time-domain profiles}
To systematically assess the methodological consistency, we first compute the QNMs for massless scalar field perturbations by applying three distinct approaches: the time-domain integration technique, the WKB approximation and the AIM.
The corresponding numerical results are presented in Table \ref{tb3:qnms3}.
The strong consensus among the results obtained from the three approaches further validates their accuracy and reinforces the robustness of each approach.
However, our computational analysis reveals that the WKB approximation demonstrates compromised convergence behavior when applied to metric \eqref{eq:lapsefunc2} at higher orders.
This limitation originates from the method's inherent difficulty in maintaining numerical fidelity for higher-order terms when addressing space-time geometries with nontrivial topological complexity \cite{Yan:2020hga, Yan:2020nvk, Konoplya:2019hlu}.
Therefore, we employ the Pad\'e approximants technique to enhance the precision of the high-order WKB calculations, enabling them to closely match the correct numerical results of QNMs.
Additionally, when extracting QNMs values from the time-domain profiles, we refrain from considering the case of $\ell=0$, since its ring-down waveform exhibits fewer oscillation peaks, complicating the extraction of accurate values.

\begin{figure}[htbp]
\centering
\includegraphics[width=1\textwidth]{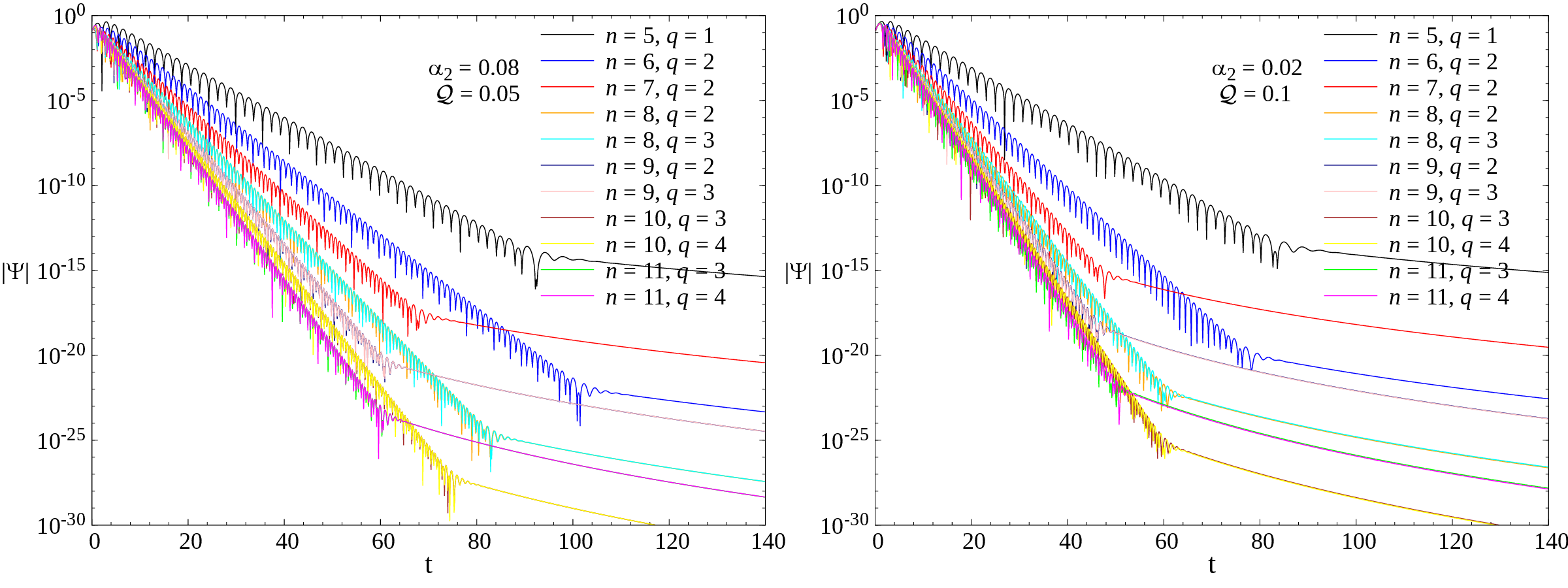}
\caption{
Time-domain profiles (plots of the ring-down waveforms and the late-time tails) of the spin-0 perturbations are governed by the effective potential $V_{\ell}^{(0)}$ in the EPYMGB space-time. 
The parameters are chosen to be $\mathsf{M}=1$, $\mathscr{G}=1$, $\ell=2$, $\mathrm{n}=0$, $v_{c}=2$, $\sigma=1$, and $m=0$.
}
\label{figure8}
\end{figure}

\begin{figure}[htbp]
\centering
\includegraphics[width=1\textwidth]{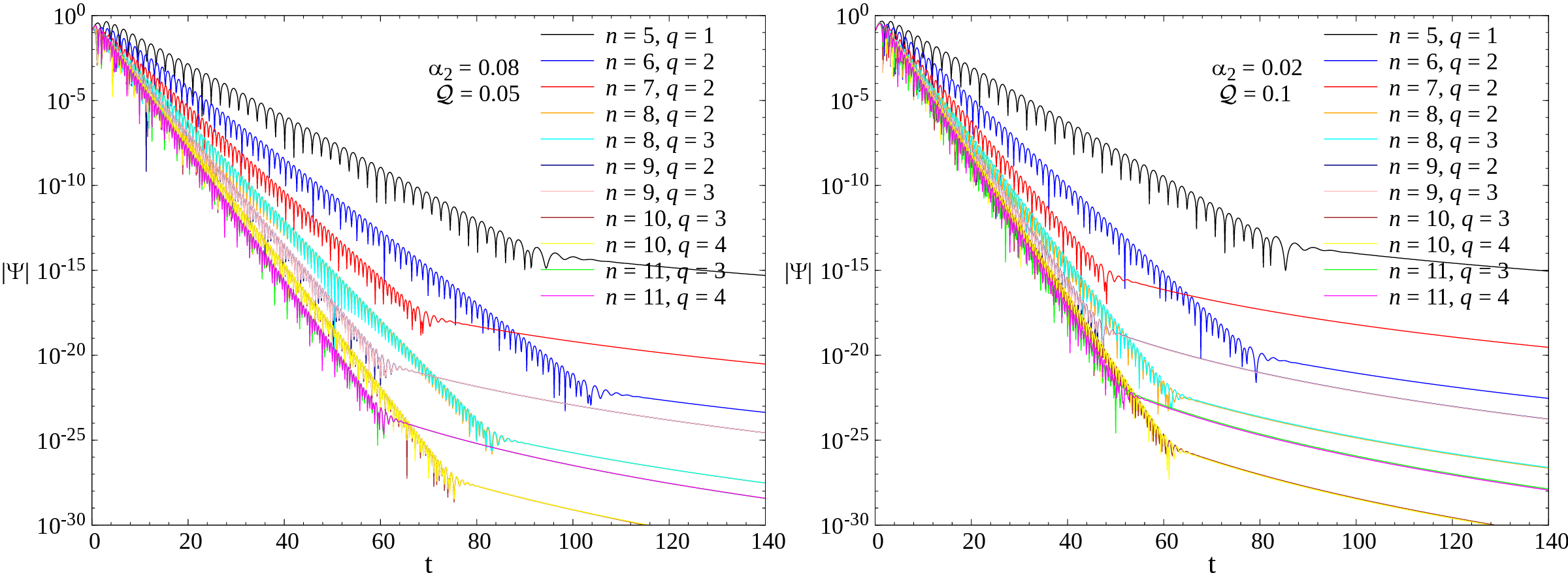}
\caption{
Time-domain profiles (plots of the ring-down waveforms and the late-time tails) of the spin-1 perturbations are governed by the effective potential $V_{\ell}^{\left(\mathrm{V}\right)}$ in the EPYMGB space-time. 
The parameters are chosen to be $\mathsf{M}=1$, $\mathscr{G}=1$, $\ell=2$, $\mathrm{n}=0$, $v_{c}=2$, and $\sigma=1$.
}
\label{figure9}
\end{figure}

\begin{figure}[htbp]
\centering
\includegraphics[width=1\textwidth]{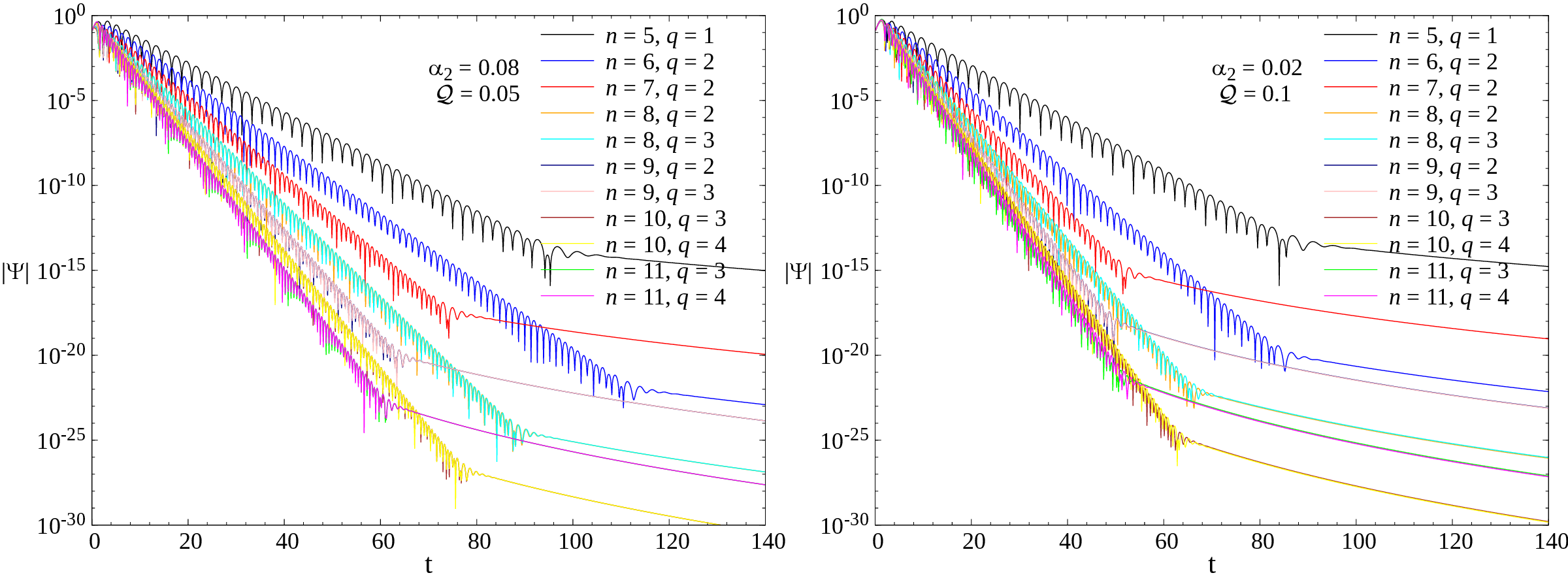}
\caption{
Time-domain profiles (plots of the ring-down waveforms and the late-time tails) of the spin-1 perturbations are governed by the effective potential $V_{\ell}^{\left(\mathrm{S}\right)}$ in the EPYMGB space-time. 
The parameters are chosen to be $\mathsf{M}=1$, $\mathscr{G}=1$, $\ell=2$, $\mathrm{n}=0$, $v_{c}=2$, and $\sigma=1$.
}
\label{figure10}
\end{figure}

\begin{figure}[htbp]
\centering
\includegraphics[width=1\textwidth]{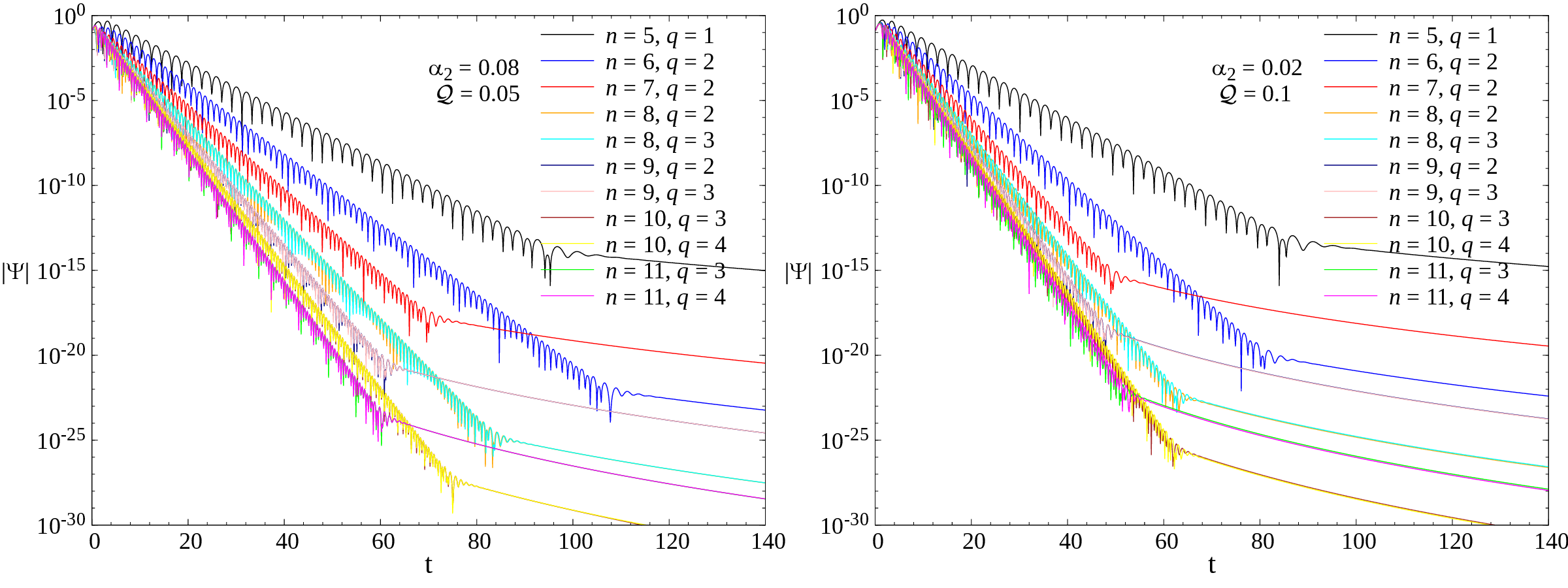}
\caption{
Time-domain profiles (plots of the ring-down waveforms and the late-time tails) of the $p$-form perturbations are governed by the effective potential $V_{\ell}^{(p)}$ in the EPYMGB space-time. 
The parameters are chosen to be $\mathsf{M}=1$, $\mathscr{G}=1$, $\ell=2$, $\mathrm{n}=0$, $v_{c}=2$, $\sigma=1$ and $p=2$.
}
\label{figure11}
\end{figure}

\begin{figure}[htbp]
\centering
\includegraphics[width=1\textwidth]{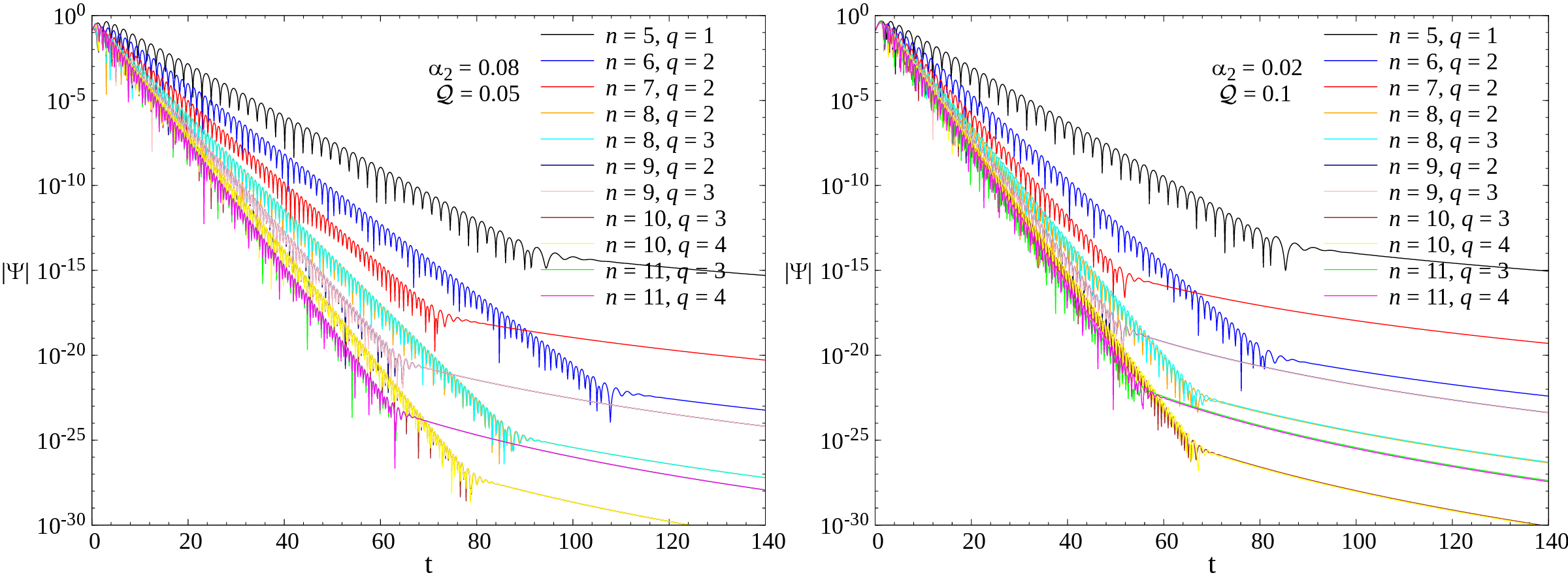}
\caption{
Time-domain profiles (plots of the ring-down waveforms and the late-time tails) of the $(p-1)$-form perturbations are governed by the effective potential $V_{\ell}^{(\tilde{p})}$ in the EPYMGB space-time. 
The parameters are chosen to be $\mathsf{M}=1$, $\mathscr{G}=1$, $\ell=2$, $\mathrm{n}=0$, $v_{c}=2$, $\sigma=1$ and $p=2$.
}
\label{figure12}
\end{figure}

\begin{figure}[htbp]
\centering
\includegraphics[width=1\textwidth]{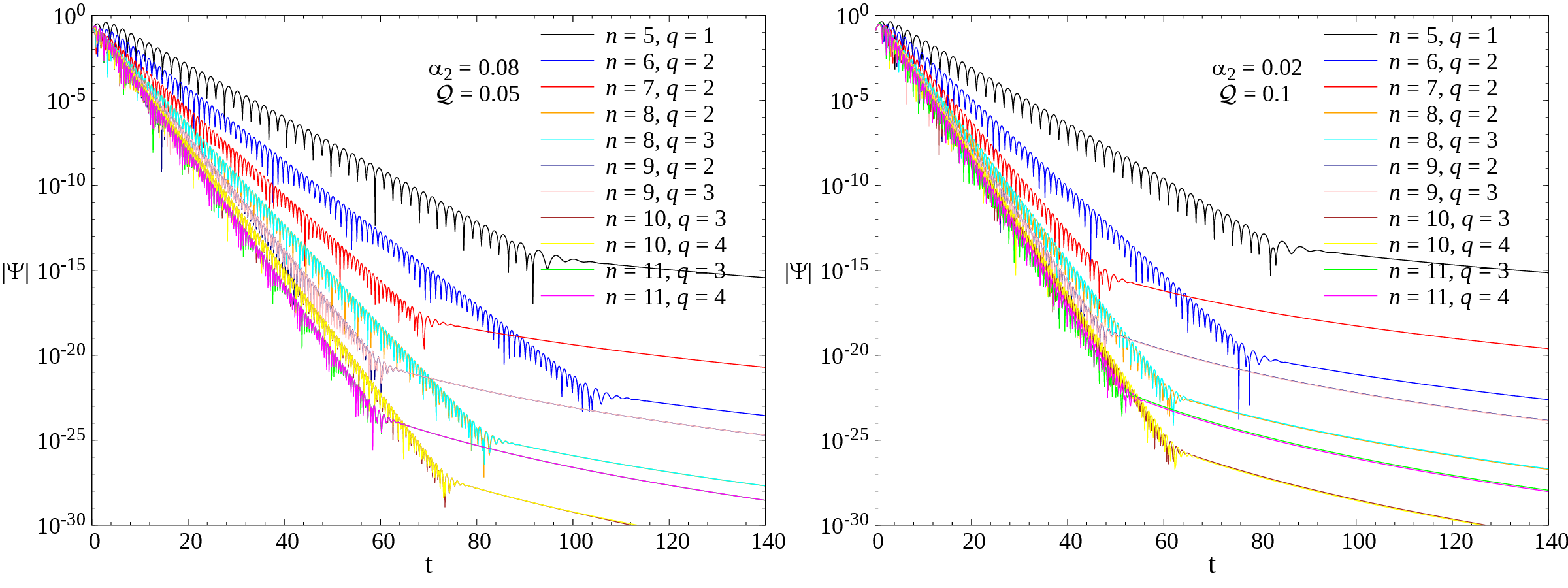}
\caption{
Time-domain profiles (plots of the ring-down waveforms and the late-time tails) of the spin-2 perturbations are governed by the effective potential $V_{\ell}^{\left(\mathrm{T}\right)}$ in the EPYMGB space-time. 
The parameters are chosen to be $\mathsf{M}=1$, $\mathscr{G}=1$, $\ell=2$, $\mathrm{n}=0$, $v_{c}=2$, and $\sigma=1$.
}
\label{figure13}
\end{figure}

\begin{figure}[htbp]
\centering
\includegraphics[width=1\textwidth]{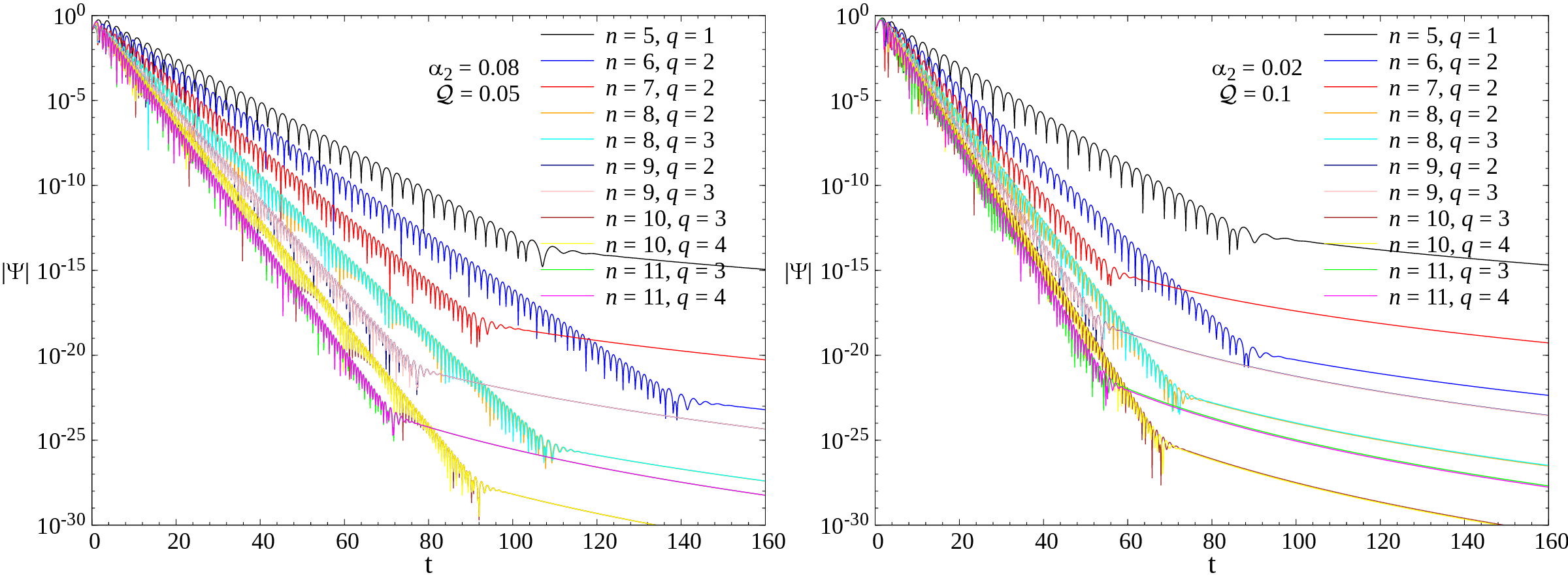}
\caption{
Time-domain profiles (plots of the ring-down waveforms and the late-time tails) of the spin-2 perturbations are governed by the effective potential $V_{\ell}^{\left(\mathrm{RW}\right)}$ in the EPYMGB space-time. 
The parameters are chosen to be $\mathsf{M}=1$, $\mathscr{G}=1$, $\ell=2$, $\mathrm{n}=0$, $v_{c}=2$, and $\sigma=1$.
}
\label{figure14}
\end{figure}

Next, we systematically compute the time-domain profiles for spin-0, spin-1, $p$-form, and spin-2 perturbations in various dimensions.
All the results are shown in Figures \ref{figure8}, \ref{figure9}, \ref{figure10}, \ref{figure11}, \ref{figure12}, \ref{figure13} and \ref{figure14}.
In these figures, we restrict our analysis to the multipole number $\ell=2$, and the width of the grid is selected as $\Delta=0.1$.
The left panels adopt an ${\alpha}_2$ values exceeding $\mathcal{Q}$ (${\alpha}_2 > \mathcal{Q}$), whereas the right panels use the opposite setting (${\alpha}_2 < \mathcal{Q}$) to establish a comparison benchmark.
The vertical coordinate ``$\left|\Psi\right|$'' is plotted on a logarithmic scale, while the horizontal coordinate ``$\mathrm{t}$'' maintains a linear scale in these graphical representations.
In numerical implementations, the time coordinate is typically rescaled by the horizon radius $r_{h}$, which scales as $r_{h}\propto\mathsf{M}^{1/(n-3)}$, so that the resulting time variable is dimensionless.
Therefore, we rescale the time coordinate as ${\mathrm{t}}/{\mathsf{M}^{1/(n-3)}}$ and set $\mathsf{M}=1$ \cite{Konoplya:2011qq}.
Furthermore, the perturbation field $\Psi$ is dimensionless, a property that follows from the linearity of the perturbation equation, which permits an arbitrary overall normalization of the field.
After significantly improving the accuracy, the late-time tail stages corresponding to each dimension can all be clearly manifested.

\begin{figure}[htbp]
\centering
\includegraphics[width=1\textwidth]{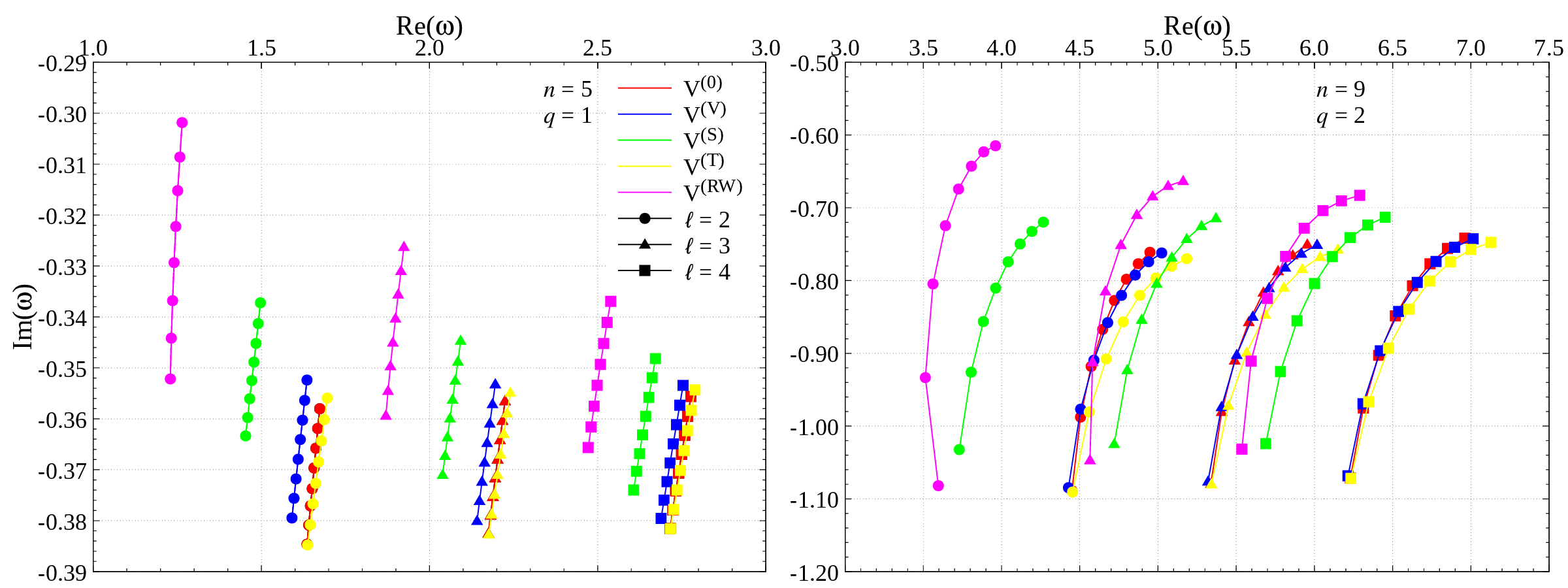}
\caption{
Influence of parameter $\alpha_{2}$ on the quasinormal frequencies corresponding to spin-0, spin-1 and spin-2 perturbations is analyzed in the special configuration of EPYMGB space-time.
The numerical results are calculated by the time-domain integration method.
The parameters are chosen to be $\mathsf{M}=1$, $\mathscr{G}=1$, $\mathcal{Q}=0.09$, $0 \leqslant\alpha_{2}\leqslant 0.07$, and $\mathrm{n}=0$.
The point $\alpha_{2}=0$ is located at the lower position, while the point $\alpha_{2}=0.07$ is at the upper position.
}
\label{figure15}
\end{figure}

\begin{figure}[htbp]
\centering
\includegraphics[width=1\textwidth]{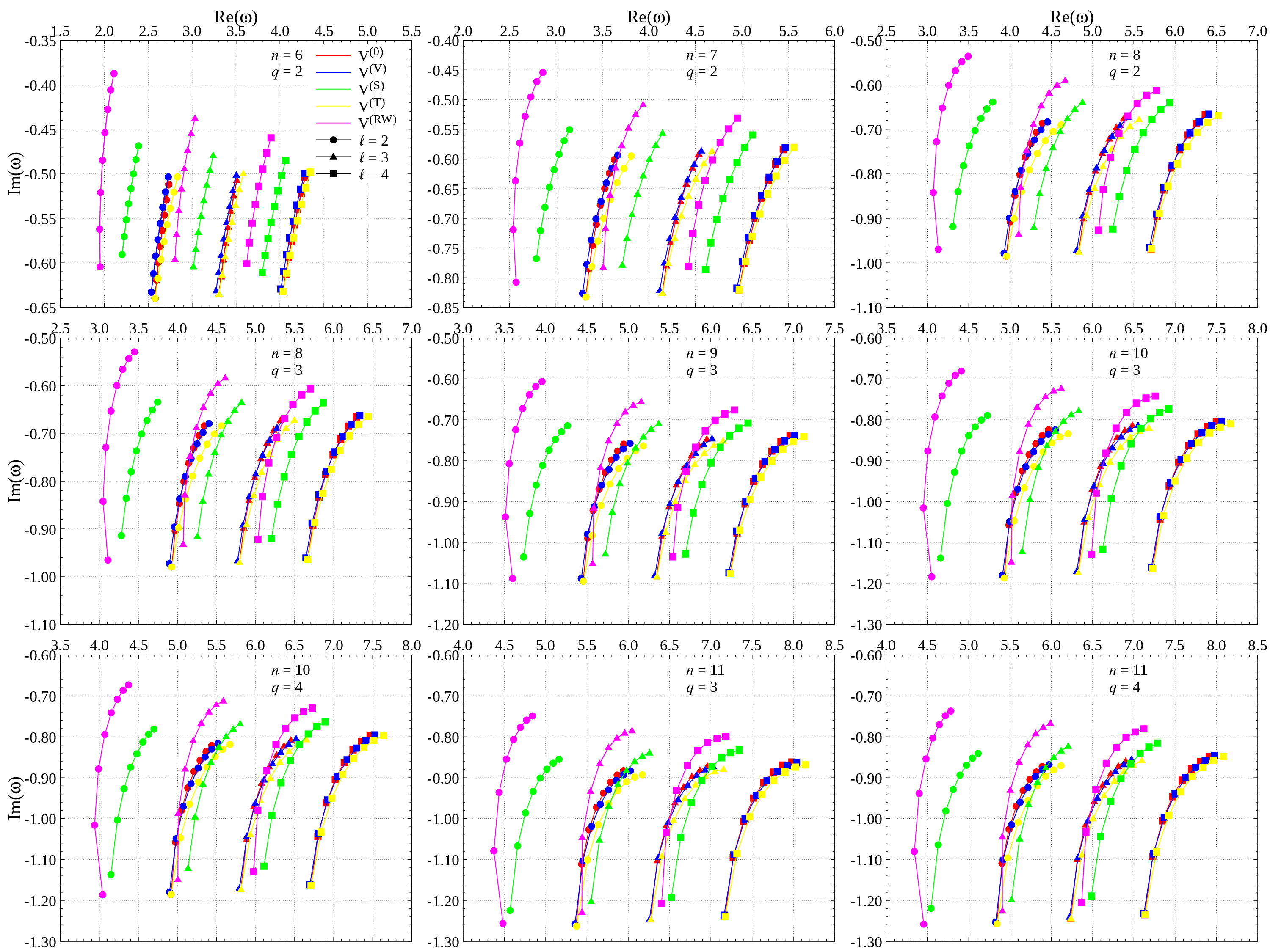}
\caption{
Similar to Figure \ref{figure15}, but with the metric corresponding to the general configuration of the EPYMGB space-time.
}
\label{figure16}
\end{figure}

\begin{figure}[htbp]
\centering
\includegraphics[width=1\textwidth]{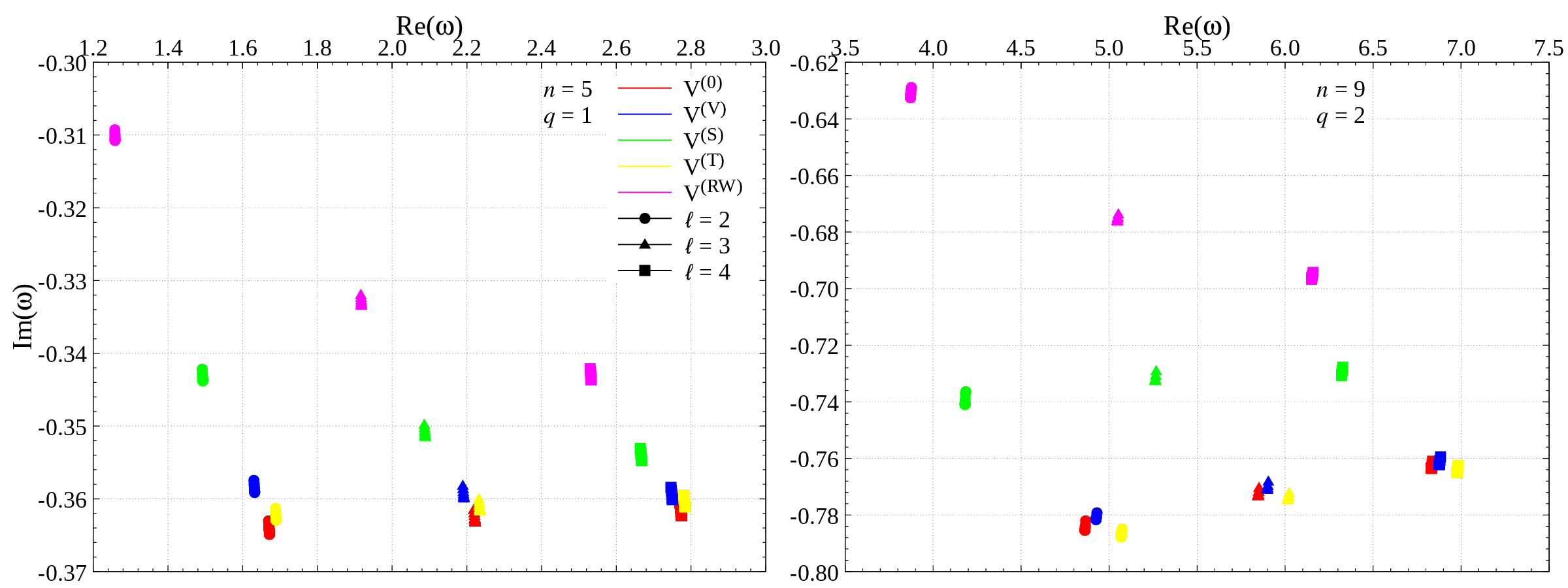}
\caption{
Influence of parameter $\mathcal{Q}$ on the quasinormal frequencies corresponding to spin-0, spin-1 and spin-2 perturbations is analyzed in the special configuration of EPYMGB space-time.
The numerical results are calculated by the time-domain integration method.
The parameters are chosen to be $\mathsf{M}=1$, $\mathscr{G}=1$, $\alpha_{2}=0.06$, $0 \leqslant\mathcal{Q}\leqslant 0.07$, and $\mathrm{n}=0$.
The point $\mathcal{Q}=0$ is located at the lower position, while the point $\mathcal{Q}=0.07$ is at the upper position.
}
\label{figure17}
\end{figure}

\begin{figure}[htbp]
\centering
\includegraphics[width=1\textwidth]{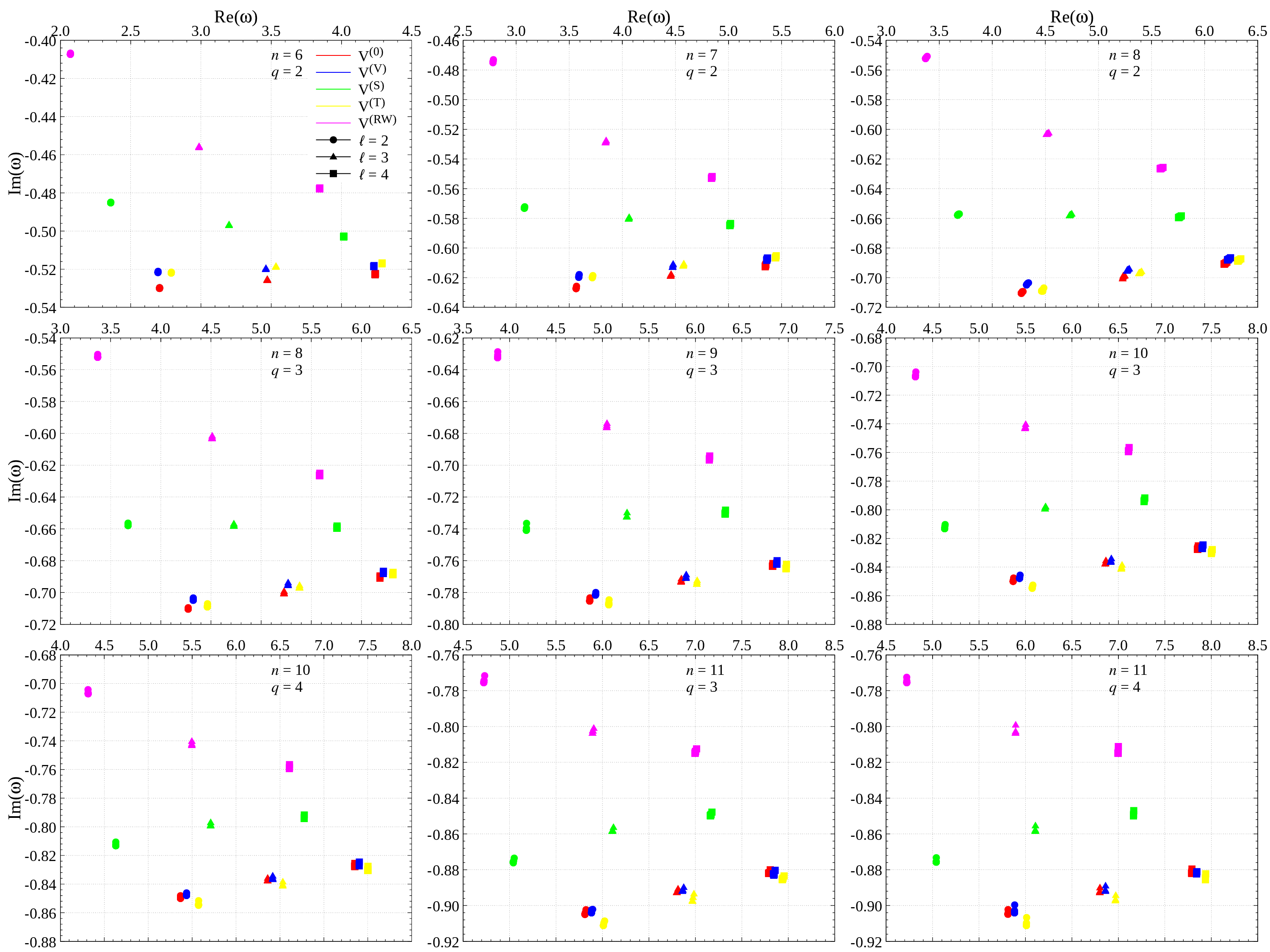}
\caption{
Similar to Figure \ref{figure17}, but with the metric corresponding to the general configuration of the EPYMGB space-time.
}
\label{figure18}
\end{figure}

To investigate the effects of the Gauss-Bonnet coupling constant ${\alpha}_2$ and the Yang-Mills magnetic charge $\mathcal{Q}$ on the real $\left(\text{Re}(\omega)\right)$ and imaginary $\left(\text{Im}(\omega)\right)$ parts of the QNMs, we computed a series of numerical results for the spin-0, spin-1 and spin-2 perturbations, as illustrated in Figures \ref{figure15}, \ref{figure16}, \ref{figure17} and \ref{figure18}.
It is worth noting that parameters ${\alpha}_2$ and $\mathcal{Q}$ were selected to vary within the same interval $\left[0,\,0.07\right]$.
Thus, it can be intuitively observed that parameter ${\alpha}_2$ has a substantially greater impact on the QNMs compared to parameter $\mathcal{Q}$.

\begin{figure}[htbp]
\centering
\includegraphics[width=1\textwidth]{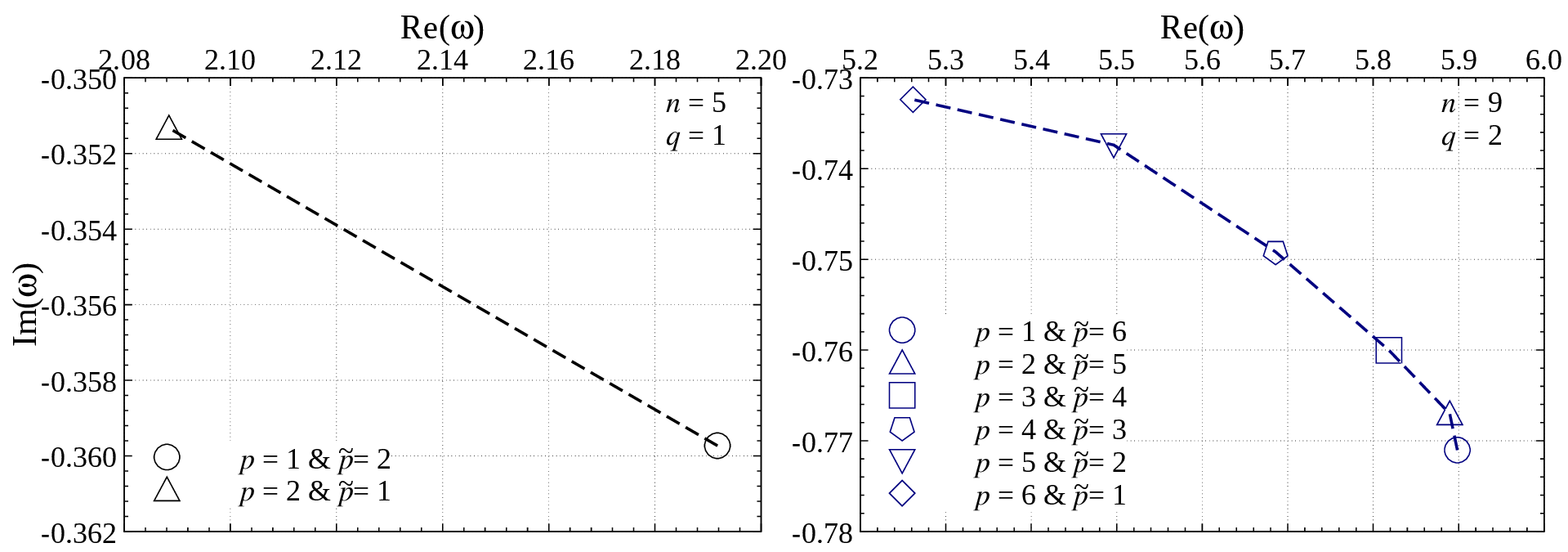}
\caption{
Influence of parameter $p$ on the quasinormal frequencies corresponding to $p$-form and $(p-1)$-form perturbations is analyzed in the special configuration of EPYMGB space-time.
The numerical results are calculated by the time-domain integration method.
The parameters are chosen to be $\mathsf{M}=1$, $\mathscr{G}=1$, $\alpha_{2}=0.06$, $\mathcal{Q}=0.02$, $\ell=3$, and $\mathrm{n}=0$.
}
\label{figure19}
\end{figure}

\begin{figure}[htbp]
\centering
\includegraphics[width=1\textwidth]{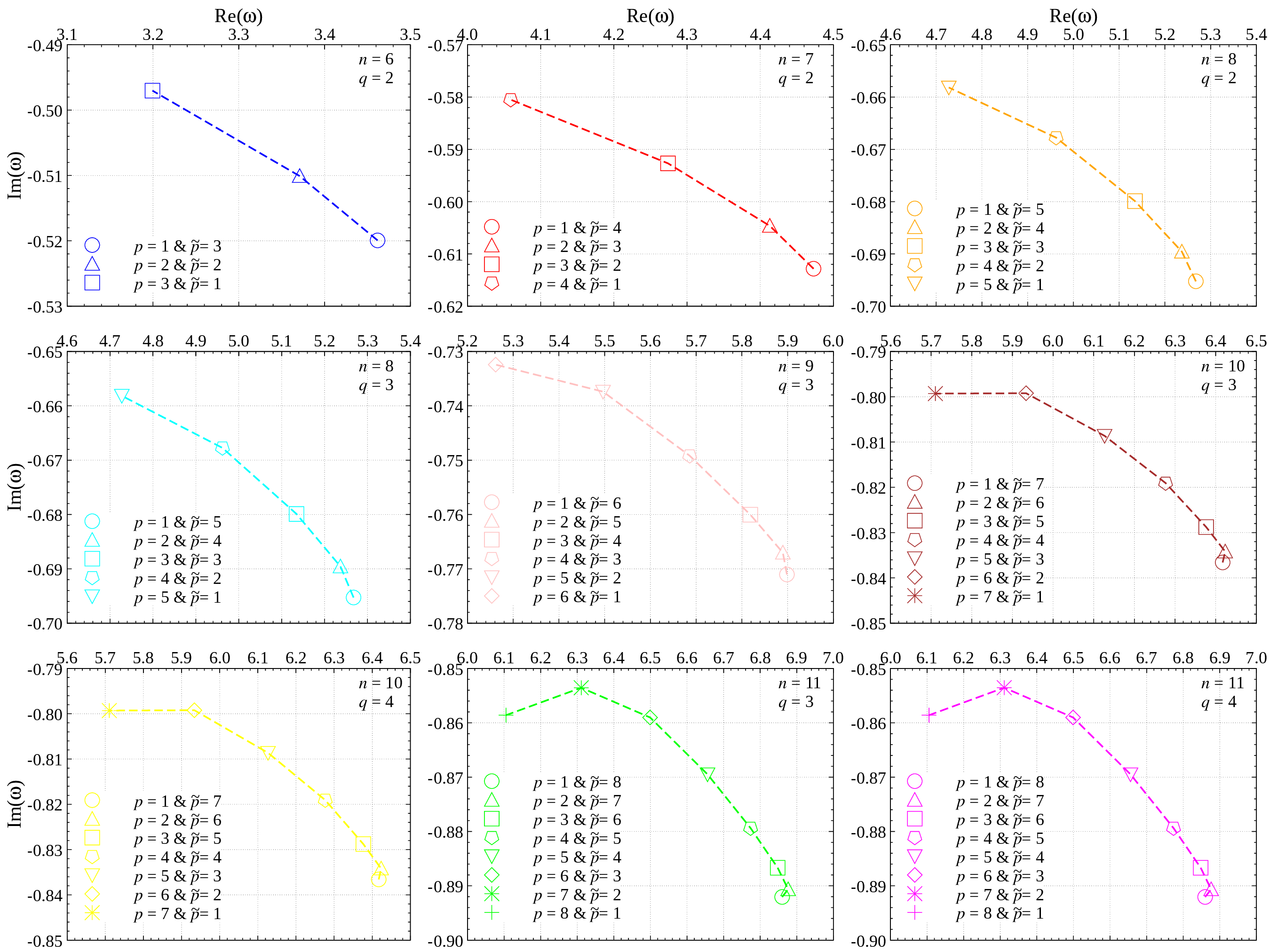}
\caption{
Similar to Figure \ref{figure19}, but with the metric corresponding to the general configuration of the EPYMGB space-time.
}
\label{figure20}
\end{figure}

Additionally, we examined the specific impact of parameter $p$ on the QNMs for both $p$-form and $(p-1)$-form perturbations, as illustrated in Figures \ref{figure19} and \ref{figure20}.
We select the allowable values of the parameter $p$ in each dimension $n$, namely $\left(1 \leqslant p \leqslant n-3\right)$, and mark the numerical results corresponding to these values with different symbols in the figures.

\begin{figure}[htbp]
\centering
\includegraphics[width=1\textwidth]{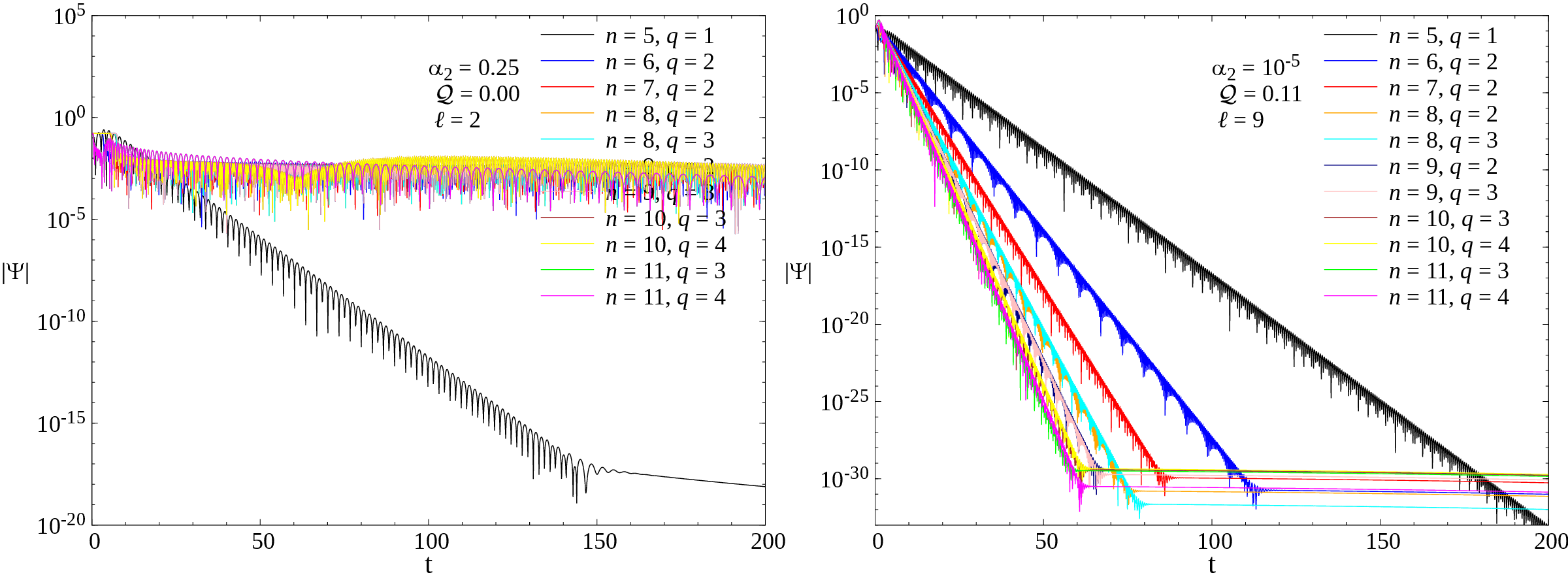}
\caption{
Time-domain profiles of the spin-2 perturbations are governed by the effective potential $V_{\ell}^{\left(\mathrm{T}\right)}$ in the EPYMGB space-time. 
The parameters are chosen to be $\mathsf{M}=1$, $\mathscr{G}=1$, $\mathrm{n}=0$, $v_{c}=2$, and $\sigma=1$.
}
\label{figure21}
\end{figure}

\begin{figure}[htbp]
\centering
\includegraphics[width=1\textwidth]{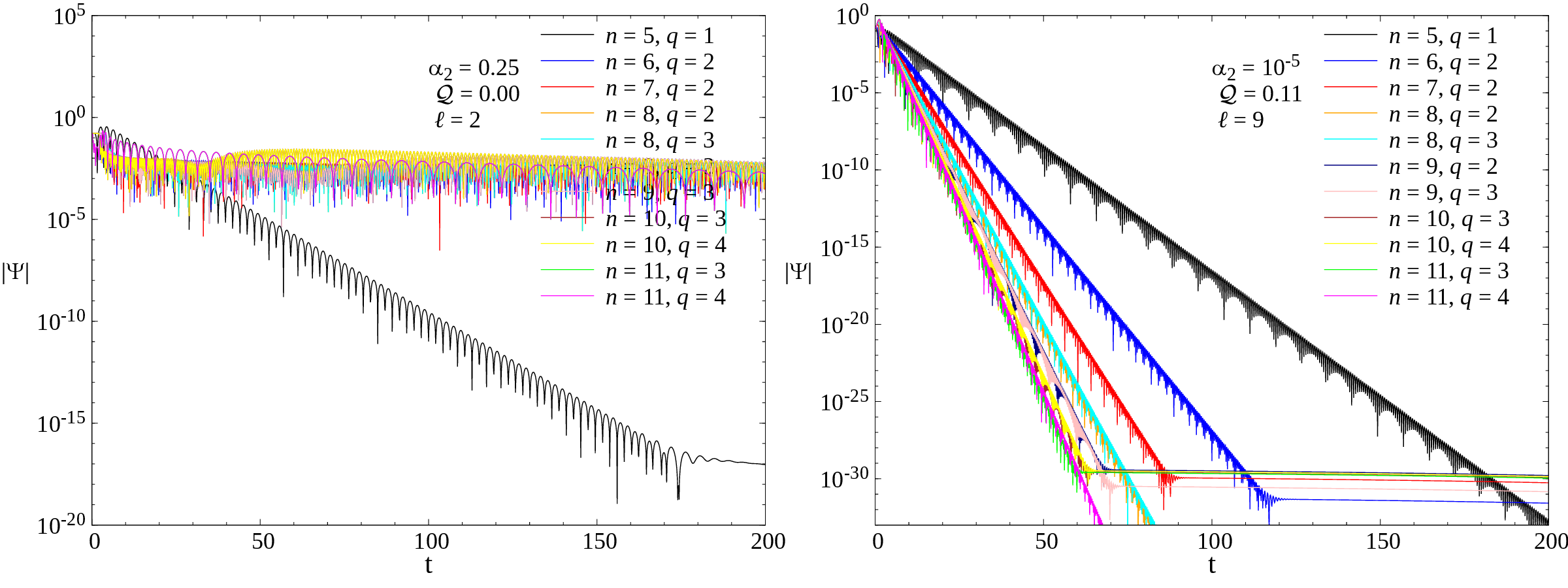}
\caption{
Time-domain profiles of the spin-2 perturbations are governed by the effective potential $V_{\ell}^{\left(\mathrm{RW}\right)}$ in the EPYMGB space-time. 
The parameters are chosen to be $\mathsf{M}=1$, $\mathscr{G}=1$, $\mathrm{n}=0$, $v_{c}=2$, and $\sigma=1$.
}
\label{figure22}
\end{figure}

Since larger values of the Gauss-Bonnet coupling constant ${\alpha}_2$ can lead to dynamical instability, we aim to identify an interval for ${\alpha}_2$ that ensures dynamical stability.
Since $\mathcal{Q}=0$ corresponds to the maximum domain of ${\alpha}_2$ values, we plotted the time-domain profiles of perturbations for the EGB black hole \eqref{eq:egb} with larger values of the parameter ${\alpha}_2$, as shown in the left panels of Figures \ref{figure21} and \ref{figure22}.
We further characterize the dynamical evolution through time-domain profiles of perturbed EYM black holes \eqref{eq:eym2}, as demonstrated in the right panels of Figures \ref{figure21} and \ref{figure22}.
We also presented the time-domain profiles corresponding to large $\ell$ values.
Since the grid width in  Eq. \eqref{eq:nwes} is set to $\Delta=0.1$, in order to ensure the effectiveness of accuracy, we select $\ell=9$ as the display.
Naturally, for larger values of $\mathcal{Q}$, perturbations also exhibit dynamical instability. 
However, since $\mathcal{Q}$ has only a minimal impact on the perturbations, we do not focus on identifying the stable region of the parameter $\mathcal{Q}$.
Therefore, we concentrate exclusively on the allowable range of the Gauss-Bonnet coupling constant ${\alpha}_2$ in the context of high-dimensional EGB black hole solutions.
By combining constraints from the high-dimensional shadow radius with the time-domain stability analysis of perturbations, we establish rigorously constrained parameter bounds, as comprehensively detailed in Table \ref{tb4:alpharange}.
Notably, the investigation into dynamical stability specifically targets gravitational perturbations (tensor modes and Regge-Wheeler modes) due to their greater physical relevance.
The new effective potentials employed in the gravitational perturbation analysis are given in Eqs. \eqref{eq:gtmvt} and \eqref{eq:gvmvrw}.
Our findings reveal perfect concordance between the constraint intervals of the tensor and Regge-Wheeler modes, with the derived parameter bounds showing complete insensitivity to variations in the parameters $q$ and $\ell$.
In Table \ref{tb4:alpharange}, the parameters are set to $\mathsf{M}=M=1$ and $\mathscr{G}=1$.

\begin{table}[tbh]\centering
\caption{
Quasinormal frequencies for the scalar field perturbation are determined via the WKB approximation, time-domain integration technique, and asymptotic iteration method (AIM).
The numerical results retain six significant digits.
The parameters are chosen to be $\mathsf{M}=1$, $\mathscr{G}=1$, ${\alpha}_{2}=0.08$, $\mathcal{Q}=0.05$, $\ell=2$, $\mathrm{n}=0$, and $m=0$.
\vspace{0.3cm}} \label{tb3:qnms3}
\begin{tabular*}{16cm}{*{4}{c @{\extracolsep\fill}}}
\hline
($n$, $q$) &  Time-domain & 13th WKB (Pad\'e) & AIM \\
\hline
($5$, $1$) & 1.68192 $-$ 0.356100 $i$ & 1.68162 $-$ 0.356965 $i$ & 1.68130 $-$ 0.356505 $i$ \\
($6$, $2$) & 2.75979 $-$ 0.496946 $i$ & 2.75729 $-$ 0.498183 $i$ & 2.75681 $-$ 0.498035 $i$ \\
($7$, $2$) & 3.67208 $-$ 0.584772 $i$ & 3.66502 $-$ 0.588699 $i$ & 3.66958 $-$ 0.584219 $i$ \\
($8$, $2$) & 4.41256 $-$ 0.675439 $i$ & 4.41263 $-$ 0.680080 $i$ & 4.39379 $-$ 0.676826 $i$ \\
($8$, $3$) & 4.40762 $-$ 0.675638 $i$ & 4.40759 $-$ 0.680503 $i$ & 4.38879 $-$ 0.676694 $i$ \\
($9$, $2$) & 5.00763 $-$ 0.758072 $i$ & 5.00352 $-$ 0.758743 $i$ & 4.98597 $-$ 0.778703 $i$ \\
($9$, $3$) & 5.00656 $-$ 0.758584 $i$ & 4.99925 $-$ 0.761088 $i$ & 4.98476 $-$ 0.779048 $i$ \\
($10$, $3$) & 5.51177 $-$ 0.828966 $i$ & 5.50400 $-$ 0.828299 $i$ & 5.49307 $-$ 0.865302 $i$ \\
($10$, $4$) & 5.51095 $-$ 0.829033 $i$ & 5.50370 $-$ 0.828276 $i$ & 5.49219 $-$ 0.865470 $i$ \\
($11$, $3$) & 5.95357 $-$ 0.887512 $i$ & 5.94422 $-$ 0.874403 $i$ & 5.94130 $-$ 0.936863 $i$ \\
($11$, $4$) & 5.95197 $-$ 0.887480 $i$ & 5.94269 $-$ 0.874567 $i$ & 5.93960 $-$ 0.936990 $i$ \\
\hline
\end{tabular*}
\end{table}

\begin{table}[tbh]\centering
\caption{Permissible range of the parameter ${\alpha}_2$ in the $n$-dimensional EGB black hole ($\mathcal{Q}=0$) is precisely determined by applying the shadow radius constraint and dynamical stability analysis.
\vspace{0.2cm}} \label{tb4:alpharange}
\begin{tabular*}{16cm}{*{4}{c @{\extracolsep\fill}}}
\hline
$n$ & $\mathrm{M87}^{\star}$ & $\mathrm{Sgr~A}^{\star}\;\;(\mathrm{Keck})$ & Stable area ($\left(\mathrm{T}\right)$ \& $\left(\mathrm{RW}\right)$) \\
\hline
5 & $0 < {\alpha}_2 \lesssim 0.373956$ & $0 < {\alpha}_2 \lesssim 0.326954$ & $0 < {\alpha}_2 \lesssim 0.299413$ \\
6 & $0 < {\alpha}_2 \lesssim 0.170079$ & $0 < {\alpha}_2 \lesssim 0.142480$ & $0 < {\alpha}_2 \lesssim 0.117488$ \\
7 & $0 < {\alpha}_2 \lesssim 0.131696$ & $0 < {\alpha}_2 \lesssim 0.106094$ & $0 < {\alpha}_2 \lesssim 0.0872425$ \\
8 & $0 < {\alpha}_2 \lesssim 0.123752$ & $0 < {\alpha}_2 \lesssim 0.0959430$ & $0 < {\alpha}_2 \lesssim 0.0888211$ \\
9 & $0 < {\alpha}_2 \lesssim 0.127387$ & $0 < {\alpha}_2 \lesssim 0.0950473$ & $0 < {\alpha}_2 \lesssim 0.109615$ \\
10 & $0 < {\alpha}_2 \lesssim 0.137810$ & $0 < {\alpha}_2 \lesssim 0.0989279$ & $0 < {\alpha}_2 \lesssim 0.155305$ \\
11 & $0 < {\alpha}_2 \lesssim 0.153603$ & $0 < {\alpha}_2 \lesssim 0.106040$ & $0 < {\alpha}_2 \lesssim 0.245830$ \\
\hline
\end{tabular*}
\end{table}

\section{Conclusions}
In this work, we investigate the exact spherically symmetric black hole solutions in Einstein-Gauss-Bonnet gravity coupled with power Yang-Mills fields.
Our focus is on two key aspects of this space-time geometry: its shadow phenomena and perturbation behavior.
We first identify the allowed region of the parameter space $\left({\alpha}_2, \mathcal{Q}\right)$ under the condition that the EPYMGB black hole possesses an event horizon.
We then elaborate on the derivation of the general formula for the shadow radius of a high-dimensional static spherically symmetric black hole.
Next, we employ the Schwarzschild-Tangherlini metric modeling to establish a novel framework that can constrain the characteristic parameters of high-dimensional black holes by leveraging the observational data of shadows.
Subsequently, we systematically present the effective potential equations governing spin-0, spin-1, $p$-form, and spin-2 perturbations within high-dimensional static spherically symmetric flat space-time.
Furthermore, we concisely outline three numerical methods utilized in determining the QNMs of black holes, with explicit focus on their implementation in spherically symmetric space-time.
Ultimately, we present the numerical results for shadow radii, quasinormal frequencies, time-domain profiles, and parameter constraints.
Following rigorous evaluation and critical analysis of these numerical results, we synthesize the core conclusions presented below:

\begin{enumerate}
\item[1.]
Through the application of the high-dimensional shadow constraint formula \eqref{eq:highdimrange} to EPYMGB black holes, we rigorously established that even in high-dimensional space-time frameworks, a physically viable domain of characteristic parameters remains definable.
We observe that the value of the shadow radius $R_{s}$ will decrease as the coupling constant ${\alpha}_2$ increases, while the valid interval of the coupling constant ${\alpha}_2$ contracts monotonically with increasing magnetic charge $\mathcal{Q}$.
Crucially, this valid interval will progressively shrink to fit within the admissible region $r_{\mathrm{sh}}$ defined by the high-dimensional shadow radius.
In other words, for each space-time dimension, the valid range of the parameter ${\alpha}_2$ in EGB black holes ($\mathcal{Q}=0$) precisely matches the maximum permissible range of ${\alpha}_2$ in EPYMGB black holes.
Overall, these results clearly indicate that under the high-dimensional shadow constraint formula, the characteristic parameters exhibit well-defined ranges limited by observational bounds, revealing the dimensional dependence of parameter constraints in extended space-time geometries.

\item[2.]
We cross-verified the high consistency of the values of QNMs calculated by three distinct numerical methods (the time-domain integration technique, the WKB approximation, and the AIM).
To address the inherent limitation of WKB in achieving higher-order convergence for complex black hole configurations, we implement the Pad\'e approximants technique, thereby resolving precision degradation in high-order term calculations.
Similarly, there are certain considerations to keep in mind when using the other two methods.
It is advisable to choose a larger value of multipole number ($\ell>1$) whenever possible for the time-domain integration method.
The selection of critical parameter values should be avoided to ensure stability and accuracy for the AIM.

\item[3.]
We provide a comprehensive display of the time-domain profiles for spin-0, spin-1, $p$-form, and spin-2 perturbations in various dimensions, from which we can intuitively obtain the following two significant features:
First, the perturbations exhibit distinct parity behavior between odd and even dimensions, with the ring-down stage of the even-dimensional perturbations persisting longer than that of the odd-dimensional perturbations.
Second, the influence of the power $q$ on the perturbations is minimal and can be regarded as negligible.
It is only in the regime of large multipole numbers $\ell$ that the perturbation duration (ring-down phase) for larger values of $q$ exceeds that for smaller $q$.
Furthermore, our analysis demonstrates that within EPYMGB space-time, spin-0, spin-1, $p$-form, and spin-2 perturbations exhibit dynamical stability in certain specific parameter regions across various spatiotemporal dimensions.

\item[4.]
Our comprehensive analysis confirms that gravitational perturbations in high-dimensional EGB space-time exhibit dynamical instability when the Gauss-Bonnet coupling constant ${\alpha}_2$ exceeds critical thresholds.
Notably, both the tensor modes and Regge-Wheeler modes share identical stability regions, with these domains demonstrating remarkable insensitivity to variations in the parameters $q$ and $\ell$.
Building on this fundamental condition, we further determine the critical bounds for the parameter ${\alpha}_2$ that maintain dynamical stability across different space-time dimensions $n$.
So far, we have compared the constraint ranges of ${\alpha}_2$ obtained from two approaches: the high-dimensional shadow radius constraint and dynamical stability analysis.
We find that the resulting ranges are in good agreement under both constraint frameworks.
Since the critical threshold for parameter instability is only weakly dependent on $\ell$, the present result may suggest that the duality between the shadow radius and eikonal perturbations (i.e., $\ell\gg1$) remains valid in high-dimensional space-times.

\item[5.]
Within equivalent parameter domains, the QNM spectrum exhibits pronounced sensitivity to the Gauss-Bonnet coupling constant ${\alpha}_2$, whereas maintaining negligible dependence on the Yang-Mills magnetic charge $\mathcal{Q}$.
This indicates that in EPYMGB space-time, the contribution of ${\alpha}_2$ to perturbations plays a dominant role, while $\mathcal{Q}$ exerts only a marginal influence.
Similarly, the effect of ${\alpha}_2$ on the black hole shadow radius is more pronounced than that of $\mathcal{Q}$.
This conclusively demonstrates that within the EPYMGB framework, the physical signature of the Yang-Mills magnetic charge $\mathcal{Q}$ in the black hole environment remains undetectable through either perturbation analysis or shadow observations.

\item[6.]
The influence of parameter $p$ on the QNMs for both $p$-form and $(p-1)$-form perturbations.
We can find that when $n=5,6,7,8,9$, the effect of parameter $p$ on the real and imaginary parts of the QNMs is monotonic, while when $n=10,11$, this effect becomes non-monotonic.
Crucially, the parameter $p$ exerts diametrically opposing effects on the value of QNMs in $p$-form and $(p-1)$-form perturbations.
This behavior stems from the inherent symmetry between the effective potentials \eqref{eq:v0vvvsvp1vp2} associated with these two perturbations within the admissible range of parameter $p$ $\left(1 \leqslant p \leqslant n-3\right)$.
\end{enumerate}

In summary, this work leads to several physical implications.
First, compared with the space-time dimensionality and the Gauss-Bonnet coupling constant, the characteristic parameters associated with non-Abelian fields are considerably more difficult to probe through shadow observations and perturbative analyses.
Second, we present a systematic investigation of various perturbations, which aids in comprehensively characterizing the dynamics of general types of fields in higher-dimensional space-times. 
Finally, the obtained parameter constraints imply that there exists a theoretical duality between the shadow radius and eikonal perturbations.
Recent studies further suggest that this nontrivial duality may extend beyond four-dimensional space-time and remain valid in higher-dimensional gravitational backgrounds \cite{Cuadros-Melgar:2020kqn, Moura:2021eln, Cai:2020igv}.
This issue will also be the focus of our future investigation.

\section*{Acknowledgements}
The author gratefully acknowledges the individuals, teams, and institutions that contributed to this research and expresses appreciation to the Shanghai Institute for Mathematics and Interdisciplinary Sciences (SIMIS) for providing valuable courses.

\section*{Data Availability Statement} 
All relevant data are within the paper.

\bibliographystyle{plain}

\bibliography{D-EPYMGB}

\end{document}